\documentclass[12pt]{article}
\usepackage{slashed}

\AtBeginDocument{
	\addtocontents{toc}{\normalsize}
}
\pdfoutput=1

\usepackage{amsmath,amssymb,amsfonts,amsthm,amscd,mathrsfs}
\usepackage{esvect}
\usepackage{xcolor}
\definecolor{darkblue}{rgb}{0.1,0.1,.7}
\usepackage[]{version}
\usepackage[]{graphicx}
\usepackage[]{latexsym}
\usepackage{geometry}
\usepackage[all,cmtip]{xy}

\usepackage[margin=10pt,font=small,labelfont=bf]{caption}
\usepackage{ifthen}
\usepackage{soul}
\usepackage{tikz}
\usepackage{array,mathrsfs,amsfonts,yfonts,dsfont,bbm,colonequals}

\usepackage{dsfont}
\usepackage{cite}
\usepackage{xspace}
\usepackage{empheq}
\usepackage{extarrows}

\usepackage{xcolor}
\usepackage[colorlinks, linkcolor=darkblue, citecolor=darkblue, urlcolor=darkblue, linktocpage]{hyperref} 

\usepackage{subcaption}
\usepackage[utf8]{inputenc}
\usepackage[nodisplayskipstretch]{setspace}
\usepackage{setspace}
\usepackage{float, graphicx}

\usepackage{enumitem}

\numberwithin{equation}{section}

\newcommand{\bD}{\textfrak{D}}
\newcommand{\bS}{\textfrak{S}}

\newcommand{\Dg}{\Delta_{\mbox{\tiny gap}}}
\newcommand{\Dir}{{\Delta_{\mbox{\tiny IR}}}}

\newcommand{\cO}{\mathcal O}

\newcommand{\reef}[1]{(\ref{#1})}
\newcommand{\be}{\begin{equation}}
\newcommand{\ee}{\end{equation}}
\newcommand{\bea}{\begin{eqnarray}}
\newcommand{\eea}{\end{eqnarray}}
\newcommand{\ba}{\begin{equation}\begin{aligned}}
\newcommand{\ea}{\end{aligned}\end{equation}}

\newcommand{\ud}{\mathrm d}

\newcommand{\Df}{{\Delta_\phi}}

\def\Xint#1{\mathchoice
   {\XXint\displaystyle\textstyle{#1}}%
   {\XXint\textstyle\scriptstyle{#1}}%
   {\XXint\scriptstyle\scriptscriptstyle{#1}}%
   {\XXint\scriptscriptstyle\scriptscriptstyle{#1}}%
   \!\int}
\def\XXint#1#2#3{{\setbox0=\hbox{$#1{#2#3}{\int}$}
     \vcenter{\hbox{$#2#3$}}\kern-.5\wd0}}

\def\dashint{\Xint-}

\def\g{\gamma}
\def\r{\rho}

\def\a{\alpha}

\def\k{\kappa}

\def\D{\Delta}
\def\G{\Gamma}

\def\ta{\tau}

\def\G{\Gamma}

\newcommand{\tE}{{\tt E}}
\newcommand{\tL}{{\tt L}}
\newcommand{\tS}{{\tt S}}
\newcommand{\tD}{{\tt D}}
\newcommand{\tH}{{\tt H}}

\usepackage{cancel}

\newcommand{\ahf}{a_h^{\mbox{\tiny free}}}

\begin{document}
	
	\vspace*{-.6in} \thispagestyle{empty}
	\begin{flushright}
	\end{flushright}
	\vspace{1cm} {\Large
		\begin{center}
			{\bf Solving 1D crossing and QFT$_2$/CFT$_1$}\\
	\end{center}}
	\vspace{1cm}
	\begin{center}
		{\bf Kausik Ghosh$^{\alpha}$, \bf Miguel F.~Paulos$^{\beta}$, No\'e Suchel$^{\theta}$}\\[1cm] 
		{
			\small
			{\em 
            ${}^{\alpha}$Department of Mathematics, King's College London, Strand, London, WC2R 2LS, United Kingdom.}\\\vspace{0.3cm}
            ${}^{\beta,\theta}${\em Laboratoire de Physique, \'Ecole Normale Sup\'erieure, \\
   Universit{\'e} PSL, CNRS, Sorbonne Universit{\'e}, Universit{\'e} Paris Cit{\'e}, \\
   24 rue Lhomond, F-75005 Paris, France}

			\normalsize
		}
		
	\end{center}
	
	\begin{center}
		{\texttt{\  ${}^{\alpha}$kau.rock91@gmail.com} 
		},{\texttt{\ ${}^{\beta}$miguel.paulos@ens.fr} 
		},{\texttt{\ ${}^{\theta}$noe.suchel@ens.fr} 
		}
	\end{center}
	
	\vspace{8mm}
	
	\begin{abstract}
 \vspace{2mm}
   We provide an effective solution of the 1D crossing equation. We begin by arguing that crossing constraints can be recast in terms of bases of sum rules associated to special sets of CFT data -- extremal solutions -- which solve these constraints in a minimal way and naturally saturate positivity bounds on the space of CFTs. We conjecture, argue and check extensively that any extremal solution behaves as a generalized free field in the UV. This allows us to reconstruct the entirety of their CFT data using a rapidly convergent ``hybrid bootstrap'' method, which combines numerics and analytics. Strikingly, as we approach special corners in the space of extremal solutions we find that their CFT data can present non-trivial structure up to arbitrarily large energies. We interpret these corners as flat space limits of QFTs in AdS$_2$, which extremal solutions naturally describe. This picture allows us to bootstrap their CFT data in these limits in terms of 2d S-matrices, and conversely provide a microscopic CFT construction of the latter. Further evidence for this QFT in AdS description of extremal solutions comes from an explicit construction of bulk QFT operators solving an AdS locality problem. Concretly we show that it is possible to canonically associate one or more such operators to any extremal solution by explicitly solving for their BOE data. In the special case where this operator is the bulk stress-tensor we combine crossing and bulk locality constraints to derive stronger bounds on the OPE and BOE data, including an exact bootstrap lower bound on the central charge $C_T\geq 1/2$.   
	\end{abstract}
	\vspace{2in}

	\newpage
	
	{
		\setlength{\parskip}{0.05in}
		\tableofcontents
		\renewcommand{\baselinestretch}{1.0}\normalsize
	}
	
	
	\setlength{\parskip}{0.1in}
 \setlength{\abovedisplayskip}{15pt}
 \setlength{\belowdisplayskip}{15pt}
 \setlength{\belowdisplayshortskip}{15pt}
 \setlength{\abovedisplayshortskip}{15pt}
 
	
	\bigskip \bigskip
    
\section{Introduction}
The principles of holography tell us that CFT spectra of scaling dimensions encode energies of quantum theories in AdS. Thus cutting open a typical CFT correlator we generically expect to find a horrendous chaotic, exponentially dense set of states exchanged in the OPE. The bootstrap program bravely attempts to reconstruct such correlators from simple assumptions, the most important of which is crossing symmetry. But how constraining is crossing? More to the point, is it sufficient to fix the exponentially dense CFT data contained in a typical correlator?

In this paper we examine this question in the context of 1d CFTs. Quite generally such CFTs arise as boundary duals\footnote{The word `dual' might irk some readers, as in such cases we don't have a nice local description of the CFT independent of the bulk as in usual AdS/CFT. But in a sense we do as the CFT can be defined by the bootstrap axioms. We would argue that this work provides evidence that this picture can be made concrete.
} of QFTs in spaces containing an AdS$_2$ factor\footnote{This includes the case where the QFT is a CFT, so that by a Weyl rescaling the 1d CFT could describe a conformal line defect, including the trivial one, i.e. dimensional reduction of a higher-D CFT to a line.}. We focus on the simplest bootstrap problem in these simplest of all CFTs: understand crossing for a single correlator of identical operators. What we find is that in this case crossing is not all-powerful: it has just enough constraining power to determine the CFT data of a single tower of ``double trace'' operators in the OPE, but no more. These act as a basis of states sufficiently flexible to soak up the constraints of crossing no matter which further states are contained in the OPE.

\begin{figure}[ht]
\centering
   
        \centering
       \includegraphics[width=0.8 \textwidth]{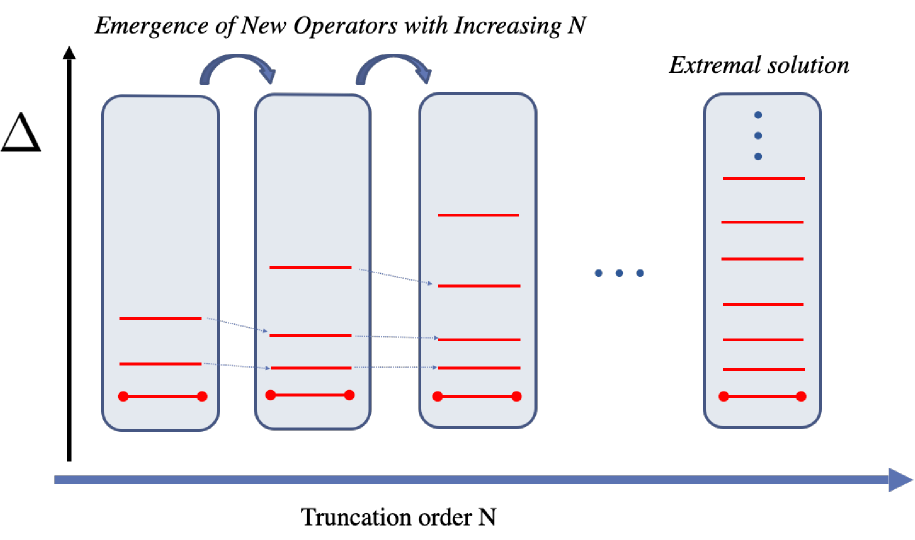}
        \caption{Extremal solutions. By saturating a CFT bound using $N$ crossing constraints we find approximate solutions with $O(N)$ operators. In this work we will provide methods for bootstrapping the $N\to \infty$ limit of this construction.
        \label{fig:extremal spectrum}
 }
    \hfill
    
\end{figure}

This begs the question: which CFT correlators {\em are} completely constrained by crossing? We call such correlators {\em extremal}. They include all CFT correlators which arise as the optimal solutions of bootstrap numerical optimisation problems (cf. figure \ref{fig:extremal spectrum}), but also more general non-unitary solutions to crossing \cite{ElShowk:2013,El-Showk:2016mxr}. 
Their defining property is that they contain the single tower of operators that we just mentioned, as well as a finite set of states that one must input by hand and which {\em define} the solution. In this work we will give an essentially complete characterization of extremal solutions, as well as efficient methods for computing the entirety of their CFT data, so that we have now an effective solution of the 1d crossing equation. \footnote{You can see these methods in action for yourself at
\url{https://cft.starfree.app/} -- a  webapp developed by Eliott Morgensztern.
}

There are two key results that allow us to do this. Firstly, based on extensive observations we make a non-trivial conjecture that all extremal solutions asymptote to a generalized free field. Denote by $\Delta^*$ the scale at which this transition occurs. We will exploit this knowledge of the UV to derive a ``hybrid bootstrap'' method for reconstructing the entire extremal solution \footnote{Our method is reminiscent, but distinct, from a proposal by N. Su~\cite{Su:2022xnj}. In particular our approach involves treating the high energy, not high spin sector, analytically. More importantly our construction still leads to rigorous bounds on all CFTs.
}. The method combines numerical solution of the data of states with $\Delta<\Delta^*$ with an analytic solution of the (infinite) states with $\Delta>\Delta^*$. Generic extremal solutions have $\Delta^*=O(1)$ and this method works extremely well. 

The second result has to do the interpretation for extremal solutions as arising from QFTs in AdS$_2$. We show that the flat space limit of such QFTs can be described by going to special corners in the moduli space of extremal solutions. In these corners $\Delta^*\to \infty$ and the hybrid method becomes difficult to apply. However, what happens now is that the extremal CFT data is well described by smooth functions. Indeed we show that the extremal data can now be packaged in a flat space S-matrix, describing scattering in the bulk QFT. Bootstrapping an extremal solution then amounts to constructing a unitarity saturating S-matrix. Conversely we can use our knowledge of such S-matrices to helps us reconstruct the CFT data in these hard to reach corners of moduli space, in what we name the ``S-matrix method'' for constructing extremal CFT correlators.

If an extremal solution is to have a QFT in AdS interpretation then it should be possible to reconstruct operators that behave locally in the AdS bulk. Thanks to previous work \cite{Levine:2023ywq,Levine:2024wqn}, we show that indeed to any extremal solution one can canonically associate one or more AdS operators which are relatively local to the extremal OPE. Schematically:
\ba
\langle \phi \phi \phi \phi\rangle\quad \mbox{extremal} \qquad \Longrightarrow \qquad \langle \phi \phi \,\Phi_{\mbox{\tiny AdS}}\rangle\quad \mbox{local}
\ea
As mentioned already in that work, this is far from a trivial statement, as it requires a delicate matching between bulk locality and crossing constraints.

While this is always possible generically, extra requirements on the bulk operator can lead to non-trivial constraints on the extremal solution. In this work we show for the first time  that crossing and bulk locality constraints can be combined to obtain stronger bounds on OPE and BOE data. This is a direct CFT analog of similar work done in the context of the S-matrix/form factor bootstrap (and to which it maps to in the flat space limit) \cite{Karateev:2019ymz}. Our construction is possible thanks to the harder UV behaviour of the crossing equation functional bases that we consider, as compared to the standard derivative basis. In particular we derive bounds on OPE and BOE data, as well as the UV central charge of the bulk theory, by setting $\Phi$ to be the bulk stress-tensor, including an exact bootstrap lower bound $C_T\geq 1/2$ under certain assumptions on the OPE.

The outline of this paper is as follows. 

\begin{itemize}
\item In section 2 we set up the bootstrap problem as well as various sets of complete bases of crossing constraints. Among these we define extremal bases and their associated extremal solutions which we conjecture to have universal asymptotic behaviour. 

\item This conjecture is used in section 3 to derive a hybrid bootstrap method for efficiently constructing any extremal solution. This method relies on the conjectured asymptotics to solve for CFT data numerically at low energies and analytically at high energies. 

\item In section 4 we give a QFT in AdS interpretation to extremal solutions. Most notably we show that the moduli space of extremal solutions includes special corners suitable for describing the flat space limit of such QFTs. In these corners the CFT data is slowly varying and can be described by a bulk flat space S-matrix. We prove this directly follows from crossing constraints, and use it to solve these constraints efficiently in these limits. We also show that local bulk operators in AdS can be canonically reconstructed from an extremal solution and determine their form factors with boundary insertions in the flat space limit. 

\item Section 5 is concerned with several applications of the methods described in the previous sections. We study in detail two families of extremal solutions: a scalar in AdS$_2$ with a $\lambda \Phi^4$ interaction, and a fermion coupled to a pseudoscalar via a Yukawa interaction. We reconstruct the spectrum of the solutions and study them in the flat space limit, recovering both S-matrices and operator form factors, with macroscopic parameters such as resonance masses determined by the microscopic CFT data.

\item In section 6 we investigate numerical bootstrap bounds which combine crossing and bulk locality constraints into a larger semidefinite problem. In particular, by reconstructing the bulk stress-tensor we are able to derive an exact lower bound on the bulk UV central charge. We also show that it is possible to derive bounds on the combined OPE/BOE data of operators of interest.
\end{itemize}
The paper is complemented by several technical appendices.

\section{Extremal Solutions to Crossing}     
\label{sec2}

\subsection{Functional bases and sum rules}
We begin by laying out some basic definitions and notation. We are interested in 1d CFT four point functions of identical operators $\phi$ of dimension $\Df$, which we write as:
\ba
\langle \phi(\infty)\phi(1)\phi(z)\phi(0)\rangle=\mathcal G(z)
\ea
The OPE allows us to write the correlator $\mathcal G$ as a sum of conformal blocks,
\ba
\mathcal G(z)=\sum_{\Delta} a_{\Delta} G_{\Delta}(z)\,, \qquad G_{\Delta}(z)&=z^{\Delta-2\Df}~ _2F_1(\Delta,\Delta,2\Delta,z)
\ea
where we defined $a_{\Delta}:=\lambda^2_{\phi \phi \mathcal O_{\Delta}}$ as the square of OPE coefficients. Crossing symmetry of the correlator can be written as.
\ba
\mathcal G(z)=\mathcal G(1-z) \qquad \Leftrightarrow \qquad \sum_{\Delta} a_{\Delta} F_{\Delta}(z)=0\,,\label{eq:crossing}
\ea
where we defined the crossing vector
\ba
F_{\Delta}(z)=G_{\Delta}(z)-G_{\Delta}(1-z)\,.
\ea
The crossing equation presents us with a continuously infinite set of constraints on a continuously infinite set of CFT data. But linearity of the equation in the $a_{\Delta}$ suggests introducing a (countable) basis in the space of crossing vectors,
\ba
F_{\Delta}(z)=\sum_{n=0}^\infty \omega_n(\Delta)\, B_{n}(z)\,\,,
\ea
where the $\omega_n$ are elements of the dual space, i.e. linear functionals,
\ba
\omega_n[B_m]=\delta_{n,m}\,, \qquad  \omega_n(\Delta):=\omega_n[F_{\Delta}]\,. \label{eq:duality}
\ea
In terms of this basis the crossing equation becomes equivalent to a set of sum rules on the CFT data:
\ba
\mathcal G(z)=\mathcal G(1-z) \qquad \Leftrightarrow\qquad \sum_{\Delta} a_{\Delta} \omega_n(\Delta)=0\,, \qquad n\in \mathds N_{0}\,.\label{eq:sumrules1}
\ea
There are many possible choices of bases. In typical bootstrap applications one sets~\cite{Rattazzi:2008pe}
\ba
B_n(z)=\frac{1}{n!}(z-z_0)^{1+2n}\,, \qquad \omega_n(\Delta)=\partial_z^{1+2n} F_{\Delta}(z_0)\,, \qquad z_0=\tfrac 12\,.
\ea
This choice has the advantages that it is simple and is manifestly a basis. However, the resulting sum rules are as difficult to solve as the initial crossing equation. In particular, if we are given a solution to crossing it is non-trivial to determine that it does indeed satisfy the infinite set of sum rules \reef{eq:sumrules1}. In the following section we will show that there exist special bases for which we can check they admit explicit solutions.

\subsection{Extremal bases}
A natural idea is to choose bases elements to be crossing vectors themselves. That is we ask for decompositions of the form:
\ba
F_{\Delta}(z)=\sum_{n=0}^\infty \tau_n(\Delta) F_{\Delta_n}(z)
\ea
There are at least two reasons why such decompositions are interesting. Firstly, the above can be interpreted as providing us with infinitely many solutions to the crossing equation, labeled by $\Delta$. Secondly, the duality conditions on the $\omega_n$ now tells us something about the functional actions:
\ba
\tau_n(\Delta_m)=\delta_{n,m}.
\ea
This provides us with a measure of control of the functional actions which we can exploit to our advantage. These arguments lead us to the following definition:

\setlength\fboxrule{1.2pt}
\setlength{\fboxsep}{8pt} 
{\hspace{-12pt}\noindent\fbox{\parbox{\textwidth-30pt}{\vspace{-6pt}\vspace{0.3cm}
{\bf Single-zero basis}\\

A {\em single-zero basis} determined by a {\em basic spectrum} $\bS:=\{\Delta_n\}_{n=0}^\infty$ is a collection of functionals $\{\tau_n^\bS\}_{n=0}^\infty$, such that the following properties hold:
\begin{itemize}
\item{\bf Duality}
\ba
\tau_n^\bS(\Delta_m)=\delta_{n,m}\,, \qquad n,m\geq 0
\ea
\item{\bf Completeness}
\ba
\mbox{Crossing} \qquad \Leftrightarrow \qquad \sum_{\Delta} a_{\Delta} \tau_n^\bS(\Delta)=0\quad n\geq 0
\ea
\end{itemize}
}
}
}
\vspace{0.3 cm}

\noindent
In this definition, we should think that {\em duality} determines functionals given a sequence $\Delta_n$, while {\em completeness} then constrains the allowed sets of $\Delta_n$. As we already mentioned, to any such basis we can associate a canonical family of solutions to crossing which we call {\em crossing blocks}:
\ba
\mathcal Q_{\Delta}^{\bS}(z)=\mathcal Q_{\Delta}^{\bS}(1-z)\,, \qquad \mathcal Q_{\Delta}^{\bS}(z):=G_{\Delta}(z)-\sum_{n=0}^\infty \tau_n^{\bS}(\Delta) G_{\Delta_n}(z)\,.
\ea
Using completeness, crossing symmetry of an arbitrary correlator can now be recast as
\ba
\mathcal G(z)=\mathcal G(1-z) \quad \Leftrightarrow \quad \mathcal G(z)=\sum_{\Delta} a_{\Delta} \mathcal Q_{\Delta}^{\bS}(z) 
\ea
There is a particularly important case of basis which arises as a degeneration of the above where an infinite number of pairs $\Delta_n,\Delta_{n+1}$ collide. In this case the functionals $\tau_n$ develop double rather than single zeros and it is convenient to go to a new basis of functionals,
\ba
\alpha_n\sim \frac{\tau_n+\tau_{n+1}}2 \,, \qquad \beta_n\sim\lim_{\Delta_{n+1}\to \Delta_n}\,\frac{\tau_{n+1}-\tau_n}{\Delta_{n+1}-\Delta_n}
\ea
The most useful cases correspond to when either all pairs collide, or when all but the first do. This leads to the following definitions:
\vspace{0.5cm}

\setlength\fboxrule{1.2pt}
\setlength{\fboxsep}{8pt} 
{\hspace{-12pt}\noindent\fbox{\parbox{\textwidth-30pt}{\vspace{-6pt}\vspace{0.3cm}
{\bf Fermionic/Bosonic double-zero basis}\\

A {\em fermionic double-zero basis} determined by a {\em basic spectrum} $\bD:=\{\Delta_n\}_{n=0}^\infty$ is a collection of functionals split into two sets, $\{\alpha_n^\bD\}_{n=0}^\infty\, \cup \,\{\beta_n^\bD\}_{n=0}^\infty$, such that the following properties hold:
\begin{itemize}
\item{\bf Duality}
\ba
\alpha_n^\bD(\Delta_m)&=\delta_{n,m}&\qquad \partial_\Delta\alpha_n^{\bD}(\Delta_m)&=0\\
\beta_n^\bD(\Delta_m)&=0&\qquad \partial_\Delta\beta_n^{\bD}(\Delta_m)&=\delta_{n,m}\\
\ea
\item{\bf Completeness}
\ba
\mbox{Crossing} \qquad \Leftrightarrow \qquad \begin{array}{ll} 
\sum_{\Delta} a_{\Delta} \alpha_n^\bD(\Delta)=0\\
\vspace{0.2 cm}\\
 \sum_{\Delta} a_{\Delta} \beta_n^\bD(\Delta)=0
\end{array}
\ea
\end{itemize}
A {\em bosonic double zero basis} satisfies almost exactly the same properties, except that the functional $\beta_0$ is identically zero. This requires dropping the righthand side of the duality conditions for $m=0$. 
}
}
}
\vspace{0.3 cm}

\noindent
For a double-zero basis, the crossing vector decomposition formula is now
\ba
F_{\Delta}(z)=\sum_{n=0}^\infty\left[ \alpha_n^{\bD}(\Delta)\, F_{\Delta_n}(z)+\beta_n^{\bD}(\Delta) \partial_{\Delta} F_{\Delta_n}(z)\right]
\ea
and accordingly the crossing block takes the form\footnote{For special sets of $\Delta_n$ these are nothing but Polyakov blocks \cite{Gopakumar:2016wkt,Gopakumar:2016cpb,Gopakumar:2018xqi,Mazac:2018ycv}.}
\ba
\mathcal Q^{\bD}_{\Delta}(z)=G_\Delta(z)-\sum_{n=0}^\infty\left[\alpha_n^{\bD}(\Delta)\, G_{\Delta_n}(z)+\beta_n^{\bD}(\Delta) \partial_{\Delta} G_{\Delta_n}(z)\right]\,.\label{eq:polyblock}
\ea
As their name indicates, double-zero functional actions $\alpha_n^{\bD}(\Delta),\beta_n^{\bD}(\Delta)$ have double zeros at prescribed locations. This makes it more likely for such functionals to be sign definite across extended ranges. Such sign definiteness property is crucial in unitary bootstrap applications where one wishes to determine bounds from positivity. We will discuss this further at the end of this subsection. 

Up to now the sets $\Delta_n$ are only constrained by the demand of completeness. This still allows for infinitely many possible choices. We thus need to introduce some new set of constraints that will uniquely specify preferred sets of $\Delta_n$. This leads us to the final set of bases that we will consider:
\vspace{0.3cm}

\setlength\fboxrule{1.2pt}
\setlength{\fboxsep}{8pt} 
{\hspace{-12pt}\noindent\fbox{\parbox{\textwidth-30pt}{\vspace{-6pt}\vspace{0.3cm}
{\bf Extremal basis/solution}\\

Consider a {\em fixed} set of CFT data $\tt S$:
\ba
\tS&=\{(\hat \Delta_i,\hat a_{i})\,, \quad i=1,\ldots K\}
\ea
The {\em extremal basis} labeled by $\tS$ is a double-zero basis $\bD=\bD(\tS)$ such that
\ba
\beta_n^\bD[\tS]:=\sum_{i=1}^K \hat a_i \beta_n^\bD(\hat \Delta_i)=0 \label{eq:extbeta}
\ea

The {\em extremal solution} determined by $\tt S$ is then the set of CFT data,
\ba
\tt E=\tt S\cup \tt D\,, \qquad \tD&=\{(\Delta_n^\tE,a_{n}^{\tE})\,, \quad n=0,\ldots \infty\}
\ea
with 
\ba
\Delta_n^{\tE}=\Delta_n^{\bD}\,, \qquad a_n^\tE=-\alpha_n^{\bD}[\tS] \label{eq:extalpha}
\ea
If all data in $\tS$ is specified then $\bD$ is fermionic. But it is possible to leave one of the $\hat a_i$ in $\tS$ undetermined, say $\hat a_1$, which we would denote as $(\hat \Delta_1,\bullet)\in \tS$. In that case $\bD$ is of bosonic type and we set
\ba
\Delta_0^\bD:=\hat\Delta_1\,, \qquad \hat a_1:=a_0^\tE=-\alpha_0^\bD[\tS] 
\ea
}
}
}
\vspace{0.3cm}

\noindent

A generic double zero basis comes bundled with the associated crossing blocks, which are special solutions containing essentially only the basic spectrum. The problem with such solutions is that they generically  include derivatives of conformal blocks. In contrast, the special property of an extremal basis is that it admits a unique special solution that does not have these derivative terms. This solution is precisely the set of CFT data $\tE$. Indeed, consider
\ba
\mathcal G^{\tE}(z):=\sum_{i=1}^K \hat a_i G_{\hat \Delta_i}(z)+\sum_{n=0}^\infty a_n^{\tE} G_{\Delta_n^{\bD}}(z)\,.
\ea
We have
\ba
\mathcal G^{\tE}(z)=\mathcal G^{\tE}(1-z)=0 \quad \Leftrightarrow \left\{\begin{array}{ll}
\alpha_n^\bD[\tE]=0\\\vspace{0.2cm}\\
\beta_n^\bD[\tE]=0
\end{array}\right.
\Leftrightarrow 
\left\{\begin{array}{ll}
\alpha_n^\bD[\tS]=-a_n^{\tE}\\\vspace{0.2cm}\\
\beta_n^\bD[\tS]=0
\end{array}\right.
\ea
where we first used completeness and then duality of the basis $\bD$. The righthand side is now indeed satisfied by
 \reef{eq:extalpha} and \reef{eq:extbeta}. Equivalently we have shown that the special combination of crossing blocks:
 \ba
 \mathcal G^{\tE}=\sum_{i=1}^ K \hat a_i \mathcal Q_{\hat \Delta_i}^{\bD}\,.
 \ea
does not include derivatives of blocks in its OPE.

The extremality equations \reef{eq:extbeta} are an infinite set of constraints that uniquely determine the basic spectrum $\bD$. For instance, imagine one could countinuously deform $\bD$ to a new set $\bD'$. Deforming the solution $\mathcal G^{\tE}$ and demanding crossing would lead to
\ba
\qquad \sum_{n=0}^\infty \left[\delta a_n F_{\Delta_n^\bD}+ a_n^{\tE}\gamma_n\partial_{\Delta} F_{\Delta_n^{\bD}}\right] =0\,,
\ea
for $\delta a_n, \gamma_n$ Regge bounded.
\footnote{By Regge bounded we simply mean that deformations of the above form must result in a new correlator which is $O(1)$ for large $z$. Essentially this translates into the statement that both $\delta a_n/a_n^{\tE}$ and $\gamma_n$ should decay for large $n$.} 
But now acting with $\beta_n^\bD,\alpha_n^\bD$ on this equation and using duality determines that these quantities must all vanish so that in fact no such deformation exists.

To summarize, we have defined special extremal bases of sum rules for the crossing equation that naturally come associated to particular sets of CFT data. As promised, for any such basis it is trivial to verify that this data is a solution. Each such basis is labeled by a particular choice of input data $\tS$ which is not fixed by crossing.

A few comments are in order. Firstly  
it is possible that not all sets $\tS$ are allowable, in the sense that there may be no real $\Delta_n$ satisfying the extremality conditions.
Conversely, in the above definition, all $a_n^E$ are assumed to be a priori strictly non-zero, but not necessarily positive. For unitary CFTs this, together with positivity of the $\Delta_n$ imposes implicit constraints on the allowed choices of $\tS$.
Secondly, here we will always take the first operator in $\tS$ to be the identity, denoted $\mathds 1$ and corresponding $(\hat \Delta,\hat a)=(0,1)$. However this is not mandatory, and it is certainly possible to find extremal solutions without the identity.

Finally, in principle $K$ could be taken to be infinite, but this would require specifying infinite sets of CFT data so here we will focus on finite $K$.

Our construction associates a canonical solution to crossing given a finite set of inputs. But in the numerical bootstrap there is also a way to do the same thing: given a finite set of assumptions on the CFT data we can solve an optimisation problem in the space of all CFTs and look for the extremal solution \cite{Rattazzi:2008pe,ElShowk:2013}. These two constructions are in fact related. We have defined two sets of extremal solutions, which are bosonic and fermionic. Then often we have the basic relation:
\begin{itemize}
\item {\bf Bosonic type:} Extremal solutions saturate an upper bound on the OPE coefficient $a_0^{\tE}$ of an operator of dimension $\Delta_0^\tE$, for all CFTs consistent with the CFT data in $\tS$ (and some minimal gap assumptions, e.g. $\Delta_g\geq \Delta_0^\tE$).
\item {\bf Fermionic type:} Extremal solutions saturate an upper bound on the gap to the first operator in the OPE, for all CFTs consistent with the CFT data in $\tS$. 
\end{itemize}
That this is the case follows from the fact that extremal solutions are associated to bases of functionals. These functionals imply bounds on unitarity CFTs thanks to the positivity properties implied by duality. In particular the above two bounds can be obtained by considering $\alpha_0^{\tE}$ and $\beta_0^{\tE}$. For an example of this the reader may refer to figures \ref{fig:basicfermionomega} and \ref{fig:basicbosonomega}.
Depending on the choice of $\tS$ it might be that some other choice of bound will lead to the associated extremal solution. Also, many different bounds are actually saturated by the same extremal solution. For instance the generalized free boson to be discussed below saturates an upper bound on the OPE of an operator of dimension $\Delta=2\Df$, but it also maximizes the gap above such an operator -- the last bound being given by the functional $\beta_1$.

\subsection{Two important examples and a conjecture}
Up to now our discussion has been quite abstract, so let us now discuss two important explicit examples. These two extremal functional bases are labeled $F,B$ for Fermionic and Bosonic and they correspond to the choices:
\ba
F&: \qquad& \tS&=\{\mathds 1\}\,& \quad \Rightarrow \quad \tD&=\{(\Delta_n^F,a_n^F)\}_{n=0}^\infty\\
B&: \qquad& \tS&=\{\mathds 1, (\hat \Delta_0=\Delta_0^B,\bullet)\}\,&\quad \Rightarrow \quad \tD&=\{(\bullet,a_0^B)\}\cup \{(\Delta_n^B,a_n^B)\}_{n=1}^\infty
\ea
with
\ba
\Delta^B_n=2\Df+2n\,, \qquad \Delta^F_n=2\Df+1+2n\,.
\ea
The fermionic basis corresponds to having as input only the existence of the identity operator, while for the bosonic basis we additionally demand the existence of an operator of dimension $\Delta_0^B=2\Df$ in the OPE (but unfixed OPE coefficient). The corresponding extremal solutions are simply generalized free field correlators
\ba
\mathcal G^{B,F}(z)=\frac{1}{z^{2\Df}}+\frac{1}{(1-z)^{2\Df}}\pm 1
\ea
with $+(-)$ for boson (fermion), from which we get
\ba
a_n^{B,F}=a^{\tt gff}_\Delta\big|_{\Delta=\Delta_n^{B,F}}\,, \qquad  a^{\tt gff}_\Delta=\frac{\sqrt{\pi } 2^{3-2 \Delta } \Gamma (\Delta ) \Gamma \left(\Delta +2 \Delta _{\phi }-1\right)}{\Gamma \left(\Delta -\frac{1}{2}\right) \Gamma \left(\Delta -2 \Delta _{\phi }+1\right) \Gamma
   \left(2 \Delta _{\phi }\right)^2}
\ea
Although it is non-trivial to prove, the bosonic/fermionic bases do satisfy completeness as required \cite{Mazac:2018ycv}, i.e.:
\ba
\mbox{Crossing} \quad \Leftrightarrow \quad \begin{array}{c}
\sum_{\Delta} a_{\Delta} \alpha_n^{B,F}(\Delta)=0\vspace{0.3 cm}\\
\sum_{\Delta} a_{\Delta} \beta_n^{B,F}(\Delta)=0
\end{array}
\ea
The relevance of these bases follows from the following conjecture:\vspace{0.3cm}

\setlength\fboxrule{1.2pt}
\setlength{\fboxsep}{8pt} 
{\hspace{-18pt}\noindent\fbox{\parbox{\textwidth-18pt}{\vspace{-6pt}\vspace{0.3cm}
{\bf  Conjecture: Asymptotic Freedom}\\

Any extremal solution $\tE$ of bosonic/fermionic type labeled by a finite set $\tS$ behaves as a free field asymptotically, in the following sense:
\ba
(\Delta_n^{\tE}, a_n^{\tE}) \quad \underset{n\to \infty} \longrightarrow \quad (\Delta_n^{B,F}, a_{n}^{B,F})\label{eq:asymptotics}
\ea
}
}
}
\vspace{0.3cm}

\noindent
As we will see, this property means that we can efficiently construct generic extremal basis/solutions using the two explicitly constructed bases. 

Unfortunately we will not be able to prove this conjecture here, but needless to say we were unable to find any counterexamples nor do we have any clue on how one might go about constructing a finite set $\tS$ which would violate it.\footnote{ Note that if $\tS$ is allowed to be infinite then the conjecture is false. For instance, one can easily construct double zero bases with basic spectra of the form $2\Df+c+2n$. For generic $c$ and finite $\tS$ these are not extremal, but they can be for infinite $\tS$. For instance the 2d Ising correlator restricted to a line has a spectrum of the form $2\Df+\frac 34+n$, so we can simply set $\Delta_n^{\tE}$ to be the states of even $n$ and $\tS$ to contain states of odd $n$.}
We can gain some intuition for why the conjecture should be true with a few simple arguments. For instance, it is easy to show that if we assume that an extremal solution has a spectrum asymptoting to dimensions of the form $2\Df+c+2n$ then $c=0$ or $1$ are the only possibilities. This is a consequence of the analysis of functional equations at large $n,\Delta$ done in section \ref{sec:smatext}, and essentially it follows from a mapping of extremal spectra to S-matrices of unit modulus. But this argument does not exclude more exotic behaviour where the extremal spectrum has $O(1)$ anomalous dimensions that never decay.

To see this is unlikely, we can give a perturbative argument. Consider a  fermionic extremal basis labeled by $\tS$:
\ba
\bD=\{\Delta_n^{\tE}\}_{n=0}^\infty\,, \qquad \tS=\mathds{1}\cup \{(\hat \Delta_i, \hat a_i)\,, \quad i=1,\ldots,K\}
\ea
and now set
\ba
\bD_\lambda:=\bD[\lambda \tS]\,,\qquad \lambda \tS:=\mathds{1}\cup \{(\hat \Delta_i, \lambda \hat a_i)\,, \quad i=1,\ldots,K\}
\ea
such that
\ba
\bD_{0}=\{\Delta_n^F\}\,, \qquad \bD_{1}=\{\Delta_n^\tE\}
\ea
In other words, $\bD_{\lambda}$ is a family of spectra which interpolate between the free fermionic one and the desired basis. The idea now is that we can construct $\bD_{\lambda}$ perturbatively in $\lambda$. For instance, to leading order in $\lambda$, the extremal solution is constructed as
\ba
\mathcal G_\lambda(z)=\mathcal G^F(z)+\lambda\left[\sum_{i=1}^K \hat a_i \mathcal Q^F_{\hat \Delta_i}(z)\right]+O(\lambda^2)
\ea
from which we read off
\ba
a_n^{\tE}=a_n^F-\lambda\left[\sum_{i=1}^K \hat a_i\alpha_n^F(\hat \Delta_i)\right]+O(\lambda^2)\\
\gamma_n^F=\Delta_n^{\tE}-\Delta_n^F=-\lambda\left[\sum_{i=1}^K \frac{\hat a_i}{a_n^F}\,\beta_n^F(\hat \Delta_i)\right]+O(\lambda^2)\\
\ea
This computation easily generalises to higher orders. The upshot is that we can check for asymptotic freedom explicitly, by examining the large $n$ behaviour of the fermionic functionals $\beta_n^F$ for fixed argument (cf. appendix \ref{app:asymp}). For instance, if all $\hat \Delta_i>1$ one finds
\ba
\gamma_n^F\underset{n\to\infty}=\frac{\sum_{p=1}^\infty\,c_p\lambda^p }{n^2}
\ea
where the above is a result to all orders in perturbation theory. This is not a proof though, since the correct order of limits we would like to take is first consider high $n\to \infty$ and then $\lambda\to 0$. And for good reason too, since perturbation theory can and does breakdown for special sets $\tS$, or in the present context for large enough $\lambda$. We will see examples of this in the application section. However, when this happens we find that this breakdown is driven by an extremal solution transitioning from bosonic to fermionic asymptotics or vice-versa. Incidentally these transition regions are precisely the interesting region for probing the flat-space limit of a bulk QFT, as we will discuss in section \ref{sec4}.

\section{Hybrid bootstrap}\label{sec3}
In this section we will describe a method for efficiently constructing an extremal spectrum $\tE$. As we will see, this actually does not require explicitly constructing the associated functional basis, although once $\tE$ is found this can be achieved straightforwardly. The idea for the method is to make use of the asymptotic freedom conjecture together with the two special bases which we have already constructed explicitly. 

\subsection{Hybrid bootstrap method}
Our goal is to construct the extremal solution $\tE$  associated to the fixed set $\tS$. To do so, we will use as a working assumption the validity of conjecture \reef{eq:asymptotics}, which tells us that the CFT data of $\tE$ becomes approximately free at sufficiently large $\Delta$. We introduce a scale $\Delta^*$ and split
\ba
\tE=\tL\cup \tH
\ea
where $\tL$ (low energy) contains states with scaling dimension below $\Delta^*$, while $\tH$ (high energy) contains all other states. The idea is to choose $\Delta^*$ so that these high energy states are sufficiently close to being free. Our strategy will be to solve for the finite set of data in $\tL$ numerically, while treating the infinite set $\tH$ analytically in perturbation theory.

The CFT data $\tE$ has to satisfy a complete set of crossing sum rules. Up to now we have constructed only two such complete sets, which were denoted as bosonic and fermionic free functionals. Below we denote simply $\hat \alpha_n, \hat \beta_n$ for the elements of each such complete set, without explicitly specifying if it is the bosonic or fermionic basis we are using. However, it will always be implicitly understood that we work with the complete basis of bosonic functionals if $\tE$ is of bosonic type, and conversely if it is fermionic. Recall that this is known a priori, as it only depends on the structure of the fixed inputs $\tS$. For convenience we will introduce the notation
\ba
B: \, h_n=2\Df+2n\qquad F:\, h_n=2\Df+1+2n\,.
\ea
After these preliminaries, the crossing sum rules which an extremal solution must satisfy can be written as:
\ba
\hat \beta_{n}[\tL]+\hat \beta_{n}[\tH]&=0\,,  \qquad \hat \alpha_{n}[\tL]+\hat \alpha_{n}[\tH]&=0\qquad h_n\leq \Delta^*\,,\\
\hat \beta_{n}[\tL]+\hat \beta_{n}[\tH]&=0\,,  \qquad \hat \alpha_{n}[\tL]+\hat \alpha_{n}[\tH]&=0\qquad h_n> \Delta^*\,,\label{eq:lowhigheqs}
\ea
The idea now is to use the duality conditions together with asymptotic freedom to simplify the functional actions on the high energy states. Recall that duality gives \footnote{For the bosonic basis, $\hat \beta_0^B=0$ and the right column duality relations do not hold when $m=0$.}
\ba
\hat \alpha_n(h_m)&=\delta_{n,m}\,,&\quad \partial_\Delta\hat \alpha_n(h_m)&=0\\
\hat \beta_n(h_m)&=0\,,&\quad \partial_\Delta\hat \beta_n(h_m)&=\delta_{n,m}
\ea
while asymptotic freedom implies
\ba
\Delta_n^{\tE}=h_n+\gamma_n\,, \qquad \gamma_n\ll 1 \quad \mbox{for}\quad h_n>\Delta^*
\ea
We have thus a small parameter which can be used 
to set up a perturbative scheme for the high energy data. We expand
\ba
\Delta_n^{\tE}=h_n+\sum_{k=1}^M\gamma_n^{(k)}\,, \qquad
a_n^{\tE}=a_n^{(0)}+\sum_{k=2}^M a_n^{(k)}\,, \qquad h_n\geq \Delta^*
\ea
up to some desired order $M$, where $\gamma_n^{(k)}=O[(\gamma_n^{(1)})^k]$. The second set of equations in \reef{eq:lowhigheqs} can now be solved for perturbatively making use of the duality conditions. We find for instance for the first two orders:
\ba
a_n^{(0)}&=-\hat \alpha_n[\tL]\\ 
a_n^{(0)}\gamma_n^{(1)}&=-\hat \beta_n[\tL]\\
a_n^{(0)}\gamma_n^{(2)}&=-\frac 12 \sum_{m=n^*}^\infty a_m^{(0)} [\gamma_m^{(1)}]^2 \partial_{\Delta}^2 \hat \beta_n(h_m)\\
a_n^{(2)}&=-\frac 12 \sum_{m=n^*}^\infty a_m^{(0)} [\gamma_m^{(1)}]^2 \partial_{\Delta}^2 \hat \alpha_n(h_m)\,. \label{eq:asymptoticres}
\ea
We can use these results to set up an iterative scheme for constructing an extremal solution. Concretely, let us rewrite the sum rules \reef{eq:lowhigheqs} as:
\ba
\hat \beta_{n}\left[\tL^{(i)}\right]+\hat \beta_{n}\left[\tH^{(i-1)}\right]&=0\,,  \qquad \hat \alpha_{n}\left[\tL^{(i)}\right]+\hat \alpha_{n}\left[\tH^{(i-1)}\right]&=0\qquad h_n\leq \Delta^*\,,\\
\hat \beta_{n}\left[\tL^{(i)}\right]+\hat \beta_{n}\left[\tH^{(i)}\right]&=0\,,  \qquad \hat \alpha_{n}\left[\tL^{(i)}\right]+\hat \alpha_{n}\left[\tH^{(i)}\right]&=0\qquad h_n> \Delta^*\,,
\label{eq:crossnum}
\ea
and define $\tH^{(0)}$ as the empty set. At step $i$ we begin by numerically solving the finite set of equations in the first line to determine an approximation to the low energy data. Next, we solve analytically the equations on the second line to find the high energy data. This is then fed back into the low energy equations and we repeat the process until convergence is found.  We find in practice that a handful of iterations suffices to achieve convergence.\footnote{Sometimes it can be necessary to input a better initial guess for $\tH^{(0)}$, e.g. using the expected form of anomalous dimensions described below.} Our scheme requires being able to compute the high energy data for all $n$. In practice we can derive analytic formulae for this data by exploiting the asymptotic expressions for the functional actions, as we will see momentarily as well as in more detail in appendix \ref{app:analg}. Finally, once the extremal solution is found it is straightforward to obtain the associated dual functional basis as we will describe in the next subsection.

To gain some intuition for the outcome of this procedure let us examine the leading order approximation of the scheme where $M=1$. For simplicity we assume that all operators in $\tS$ have $\Delta>1$. The high energy CFT data is given by%
\ba
\gamma_n^{(1)}&=-\hat \beta_n[\tL]/a_n^{(0)}\,, \qquad a_n^{(0)}&=-\hat \alpha_n[\tL]
\ea
These equations are valid for all $h>\Delta^*$, but let us now take $h\gg 1$. With large $n$ holding $\Delta>1$ fixed the functional actions simplify and we find (cf. appendix \ref{app:asymp}):
\ba
\hat \beta_n(\Delta)&\sim\frac{a_{h_n}^{\tt GFF}}{h_n^2}\,  \xi(\Delta)\,, \qquad \hat \alpha_n(\Delta)&\sim \frac 12\partial_n \left(\frac{a_{h_n}^{\tt GFF}}{h_n^2}\right)\,  \xi(\Delta)
\ea
where $\xi(\Delta)$ is a certain functional.\footnote{This functional however can only act on super Regge bounded correlators.}
This gives us to leading order in the large $h$ expansion:\footnote{A better approximation for $a_n^{(0)}$ is actually
\ba
a_n^{(0)}=a_{h_n}^{\tt GFF}+\frac 12\partial_n (a_{h_n}^{\tt GFF} \gamma^{(1)}_n)\,.
\label{eq:betterasymp}
\ea
}
\ba
\gamma_n^{(1)}\sim -\frac{\xi[\tL]}{h_n^2}\,, \qquad
a_n^{(0)}\sim a_{h_n+\gamma_{n}^{(1)}}^{\tt GFF} \label{eq:exactasymp}
\ea
These expressions provide us with an approximate high energy spectrum {\em for all} $h$ above $\Delta^*$ which is self-consistent with the hypothesis of asymptotic freedom. However, they do not imply it. To see this, we note that the functional $\xi(\Delta)$ has large $\Delta$ behaviour
\ba
a_{\Delta}^{\tt GFF}\, \xi(\Delta)\underset{\Delta\gg 1}\propto \sin^2 \left[\frac\pi 2(\Delta-h_0)\right]\,\times \Delta
\ea
and thus what one must really check for asymptotic freedom is whether $\xi[\tL]$ grows with $\Delta^*$ or if it eventually saturates. It is only in the latter case that our scheme makes sense, as results then become independent of the arbitrary choice of $\Delta^*$ (as long as sufficiently large). As an extreme example, if we tried to apply our scheme using a bosonic basis to bootstrap a fermionic solution, we would find that $\xi[\tL]\sim O[(\Delta^*)^2]$ and we would never get convergence in $\Delta^*$. More generally, the method described here is particularly useful if $\Delta^*$ is not too large, as we must still solve numerically for low energy data up to that scale. As we will see later, in order to extract the flat space limit of CFT correlators describing QFT in AdS we will have to consider extremal solutions where $\Delta^*\gg 1$. In this case we will show that there is an alternative method to construct the extremal solution, where the high energy data is solved for in terms of a bulk S-matrix.

\subsection{Constructing the extremal bases}
Let us now explain how to construct the extremal functional bases dual to an extremal solution~$\tE$. The dual basis is useful in particular if one wishes to prove bounds which hold for any CFT -- bounds saturated by the extremal solution. We follow the same logic as in the construction of the extremal spectrum: we again do the split between low and high energies, and focus on the construction of the low energy extremal functionals. Denote one such functional as $\omega$. Using completeness of the free functional basis we can write
\ba
\omega=\omega_{\tL}+\omega_{\tH}=\sum_{m<n^*} \left[ A_m \hat \alpha_m+ B_m \hat \beta_m\right]+\sum_{m\geq n^*} \left[ A_m \hat \alpha_m+ B_m \hat \beta_m\right]
\ea
To determine the coefficients appearing above we have to impose the correct duality conditions, which are for all but one $n$ of the form
\ba
\omega(\Delta_n^{\tE})=\partial_{\Delta} \omega(\Delta_n^{\tE})=0
\ea
We now write again, for $m\geq n^*$:
\ba
A_m=A_m^{(2)}+ A_m^{(3)}+\ldots\,,\\
B_m= B_m^{(1)}+B_m^{(2)}+\ldots\,,\\
\ea
and get solutions:
\ba
A_m^{(2)}&=\frac 12  \gamma_m^2\,\partial_{\Delta}^2 \omega_{\tL}(h_m)\\
B_m^{(1)}&=-\gamma_m\partial_{\Delta}^2 \omega_{\tL}(h_m)\\
A_m^{(3)}&=\frac 13\, \gamma_m^3\partial_{\Delta}^3 \omega_{\tL}(h_m)+\frac 12 \sum_{p=n^*}^\infty B_p^{(1)} \gamma_m^2\partial^2_{\Delta}  \hat \beta_p(h_m)\\
B_m^{(2)}&=-\gamma_m^2\frac 12\, \partial_{\Delta}^3 \omega_{\tL}(h_m)- \sum_{p=n^*}^\infty B_p^{(1)} \gamma_m \partial^2_{\Delta} \hat \beta_p(h_m)
\ea
While it is possible to go further, below we will stick to this approximation. Note that $\hat\beta_m$ is small in the vicinity of $h_m$, so that we are correct in keeping these terms but not $B_m^{(3)}$. The strategy is now as for the construction of the extremal solution: we write
\ba
\omega=\omega_{\tL}^{(i)}+\omega_{\tH}^{(i-1)}\,, \qquad \omega_{\tH}^{(0)}=0
\ea
and proceed iteratively, imposing duality conditions to determine $\omega_{\tL}^{(i)}$ numerically and $\omega_{\tH}^{(i)}$ analytically. At the end of this construction, the functional has double zeros at the correct locations all the way to infinity.

We conclude this section with a small technical point. If we want to obtain fully rigorous numerical bounds is that we want our basis functionals to be asymptotically sign definite. This is true for any basis functional $\hat \alpha_m$ and $\hat \beta_m$ and any finite linear combination of them: in particular they have exact double zeros at $\hat \Delta_m$. However, after dressing our functionals have only approximate double zeros instead at $\Delta_m=h_m+\gamma_m$, since we had to treat $\gamma_m$ as a small parameter in solving the equations. Thus in general these will only be local minima, and we would like them to be as close to the real axis as possible but not below it. To deal with this we simply add small positive contribution to the dressed functionals in the vicinity of the local minima:
\ba
\omega^*=\omega+\sum_{m=N+1}^\infty \left[(A_m+r_m)\hat \alpha_m+B_m \hat \beta_m\right]
\ea
with $A_m, B_m$ determined as before. In practice we can choose $r_m$ to be a small constant for any $m,\omega$.

\subsection{Applications}
\label{sec:applications}
\subsubsection{Generic $\tS$}
In this section we will demonstrate how to apply the hybrid bootstrap method described above in several cases.

We begin by constructing a fermionic type extremal solution. As discussed, we claim such solutions correspond to sets $\tS$ containing operators for which we specify both dimensions and OPE coefficients, and that they are naturally associated to bounds on the maximal gap above operators in $\tS$. For the numerical results shown below 
we have chosen
\ba
\tS =\{(\Delta_i,a_i)\} =\left\{\mathds 1,(\tfrac{3}{2},2),(\tfrac{7}{4},1),(\tfrac{21}{10},\tfrac{1}{4})\right\}\,.
\ea
and set $\Df=1$. There is nothing particular about this choice, and our construction works equally well for others. 

We apply the method described by equations (\ref{eq:crossnum}), working with the free fermionic functional basis.
To solve the equations for the low lying data we simply use Newton's method with an initial guess: at step $i$ we use the solution constructed at step $i-1$ as our initial guess, and at step 1 we take it to be described by a generalized free fermion. In this example we set $\Delta^*=h_{N_{\tL}}$ with $N_{\tL} = 10$, so that the first ten operators other than those in $\tS$ are solved for numerically, while higher states are determined from \reef{eq:asymptoticres}. Ten iterations of the dressing procedure were performed and Newton's method was run in five steps each time. 

\begin{figure}[t!]
\begin{tabular}{cc}
\includegraphics[width=0.45\textwidth]{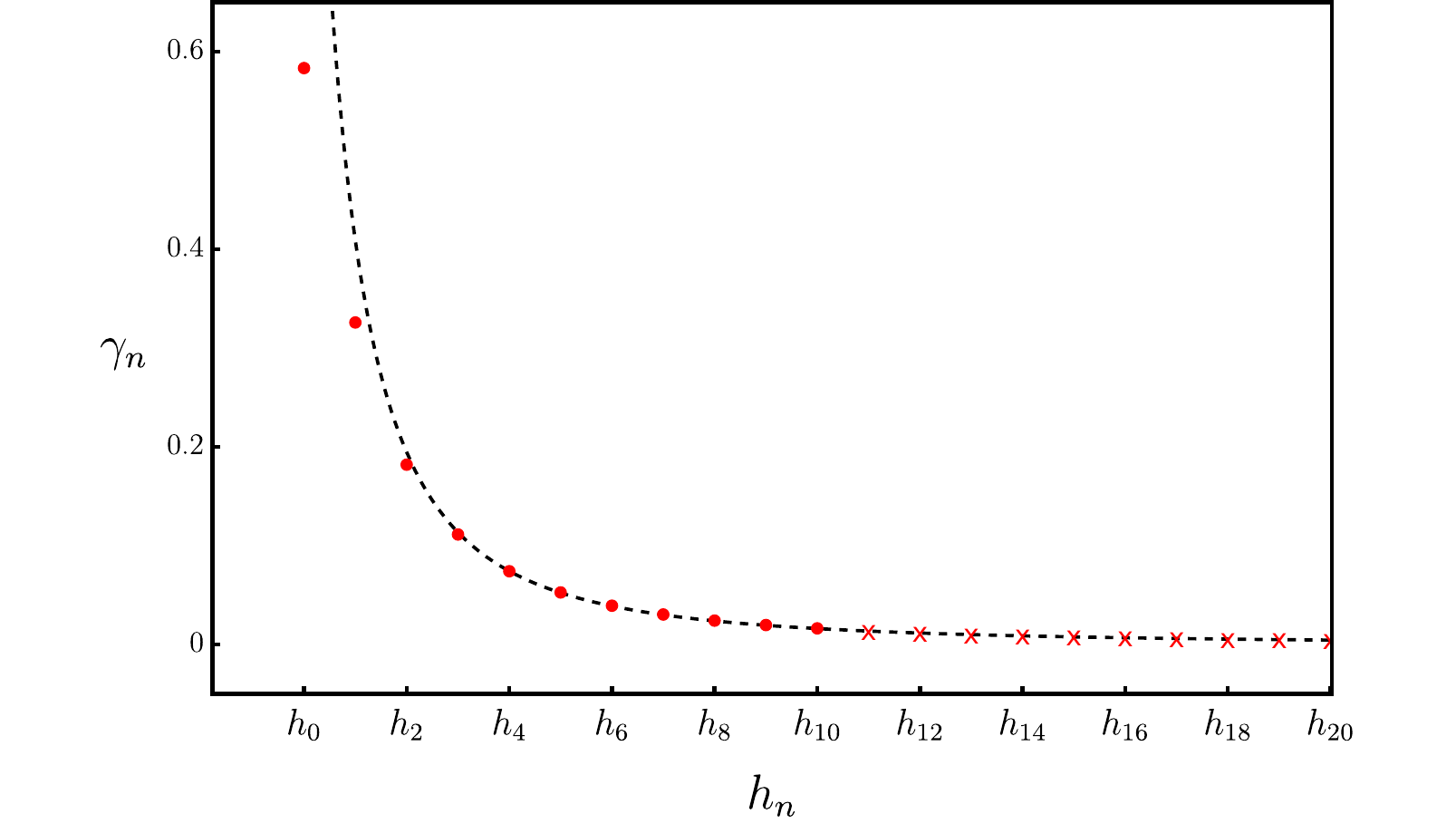}
&
\includegraphics[width=0.47\textwidth]{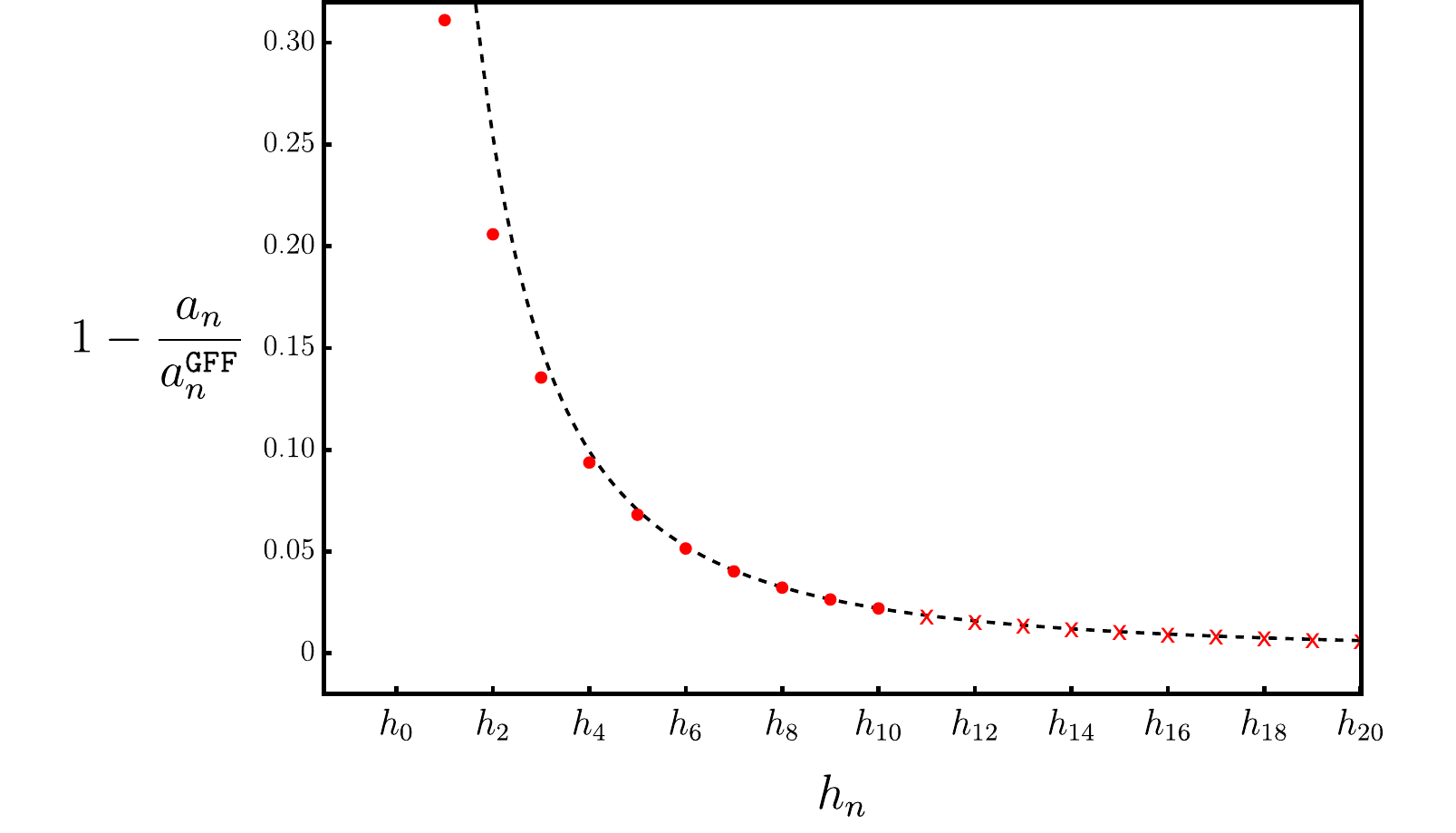}
\end{tabular}
\caption{CFT data for the fermionic solution. Red circles are computed numerically, while crosses are obtained in perturbation theory using (\ref{eq:asymptoticres}). Left: anomalous dimensions for the fermionic solution. The black line scorrespond to the prediction from (\ref{eq:exactasymp}). Right: OPE data for the fermionic solution. The black line correspond to the prediction from (\ref{eq:betterasymp}).\label{fig:basicfermion}}
\end{figure}

The resulting anomalous dimensions and OPE coefficients are shown in figure \ref{fig:basicfermion}. As we can see the CFT data rapidly converges to that of a generalized free fermion, in agreement with the asymptotic freedom conjecture. Note that this is not an artifact of our choice of basis: if this was the case the low and high energy data would not connect smoothly. The fact they do signals that our method has converged, providing a holistic approximation to the entire extremal solution.
\begin{figure}[h!]
    \centering
    \hspace{-1cm}
    \includegraphics[width=0.6\linewidth]{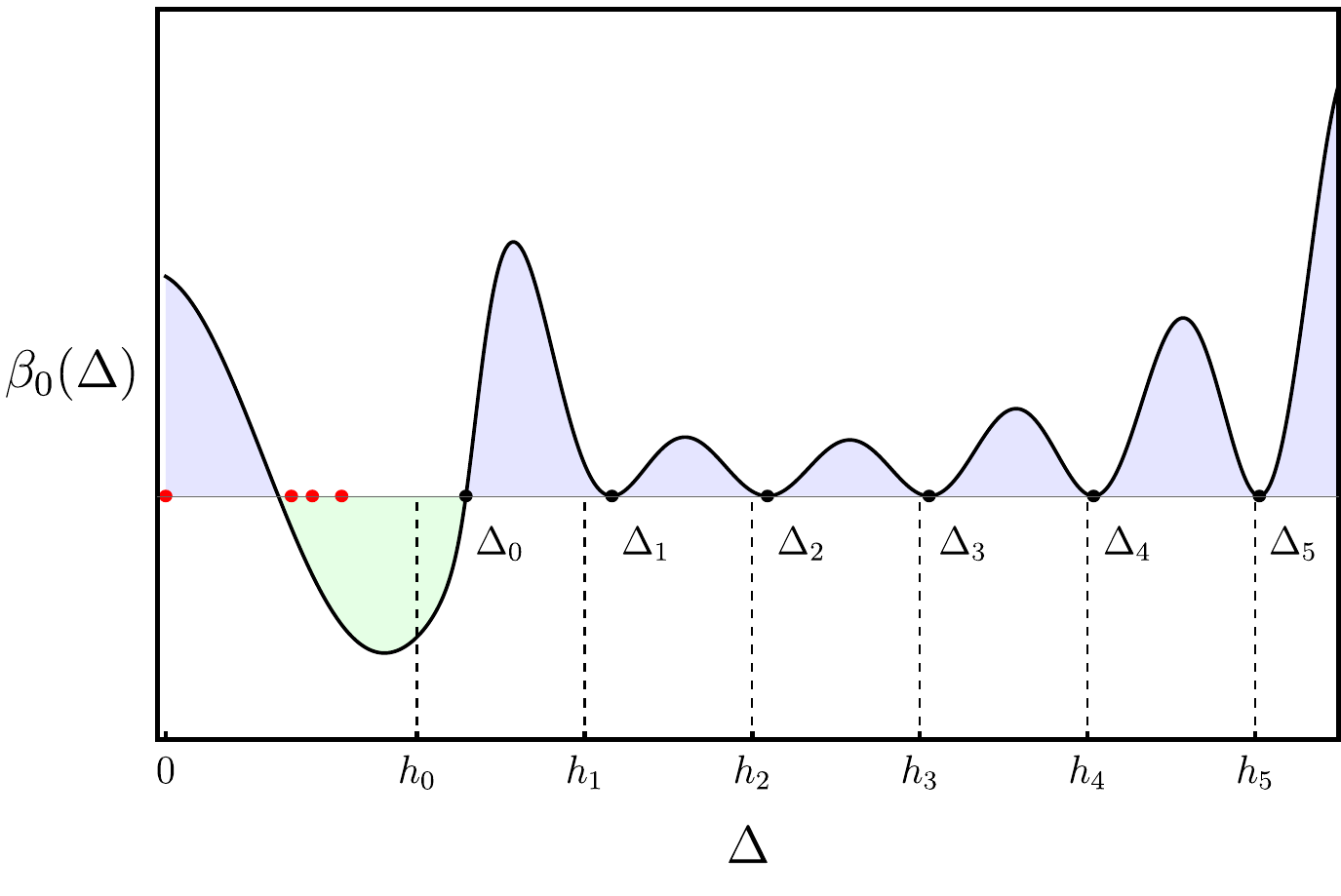}
    \caption{Interacting fermionic solution and its associated extremal functional. The dots represent the positions of the operators: in red is the set $\tS$, while in black is the set $\tD$. The dashed lines mark out the dimensions of operators in the generalized free fermion solution. The contribution from the identity cancels with those of other operators in $\tS$ so that $\beta_0[\tS]=0$. The positivity properties of $\beta_0$ imply the existence of an operator in the green region for any CFT. This bound is optimal for CFTs whose spectrum includes the states in $\tS$, as it is saturated by the corresponding extremal solution.
    \label{fig:basicfermionomega}
    }
\end{figure}
Once we find the extremal spectrum we can also reconstruct the associated basis of extremal functionals. As an example, we show the curve for $\beta_0(\D)$ in Figure \ref{fig:basicfermionomega}. This functional is constructed to satisfy $\beta_0\left[\tS\right]=0$ and have double zeros on all other operators in the extremal solution except the first one. We can see the resulting functional is actually positive above $\Delta_0$. Thus, for any solution to crossing containing the states in $\tS$ this functional gives a rigorous upper bound on the dimension of the first operator in the OPE above states in $\tS$, a bound which is saturated by the extremal solution we have constructed.

\begin{figure}[t!]
\begin{tabular}{cc}
\includegraphics[width=0.46\textwidth]{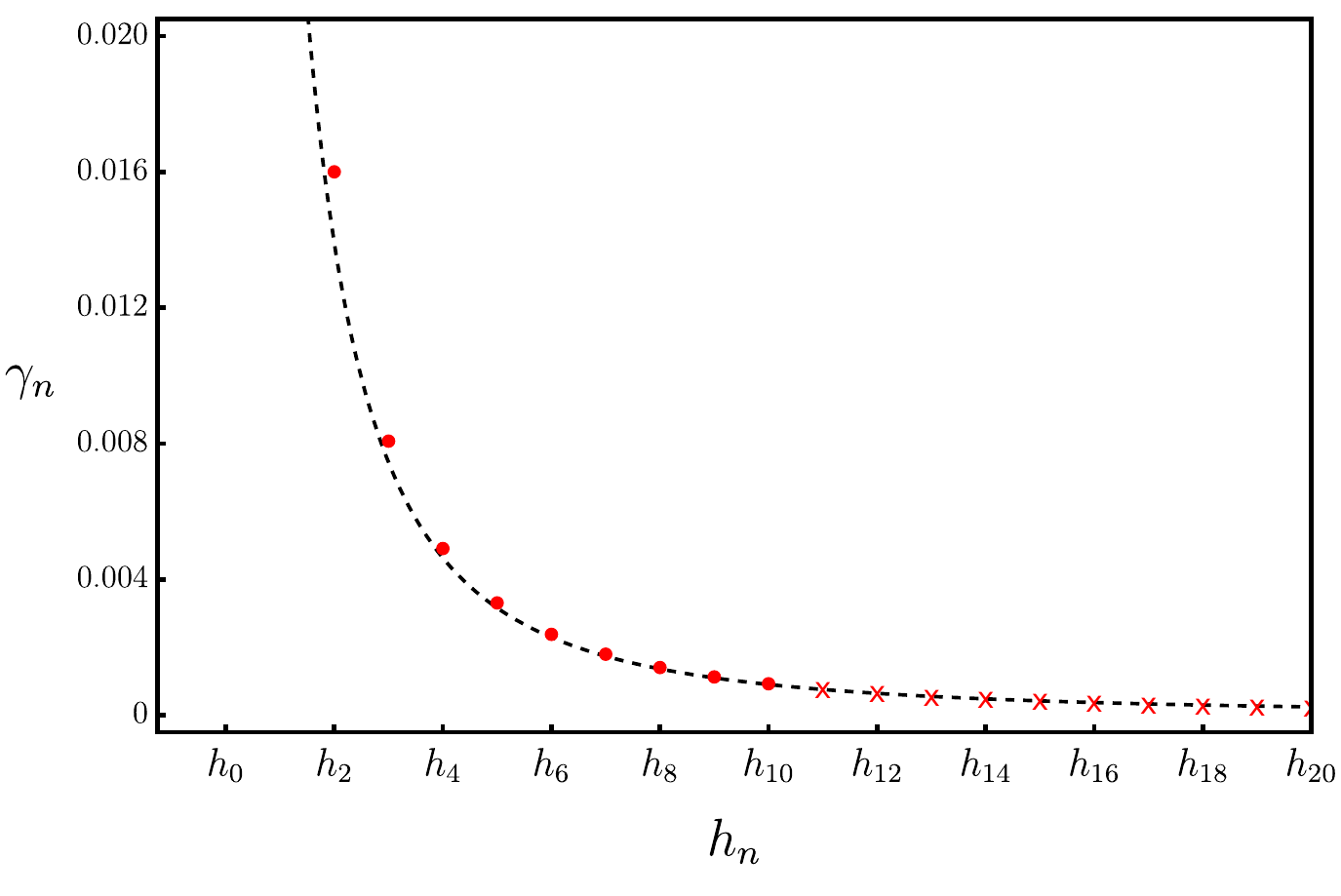}
&
\includegraphics[width=0.46\textwidth]{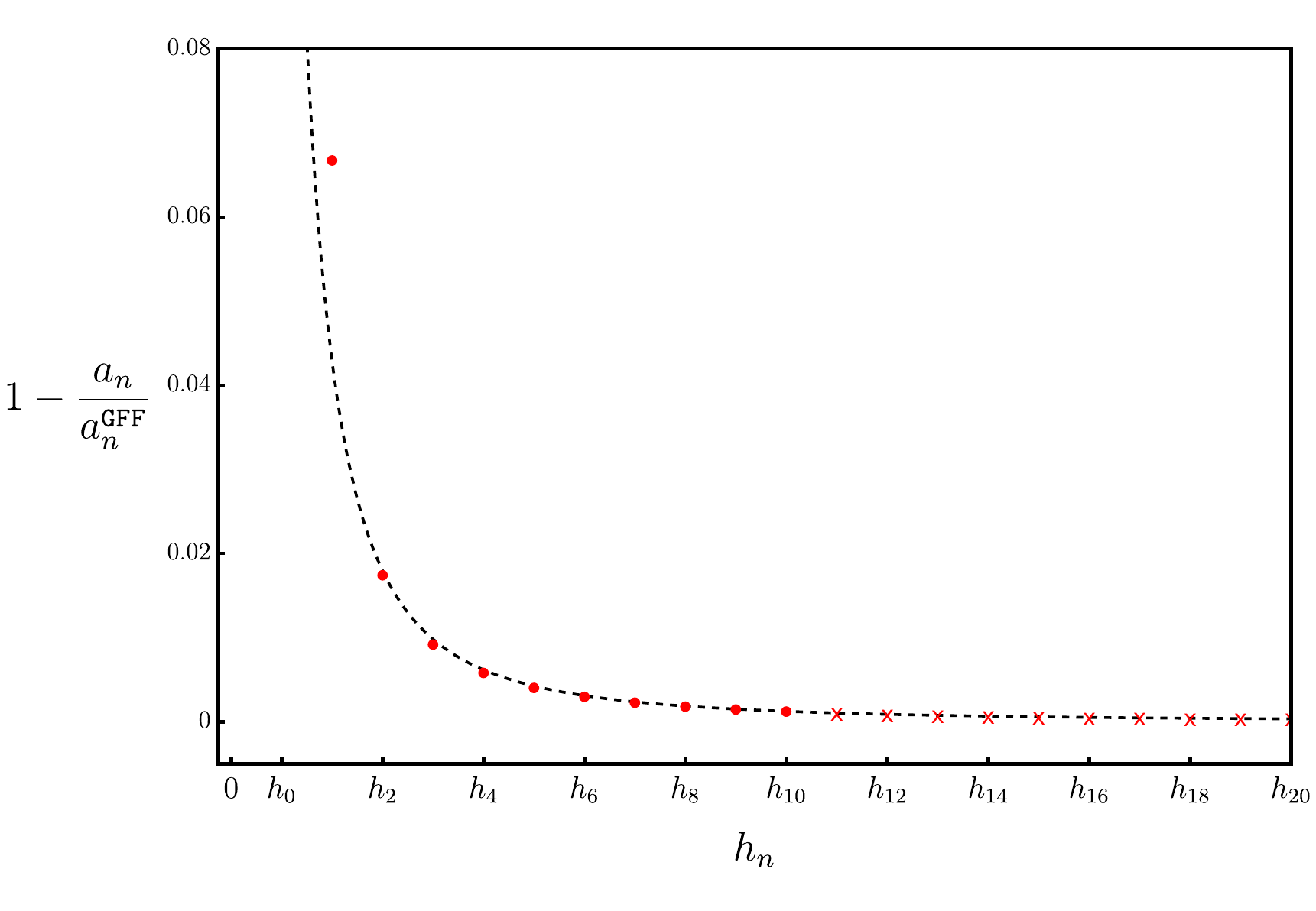}
\end{tabular}
\caption{CFT data for the bosonic solution. Red circles is computed numerically, while crosses are obtained in perturbation theory using (\ref{eq:asymptoticres}). Left: anomalous dimensions for the fermionic solutions. The black line scorrespond to the prediction from (\ref{eq:exactasymp}). Right: OPE data for the fermionic solution. The black line correspond to the prediction from (\ref{eq:betterasymp}).\label{fig:basicboson}}
\end{figure}

The next case we consider is a bosonic solution. As our set $\tS$ we take the same states as before but now we also take the dimension of the first operator in $\tD$ to be fixed:
\ba
\tS =\{(\Delta_i,a_i)\} =\left\{\mathds 1,(\tfrac{3}{2},2),(\tfrac{7}{4},1),(\tfrac{21}{10},\tfrac{1}{4}),(\hat \Delta_0,\bullet)\right\}\,\,, \qquad \hat \Delta_0= 2\Df+4/5
\ea
with $\Df=1$ once again. As we explained, in the extremal solution the OPE coefficient $a_0$ is not a free parameter. Here we expect such a solution to be of bosonic type, so that it is convenient to apply the hybrid bootstrap method with the generalized free boson functional basis.
Resulting anomalous dimensions and OPE coefficients are shown in figure \ref{fig:basicboson}. In this case $\alpha_0\left[\tS\right]=-a_0^{\tE}$, and $\alpha_0(\D)\geq0$ for $\D\geq \hat \D_0$.

\begin{figure}[h!]
    \centering
    \includegraphics[width=0.6\linewidth]{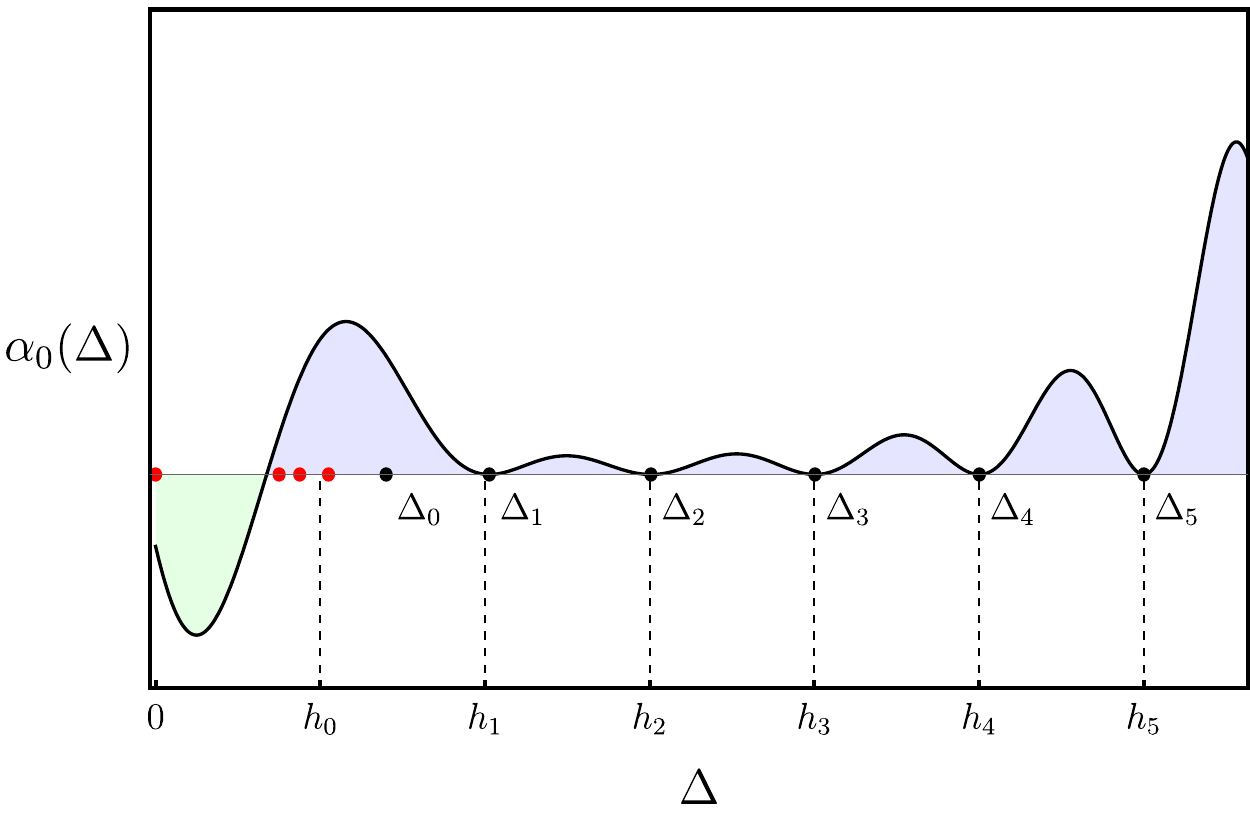}
    \caption{Bosonic extremal solution and the  associated extremal functional $\alpha_0$. The dots represent the positions of the operators: in red is the set $\tS$, while in black is the set $\tD$. As $\D$ increases, the operators approach the generalized free boson values.}
    \label{fig:basicbosonomega}
\end{figure}
As before, we can build the dual set of extremal functionals. We show the $\alpha_0$ functional in figure \ref{fig:basicbosonomega}.
This functional implies an optimal upper bound on the OPE coefficient of an operator of dimension $\hat \Delta_0$ in any CFT containing the spectrum in $\tS$, a bound which is saturated by the extremal solution:
\ba
a_0\leq -\alpha_0\left[\tS\right]\equiv a_0^{\tE}
\ea
As was the case for the fermionic solution, both the OPE data and scaling dimensions rapidly approach those of a generalized free boson.

\subsubsection{Finetuned $\tS$}
\label{sec:appfinetune}
In both examples discussed above we saw that the extremal spectrum rapidly converged towards its expected free asymptotics. This is what we observe to be the case for generic choices of the data in $\tS$. However, it is possible to finetune this data so as to make the approach towards free asymptotics arbitrarily slow.

To see how, consider for instance a fermionic solution with $\tS=\{\mathds 1,(\hat \D_0,\hat a_0)\}$. We know that not every value of $\hat a_0$ is allowed: after all there is a maximum for it corresponding to the extremal solution labeled by $\tS=\{\mathds 1,(\hat \D_0,\bullet)\}$, which is bosonic. Therefore, our solution is of fermionic type for $0\leq \hat a_0< a_0^{\tE}$ (with $\hat a_0 = 0$ corresponding to the free fermion), but of bosonic type when $\hat a_0 = a_0^{\tE}$. There is thus some kind of discontinuity. What happens is that away from the maximum $a_0^{\tE}$, the solution always asymptotes to the generalized free fermion as it should, but it takes longer and longer to do so as we approach this maximum. Concretely, there is an order of limits issue:
\ba
0 = \lim_{a_0\to a_0^{\tE}}\lim_{n\to\infty}\gamma_n \neq \lim_{n\to\infty}\lim_{a_0\to a_0^{\tE}}\gamma_n = 1
\ea
A similar situation can be arranged for bosonic type solutions. Consider now instead $\tS=\{\mathds 1,(\hat \Delta_0,\bullet)\}$. The dimension $\hat \D_0$ cannot be arbitrary but in fact has a maximum value $\D_0^{\text{max}}=2\Delta_{\phi}+1$ for which the solution is in fact the free fermion. As we approach this maximum value there is again a discontinuity and the solution takes longer and longer to asymptote to a free boson as $\hat \D_0$ approaches its maximum. 
 
The takeaway is that the method we have described becomes more and more difficult to use as we perform finetunings such as these on our sets $\tS$, or more precisely as we approach the boundaries of the set of all $\tS$ consistent with unitarity. It is in this sense that the choices we made earlier for the sets $\tS$ were generic, as they were well away from these boundaries. To find extremal solutions with $\tS$ close to these edge cases we have to use more and more functionals in order to reach the scale $\Delta^*$ which is being pushed off to infinity. 

\begin{figure}
        \centering
    \hspace{-1cm}
    \begin{tabular}{lr}
\includegraphics[width=0.47\linewidth]{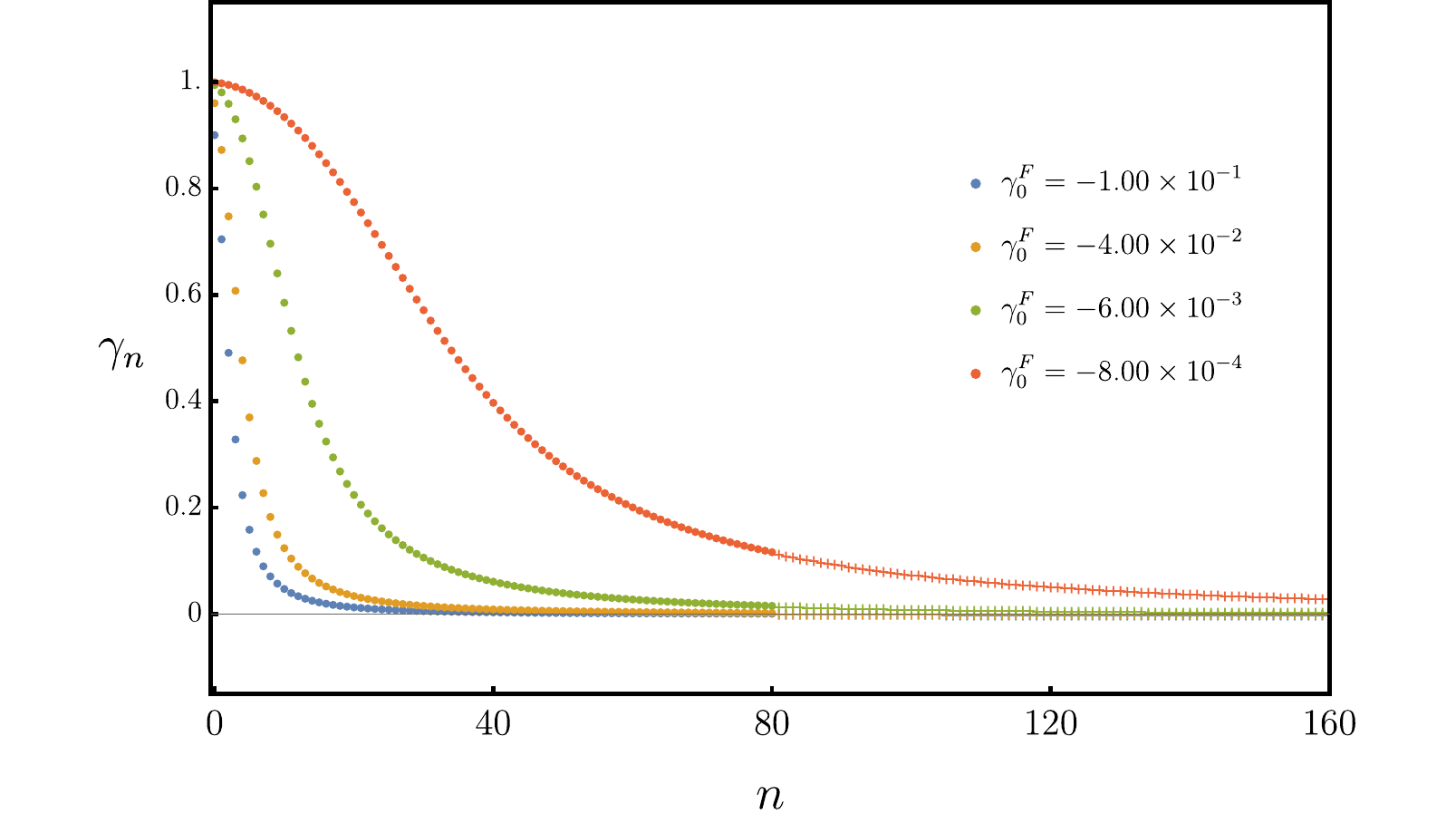} &
    \includegraphics[width=0.47\linewidth]{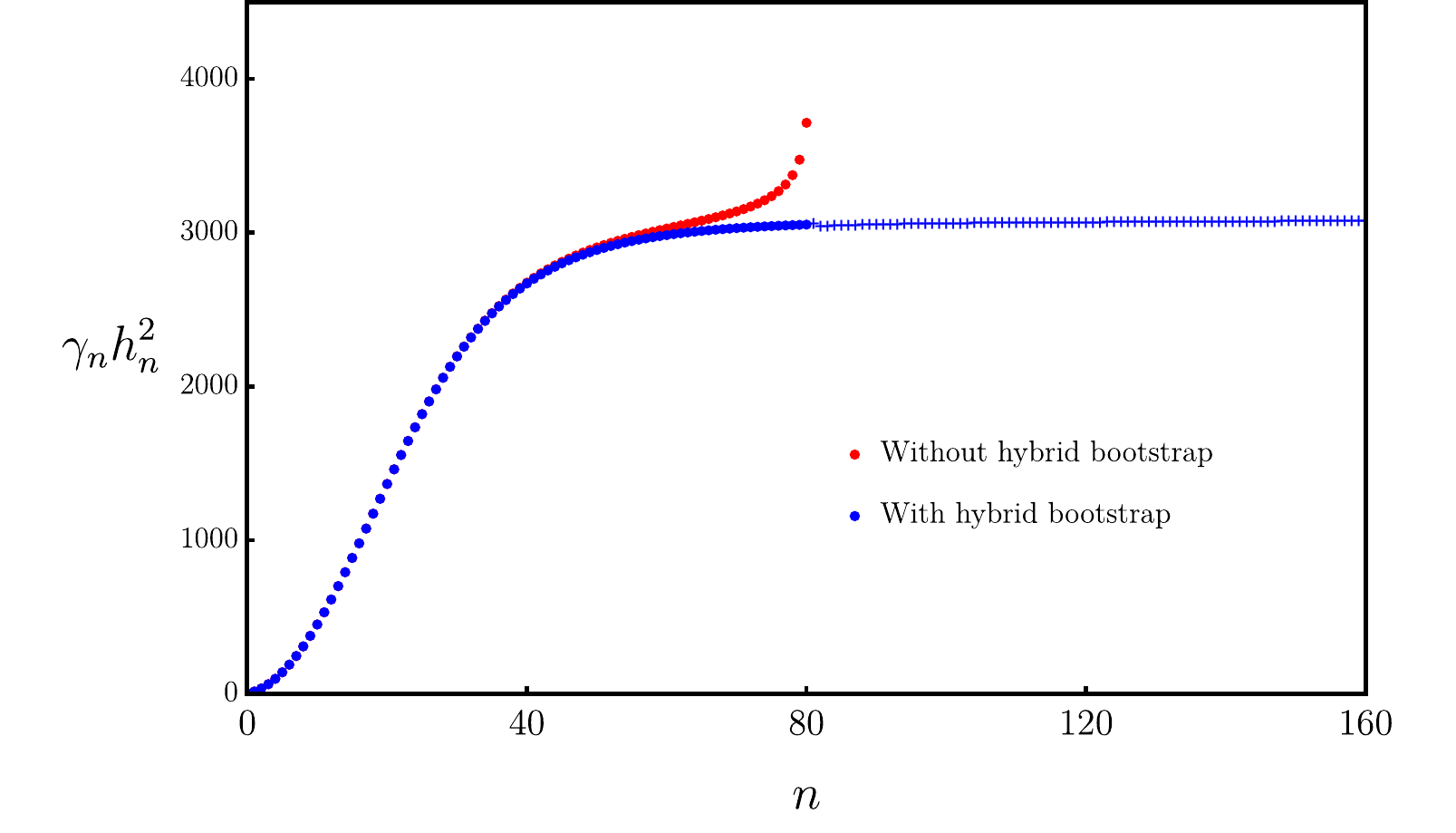}
    \end{tabular}
   
    \caption{Left: Anomalous dimensions for different values of $\gamma_0^F$. Dots/dashed curves represent the low/high energy CFT data computed numerically/analytically with the hybrid bootstrap method. In the finetuning limit $\gamma_0^F\to 0$ the CFT spectrum contains structure up to arbitrarily large scales.    
    Right: Extremal spectrum obtained with and without the hybrid bootstrap procedure for $\gamma_0^F=-8\times 10^{-4}$.}
    \label{fig:nearflatgammaversus}
\end{figure}

To investigate this question, let us for definiteness consider the family $\tS=\{\mathds 1,(\hat \D_0,\bullet)\}$, and set $\gamma_0^F:=\hat \D_0-1-2\Df$, so that $\gamma_0^F\ll 1$ is the finetuning limit. On the left of figure~\ref{fig:nearflatgammaversus} we show the spectrum of anomalous dimensions as we lower $\gamma_0^F$. As we can see, although these curves eventually do decay, they take longer and longer to do so, so that there is structure in the spectrum up to larger and larger scales. This means that in our method if we hold our cutoff $\Delta^*$ fixed, the anomalous dimension at the cutoff increases as $\gamma_0^F$ is lowered, so that eventually our approximations break down. In passing, note that before this happens is precisely when our method is most useful, as the corrections to the spectrum which is obtained without UV completion are then the largest. This is displayed on the right of the Figure \ref{fig:nearflatgammaversus}. 

\begin{figure}
    \centering
    \hspace{-1cm}
    \includegraphics[width=0.6\linewidth]{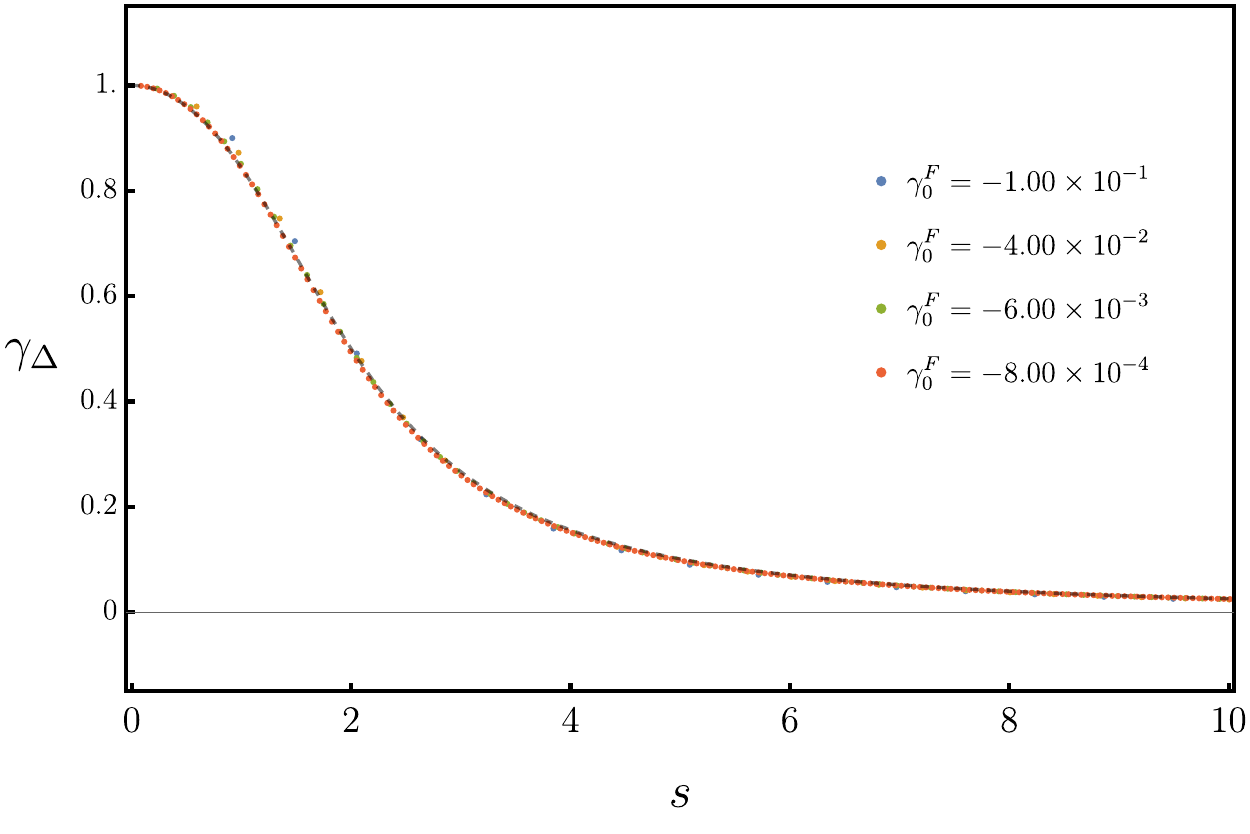}
    \caption{Extremal spectra plotted as a function of $s=-\gamma_0^F \Delta^2$. As $\gamma_0^F$ shrinks all spectra fall into the same smooth curve described by equation \reef{eq:smooth}}
    \label{fig:nearflatgammarescaled}
\end{figure}

At this point we can make two observations. Firstly, as we finetune we see that the spectrum curves seem to be approaching smooth functions of the label $n$. Secondly, although the scale $n$ at which the curves bend is varying with $\gamma_0^F$, the curves themselves all seem to be very similar. This suggests rescaling the $x$ axis and replotting, which we do in figure \ref{fig:nearflatgammarescaled}. Remarkably the curves now all look identical. In this way of parametrising them, lowering $\gamma_0^F$ just corresponds to densifying the number of points until we approach a smooth function. In the next section we will argue that this function turns out to be determined by a 2d integrable S-matrix, in particular:
\ba
\gamma(s) = -\frac{1}{\pi}\arg(S(s))\,,\qquad S(s) = \frac{s-im}{s+im} \label{eq:smooth}
\ea
with $s=-\gamma_0^F\D^2$. 

In fact, we will understand that the finetuning limits we have been mentioning here correspond to flat space limits of dual QFTs in AdS$_2$. But for the purposes of this section, these results bring with them a hopeful message: although in these limits the hybrid bootstrap method eventually breaks down, we may hope to replace it by working with a smooth description of the CFT data in terms of an S-matrix, and this is exactly what we will do.

\section{Bulk reconstruction and the flat space limit} \label{sec4}
A very general procedure for constructing 1d CFTs is to consider QFTs in spaces of the form AdS$_2\times \mathcal M$. In this section we will explore how extremal solutions are intimately related to such constructions.

\subsection{The flat space limit}
A QFT in AdS$_2$ induces a family of 1d CFTs living on its boundary.\footnote{Actually, many such families depending on choices of boundary conditions, curvature couplings and so on.} This family is parameterized by a dimensionless coupling $\Dg\equiv m_{\mbox{\tiny QFT}} R_{\mbox{\tiny AdS}}$ which can be taken to be some typical mass scale of the QFT multiplied by the AdS radius. Now, while any QFT in flat space can be thought of as an RG flow between a UV and an IR CFT, when we place it in AdS we introduce an effective IR cutoff which prevents us from probing the full flow. Nevertheless, by adjusting the free parameter $\Dg$ we can smoothly interpolate between the two extreme cases where the bulk QFT becomes the UV or IR CFT$_{2}$ in AdS$_2$. Incidentally, note that in these limits the bulk theory becomes a CFT in AdS which is Weyl equivalent to a CFT with a boundary, so that the whole setup becomes equivalent to a BCFT$_2$.

In order to access the flat space physics of the QFT we must take a double scaling limit: we must certainly take $m_{\mbox{\tiny QFT}}$ to be very large, so that the entire RG flow can happen in a very small patch of AdS of negligible curvature; but at the same time, this means that to probe this flow we must consider high energy observables, sensitive to distances much smaller than the AdS scale. Thus the appropriate limit is:
\ba
E R_{\mbox{\tiny AdS}} \to \infty\,, \quad m_{\mbox{\tiny QFT}} R_{\mbox{\tiny AdS}} \to \infty\,, \qquad E/m_{\mbox{\tiny QFT}}\ \mbox{fixed}\,.
\ea
Since energies in AdS space correspond to scaling dimensions of operators in the boundary CFT, it follows that the flat space limit is obtained by probing the high scaling dimension region of CFT spectra, in the limit where the bulk QFT is approaching the IR BCFT. 

The question of how precisely one can extract flat space observables from CFT data has been adressed in a number of works e.g.\cite{Polchinski:1999ry,Raju:2012zr,Giddings:1999qu,Gary:2009ae,Penedones:2010ue, Fitzpatrick:2011hu,Fitzpatrick:2011dm,QFTinAdS,Komatsu:2020sag,Hijano:2019qmi,Cordova:2022pbl}. Here we are interested in the case where the QFT has massless states in flat space, so that the IR BCFT mentioned above is non-trivial. 
One observable of interest is the flat space S-matrix which describes scattering of the massless states. Concretely, suppose that a CFT operator $\phi$ has overlap with a massless state in the bulk. For $d>1$ there is a recipe proposed by Penedones in \cite{Penedones:2010ue} for extracting the S-matrix from the Mellin amplitude, but this recipe cannot be extended to $d=1$. Here we propose that the on-shell 2-to-2 flat space S-matrix may be extracted from the {\em phase shift formula} \cite{QFTinAdS}:
\ba
S(s)=\lim_{\Dg\to \infty}\frac{1}{(\Dg)^{\beta}} \sum_{|\Delta- \sqrt{s} \Dg|=(\Dg)^\beta} \left(\frac{a_{\Delta}}{a_{\Delta}^{\tt gff}}\right)\, e^{-i \pi(\Delta-2\Df)}
\ea
where we should take $0<\beta<1$. Essentially this formula is an average of phase factors for each exchanged primary, weighted by the OPE density, around a mean energy $\sqrt{s}$.
This formula was proven to hold when the QFT is gapped 
\cite{Komatsu:2020sag,Cordova:2022pbl}, which is the case where in the flat space limit we have $\Df\to \infty$. It is however very natural to suppose the above also holds in the gapless case. After all, when the theory is massive the formula is valid at all energies, including energies much higher than the mass gap where it becomes irrelevant. Alternatively, we can imagine a situation where there is a gapless QFT which subsequently gets deformed by a mass term, in such a way that the scale of the flow is much larger than the mass term, which in turn is much larger than the AdS scale. Either way, the above formula should be correct.

\subsection{S-matrices from extremal correlators}\label{sec:smatext}
We will now show that the construction of extremal solutions is deeply connected to the construction of 2d integrable S-matrices. In particular, in the flat space limit, we will show that these problems become essentially the same.

First let us discuss kinematics. We want to consider extremal correlators which access the flat space limit of the bulk QFT. Interesting dynamics occurs at scales $\Delta\sim \Dg$, i.e. at energies comparable to the QFT scale $m_{\mbox{\tiny QFT}}\sim \Dg/R_{\mbox{\tiny AdS}}$, and in the flat space limit we have $\Dg\gg 1$. Since the RG flow is happening in AdS it is IR regulated. To keep track of this IR cutoff it is convenient to introduce a scale $1\ll\Dir\ll \Dg$, which we can think of as an IR cutoff on the QFT below which curvature effects become relevant. In particular, the discreteness of the spectrum of the theory becomes apparent only below this scale. 

Let us consider then an extremal solution, whose spectrum consists of the operators in $\tS$ together with a tower with dimensions $\Delta_n=2\Df+2n+\gamma_n$ with $\lim_{n\to\infty}\gamma_n\in \mathbb{Z}$. Let us make the assumption that %
its CFT data is described by slowly varying functions:
\ba
\gamma_n\underset{n\sim \Dg}{=}\gamma(s_n)\,, \qquad \frac{a_{\Delta_n}}{a_{\Delta_n}^{\tt GFF}}\underset{n\sim \Dg}{=}\r(s_n)
\ea
with
\ba
s_n=s(h_n) \,, \quad s(\Delta)=\left(\frac{\Delta}{\Dg}\right)^2\,.\quad 
\ea
Thanks to this assumption we can apply the phase shift formula to the extremal solution to obtain
\ba
S(s)= \rho(s)e^{-i \pi \gamma(s)}
\ea
Thus $S(s)$ neatly packages the entire CFT data of the extremal solution. We will now apply the functional sum rules of the bosonic/fermionic free functional bases to constrain the possible functions $\gamma(s),\rho(s)$ and check that the above does behave as an S-matrix. 
We focus on sum rules for functionals with label $h \sim \Dg$, and introduce the shorthand notations:
\ba
s_h:=s(h)=\left(\frac{h}{\Dg}\right)^2 \,, \qquad s_{\mbox{\tiny IR}}:=s(\Dir)=\left(\frac{\Dir}{\Dg}\right)^2\,.
\ea
We will analyse the $\beta$ and $\alpha$ functionals separately. In fact it turns out be convenient to work with slightly modified $\alpha$ functionals, $\tilde \alpha_h=\hat \alpha_h-c_h \hat \beta_h$ which have improved large $\Delta$ behaviour, see appendix \ref{app:asymp}. Using the asymptotic expressions for the functional actions in that appendix, we find:
\ba
\sum_{\Delta} \frac{a_{\Delta}}{a_{h}^{\tt GFF}} \tilde \alpha_h(\Delta)&=0 \\
\Leftrightarrow -\sum_{\Delta}^{\Dir} \frac{a_{\Delta}}{a_{h}^{\tt GFF}} \tilde \alpha_h(\Delta)&=\frac{8}{\pi^2\, \Dg}\,\dashint_{s_{\mbox{\tiny IR}}}^\infty \ud s \frac{ s_h^{\frac 52}}{(s^2-s_h^2)^2} \mbox{Re}\bigg(|S(s)|\mp S(s)\bigg)+|S(s_h)|\,,
\ea
Here we have approximated the sum over states above $\Dir$ by a continuous integral and the $-(+)$ sign corresponds to the bosonic/fermionic basis. Note that the second term on the right arises from summing over states with $\Delta=h+\gamma(s_h)+2k$ with $k$ integer, which is ommited by the principal value prescription on the integral. In the flat space limit we take $\Dg\to \infty$ so that the entire integral is suppressed. As for terms in the sum on the lefthand side they are also suppressed for large $h$, with the exception of the identity for which $\tilde \alpha_h(0)=-a_h^{\tt GFF}$. Overall this then leads to the equation:
\ba
\rho(s)=|S(s)|=1
\ea
Thus to leading order the OPE data of the extremal solution coincides with the GFF one and the function $S(s)$ has unit modulus for real argument. Moving on to the $\beta$ functional equations, using $|S(s)|=1$ and focusing on leading terms we now get:

\ba
\sum_{\Delta} \frac{a_{\Delta}}{a_{h}^{\tt GFF}} \hat \beta_h(\Delta)=0 \Leftrightarrow \frac{2}{\pi}\,\dashint_{0}^\infty \ud s \frac{ s_h}{s^2-s_h^2} \mbox{Re}\bigg(S(s)\mp 1\bigg)=\mbox{Im}\, S(s_h) \label{eq:functionaldisp}
\ea
The $\beta$ sum rules have become an integral equation for the anomalous dimensions encoded in $S(s)$. It is easy to construct solutions to this equation. Let us introduce a function $\mathcal S(s)$ satisfying
\begin{enumerate}
\item $\mathcal S(s)$ is analytic on the upper half-plane and $\mathcal S(s^*):=\mathcal S^*(s)$.
\item It is crossing symmetric:
$
\mathcal S(s)=\mathcal S(-s)
$
\item It is bounded on the upper half-plane, $|\mathcal S(s)|\leq 1$
\end{enumerate}
Note that these are precisely those properties satisfied by a 2-to-2 S-matrix. Any such function satisfies a dispersion relation (from Cauchy's formula):\footnote{Here we set
\ba
\mathcal R f(z):= \lim_{\epsilon \to 0} \frac 12 [f(z+i\epsilon)-f(z-i\epsilon)]\,, \quad \mathcal I f(z):=\lim_{\epsilon \to 0} \frac 1{2i} [f(z+i\epsilon)-f(z-i\epsilon)]\,,
\ea
}
\ba
\mathcal S(s)=\pm 1+\frac{2}{i\pi}\int_0^\infty \ud s' \frac {s}{[(s')^2-s^2]} \mathcal R\left[\mathcal S(s')\mp 1\right] \qquad \mbox{for}\ \mbox{Im}\, s>0 \label{eq:dispS}
\ea
Setting now
\ba
S(s)=\lim_{\epsilon \to 0} \mathcal S(s+i\epsilon)\,, \qquad s\in \mathds R_{>0}
\ea
it is easy to see that \reef{eq:functionaldisp} is indeed satisfied from \reef{eq:dispS}.

The S-matrices suitable for describing extremal solutions must satisfy crossing and saturate unitarity. A general ansatz for such S-matrices is given as a product of Castillejo-Dalitz-Dyson (CDD) factors:
\ba
S(s)=\prod_{i=1}^P\, \frac{s- \mu_i}{s-\mu_i^*}
\ea
with $\mbox{Im}\,\mu_i>0$ and further constrained so that $S^*(-s)=S(s)$.

\subsection{Extremal correlators from S-matrices}
\label{sec:smatmethod}
So far we have discussed only those functional sum rules with $h\sim \Dg$. As a consequence all information about the fixed inputs $\tS$, contained in the sum over states below $\Dir$, has dropped out.
At this level, any $\mathcal S(s)$ satisfying the assumptions spelled out above can in principle be associated to an extremal solution. In order to determine which precise S-matrix must be associated to a particular extremal solution, we must match the high energy with the low energy CFT data by considering the sum rules with $h=O(1)$. 
Let us therefore take a closer look at sum rules with $h<\Dir$. They take the form:
\ba
-\sum_{\Delta}^{\Dir} \frac{a_{\Delta}}{a_h^{\tt GFF}}\, \check \alpha_h(\Delta)&=\frac{A_h}{\pi^2 \Delta^6_{\mbox{\tiny gap}}}\, \int_{s_{\mbox{\tiny IR}}}^\infty \,\ud s \frac{1}{s^4-s_h^4}\mbox{Re}[1\mp S(s)]\\
-\sum_{\Delta}^{\Dir} \frac{a_{\Delta}}{a_h^{\tt GFF}}\, \hat \beta_h(\Delta)&=\frac{B_h}{\pi^2 \Delta^2_{\mbox{\tiny gap}}}\, \int_{s_{\mbox{\tiny IR}}}^\infty \,\ud s \frac{1}{s^2-s_h^2}\mbox{Re}[1\mp S(s)] \label{eq:ab}
\ea
As $s\to 0$ the S-matrix must go to $\pm 1$. We can then choose to work with a functional basis such that the righthand side of these equation is suppressed, namely the fermionic/bosonic basis if $S=-1$ or $+1$ respectively. Therefore these equations can be solved numerically to leading order with zero righthand side. The matching at this order is trivial: this numerical solution will asymptote to a nearly free spectrum as $\Delta\to \Dir$. At subleading order however there is a non-trivial matching between a candidate S-matrix on the righthand side and the low energy CFT data. More generally we can use these equations as follows:

{\bf Step 0:}
\begin{enumerate}
\item Write down a trial S-matrix made up of a finite number $P$ of CDD factors on the right hand side. This finite number should line up with the number of parameters in the set $\tS$, dimensions and OPE coefficients. 
\item Solve equations \reef{eq:ab} numerically for the IR data (setting the right hand side to zero) excluding the first $P$ functionals (distributed equally among $\alpha$ and $\beta$ type functionals).
\end{enumerate}
Then we iterate:

{\bf Step i:}
\begin{enumerate}
\item Plug in the IR CFT data determined at step $i-1$ into the first $P$ functional equations to determine the parameters in $S(s)$, determining a solution $S^{(i)}(s)$.
\item Plug $S^{(i)}(s)$ into the remaining functional equations and solve for the IR CFT data.
\end{enumerate}
In practice convergence of this method is achieved in a handful of steps. The output is the full set of CFT data describing the extremal solution, in the IR as given by a finite set of discrete data, in the UV as an S-matrix. In the present approach $\Dir$ plays the role of $\Delta^*$ in the hybrid bootstrap method of section \ref{sec3}, in the sense that it represents the scale splitting the IR states, which we solve for numerically, from UV ones, which we solve for analytically. Unlike the previous method however, this is not the scale where the spectrum is weakly coupled, but rather where a continuous description in terms of an S-matrix takes over. We can trust the results of our method by 1. checking that the output is unchanged as $\Dir$ is changed 2. checking that the final set of CFT data satisfies exact sum rules up to corrections suppressed by $1/\Dg^2$.

\subsection{Bulk operator reconstruction}
\label{sec:bulk}
We now switch gears to discuss another aspect of the relation between extremal solutions and QFTs in AdS$_2$. Concretely we will explain how to any extremal solution it is possible to canonically associate one or more operators which behave as local fields in a dual AdS$_2$ space. This construction was essentially already described in previous work \cite{Levine:2023ywq,Levine:2024wqn}, so that the contents of this subsection are mostly a review.

Local fields in AdS can be described in terms of the CFT living at the boundary of the latter using what is known as the boundary operator expansion (BOE). Introduce the AdS$_2$ metric,
\ba
\ud s^2=\frac{\ud u^2+\ud x^2}{u^2}
\ea
A bulk field $\Psi(y,x)$ (which need not be elementary) can be written as
\ba
\Psi(u,x)=\sum_{\Delta} \mu^{\Psi}_{\Delta}\, \mathcal C(u^2 \partial^2_x) \mathcal O_{\Delta}(x)
\ea
with $\mathcal C$ a known function fixed by the AdS symmetry, and where the sum runs over primaries of the boundary 1d CFT. Below we will be interested in what we call AdS form factors, which are three point correlators of one bulk field and two boundary operators. We can write them as
\ba
\langle \Psi(u,x) \phi(x_1)\phi(x_2)\rangle=\frac{1}{x_{12}^{2\Df}}\, \mathcal H(w)\,,\qquad w=\frac{(x_1-x_2)^2 u^2}{[(x-x_1)^2+u^2][(x-x_2)^2+u^2]}\,.
\ea
Applying the BOE decomposition leads to the representation
\ba
\mathcal H(w)=\sum_{\Delta} b_{\Delta}\, H_{\Delta}(w)\,, \qquad H_{\Delta}(w)=\mathcal N_{\Delta} w^{\frac{\Delta}2}\, _2F_1(\tfrac{\Delta}2,\tfrac{\Delta}2,\Delta+\tfrac 12,w)\label{eq:H}
\ea
where we set $b_{\Delta}:=\mu^{\Psi}_{\Delta} \lambda^{\phi \phi}_{\mathcal O_{\Delta}}$ and the normalisation factor is chosen by convenience to be
\ba
\mathcal N_\Delta=\frac{\Gamma(\tfrac{\Delta}2)^2}{\pi \Gamma(\Delta+\tfrac 12)}\,.\label{eq:N}
\ea
For spacelike separated operators we have $w\in (0,\infty)$. In this kinematical region the form factor $\mathcal F(w)$ should be analytic, but this is not guaranteed by the BOE which converges only for $w<1$. Instead, the coefficients $\mu_{\Delta}^{\Psi}$ must be delicately tuned so as to avoid unphysical singularities \cite{Kabat:2016zzr}. These tunings are ensured by sum rules that these coefficients must satisfy:
\ba
\sum_{\Delta} b_{\Delta} \theta_n^{\mathcal A}(\Delta)=0\,, \qquad n=1\,\ldots, \infty \label{eq:sumrulelocal}
\ea
In \cite{Levine:2024wqn} it was shown that there are a great many possible choices for the functional actions $\theta_n^{\mathcal A}(\Delta)$. Given any spectrum of the form
\ba
\mathcal A:=\{\Delta_n^{\mathcal A}=2\alpha+2n+\gamma_n\}_{n=1}^\infty\,,\qquad  \gamma_n\underset{n\to \infty}=O(n^{-\epsilon})\,, \quad \epsilon>0 \label{eq:allowedA}
\ea
then a complete set of sum rules is obtained by setting:
\ba
\theta_n^{\mathcal A}(\Delta)=\prod_{m=1,m\neq n}^\infty \frac{(\Delta-\Delta^{\mathcal A}_m)(\Delta+\Delta^{\mathcal A}_m-1)}{(\Delta_n^{\mathcal A}-\Delta_m^{\mathcal A})(\Delta_n^{\mathcal A}+\Delta^{\mathcal A}_m-1)} 
\label{eq:localfs}
\ea
Note that in particular we have the duality conditions
\ba
\theta_n^{\mathcal A}(\Delta_m^{\mathcal A})=\delta_{n,m}\,, \qquad n,m\geq 1\,.
\ea
The sum rules \reef{eq:sumrulelocal} completely encapsulate the constraints of locality, and moreover they constrain the large $w$ behaviour of a form factor. In particular applying sum rules with a given value of $\alpha$ demand $\mathcal F(w)=O(w^{\alpha+1-\eta})$ for some $\eta>0$.

Suppose we are given an extremal spectrum. Is it possible to construct a local AdS operator whose BOE expansion consists of the operators in that solution? I.e. can we find solutions to the equations:
\ba
\sum_{i=1}^K \hat b_i \theta^{\mathcal A}_n (\hat \Delta_i^{\tE})+\sum_{m=0}^\infty b_m \theta^{\mathcal A}_n(\Delta_m^{\tE})=0\,, \qquad n=1\,\ldots,\infty
\ea
Thanks to the freedom in the choice of $\mathcal A$ the answer is in general yes. This can be seen by simply choosing $\mathcal A$ to essentially coincide with the spectrum of operators in the extremal solution. In particular let us choose $\mathcal A=\{\Delta^{\tE}_m\}_{m=1}^\infty$. This choice is compatible with \reef{eq:allowedA} thanks to the asymptotic freedom conjecture. Then the locality sum rules have the manifest solution
\ba
b_m=-b_0 \theta^{\mathcal A}_m(\Delta_0^{\tE})-\sum_{i=1}^K \hat b_i \theta_m^{\mathcal A}(\hat \Delta_i^{\tE}) \label{eq:solbs}
\ea
in terms of $K+1$ parameters. Defining local blocks as
\ba
\mathcal L_{\Delta}^{\mathcal A}(w)=H_{\Delta}(w)-\sum_{n=1}^\infty \theta_n^{\mathcal A}(\Delta) H_{\Delta_n^{\mathcal A}}(w)
\ea
then the most general form factor with a BOE matching an extremal solution is written
\ba
\mathcal H(w)=b_0 \mathcal L_{\Delta_0^{\tE}}^{\mathcal A}(w)+\sum_{i=1}^K \hat b_i \mathcal L_{\hat \Delta_i^{\tE}}^{\mathcal A}(w)
\ea
This solution contains a number of free parameters. One such parameter simply encodes the overall normalisation of the bulk operator and cannot be fixed by locality. The other parameters must be input by hand as they reflect different possibilities in our construction of a local bulk operator. 
To understand these ambiguities, and for definiteness, let us suppose that the extremal solution is of bosonic type. Then the extremal spectrum consists of $K$ `bound states' or `single trace' operators, together with an infinite tower of double trace operators. Let us denote bulk fields dual to such bound states as $\Psi_i$ and $\Phi$ as the dual to the CFT operator $\phi$. Then the candidate local operators are $\Psi_i, \Phi^2$ and (AdS bulk) derivatives thereof. In terms of BOE content these operators can be {\em defined} as solutions to locality above where all but one coefficient on the RHS is switched off. 

\subsubsection{An example}
As an important example of this construction, let us describe the extremal form factors associated to the GFF correlators of external dimension $\Df$. In this case we choose 
\ba
\Delta_m^{\mathcal A}=h_m:=2\alpha+2m\,, \qquad \Rightarrow \quad \left\{ \begin{array}{cc} \alpha=\Df & \mbox{Boson}\\ \alpha=\Df+\frac 12& \mbox{Fermion} \end{array}\right.
\ea
The functional actions corresponding to these choices are denoted $\theta^{(\alpha)}(\Delta)$ and they admit closed form expressions,
\ba
\theta_n^{(\alpha)}(\Delta)=\frac 2\pi \frac{\left(\frac{h_n-2\alpha}2\right)_{\frac 12+2\alpha}}{\left(\frac{\Delta-2\alpha}2\right)_{\frac 12+2\alpha}}
\,
\frac{(2h_n-1)\,\sin\left[\tfrac \pi 2 (\Delta-h_n)\right]}{(h_n+\Delta-1)(\Delta-h_n)}
\ea
In particular certain local blocks can be written down in closed form:
\ba
\mathcal L^{(\alpha)}_{h_0}(w)=w^{\alpha}&=\sum_{n=0}^\infty \theta_n^{(\alpha)}(h_0) H_{h_n}(w)\\
&=
\sum_{n=0}^\infty (-1)^n b_{h_n}^{\tt gff} H_{h_n}(w)
\ea
where by definition
\ba
b_{\Delta}^{\tt gff}:= \frac{2\,\left(\frac{\Delta-2\alpha}2\right)_{\frac 12+2\alpha}}{\Gamma\left(\frac 12+2\alpha\right)}
\,
\frac{(2\Delta-1)}{(\Delta+2\alpha-1)(2\alpha-\Delta)} \label{eq:bgff}
\ea
which is the analytic continuation in $n$ of $(-1)^n\theta_n^{(\alpha)}(h_0)$.

\subsection{Form factors from extremal correlators}
We now describe the flat space limit of form factors associated to an extremal solution. In this case we have \cite{Levine:2023ywq}:
\ba
F(s)=\lim_{\Dg\to \infty}\frac{1}{(\Dg)^{\beta}} \sum_{|\Delta- \sqrt{s} \Dg|=(\Dg)^\beta} \left(\frac{b_{\Delta}}{b_{\Delta}^{\tt gff}}\right)\, e^{-i \frac{\pi}2(\Delta-2\Df)}\label{eq:formfactor}
\ea
with $F(s):=\langle k_1,k_2| \Psi(0)\rangle$. The averaging kernel is chosen so that for free theory we reproduce the expected outcome, essentially because the average becomes trivial. From the discussion in section \ref{sec:bulk} it is sufficient to consider the limit for a single local block of fixed dimension $\hat \Delta$, denoting the result as $F_{\hat \Delta}$. 
When applying it to an extremal solution the averaging is trivial, and we get:
\ba
F_{\hat \Delta}(s) = e^{-i \pi \frac{\gamma_n}2}\, \frac{b_{\Delta_n} (-1)^n}{b^{\tt gff}_{\Delta_n}} \bigg|_{\Delta_n=\Dg\sqrt{s}}
\ea
Now note
\ba
b_{\Delta_n}(-1)^n&=(-1)^n\theta^{\mathcal A}_n(\hat \Delta)\\
&= \prod_{m=1,m\neq n}^\infty \left|\frac{(\hat \Delta-\Delta_m)(\hat \Delta+\Delta_m-1)}{(\Delta_n-\Delta_m)(\Delta_n+\Delta_m-1)}\right|
\ea
Taking logarithms and approximating the resulting series as an integral we can find
\ba
\log[b_{\Delta_n}(-1)^n]&=\Dg^2 \dashint_{s_1}^{\infty} \frac{\ud s'}{2\sqrt{1+4\Dg^2 s'}}\,  \log \left|\frac{s'-\hat s}{s'-s}\right| \\
&-\frac 12 \dashint_{s_1}^\infty \ud s' \gamma'(s')\,\log \left|\frac{s'-\hat s}{s'-s}\right|+\ldots
\ea
with 
\ba
s=\left(\frac{\Delta_n}{\Dg}\right)^2\,, \qquad \hat s=\left(\frac{\hat \Delta}{\Dg}\right)^2\,,\qquad s_1=\left(\frac{\Delta_1}{\Dg}\right)^2
\ea
and where the ellipsis represents contributions which are independent of $s$. The first integral can be done straightforwardly, while on the second we can do integration by parts. Taking the large $\Dg$ limit we get
\ba
\log[b_{\Delta_n}(-1)^n]&=\frac 12 \dashint \frac{\ud s'}{s'} \frac{s}{s-s'} \left[\gamma(s')-\gamma(0)\right]+\left(\Delta_1-\frac 52\right)\log(\Delta_n) 
\ea
On the other hand we have
\ba
\log[b_{\Delta_n}^{\tt gff}]\underset{n\to \infty}{=} \left(\Delta_0^{\tt gff}-\frac 12\right)\, \log(\Delta_n)+\ldots
\ea
It is then easy to see that
\ba
F_{\hat \Delta}(s)=\lim_{\epsilon \to 0} \mathcal F_{\hat \Delta}(s+i\epsilon)\,,
\ea
with
\ba
\mathcal F_{\hat \Delta}(s)=C(\Df,\hat \Delta) \,
(-s)^{\frac{\gamma(0)}2}\exp\left[\int_0^\infty \frac{\ud s'}2 \left[\gamma(s')-\gamma(0)\right] \frac{s}{s'(s'-s)}\right] \label{eq:formfactorflat}
\ea
for some constant $C$. This expression is manifestly analytic way from $s>0$, i.e. it is local, and it is easily checked that
\ba
\lim_{\epsilon\to 0^+}\, \frac{\mathcal F_{\hat \Delta}(s+i\epsilon)}{\mathcal F_{\hat \Delta}(s-i\epsilon)}=e^{-i\pi \gamma(s)}=S(s)
\ea
which is Watson's equation for a form factor. We conclude that in the flat space limit our solution for the AdS form factor precisely agrees with expectations.

\section{Applications to QFT$_2$/CFT$_1$}
\label{sec5}
In this section we will bootstrap extremal solutions which represent RG flows in AdS and extract their flat space limit.
\subsection{The OPE max/$\Phi^4$ family}
\label{sec:single}
We begin by considering what is perhaps the simplest example of a family of extremal solutions which can be interpreted as describing a QFT in AdS$_2$. It has been previously studied in \cite{Mazac:2018ycv,Paulos:2019fkw,Levine:2023ywq}. This family is defined by the choice:
\ba
\tS=\{\mathds 1,(\hat \Delta_0,\bullet)\}
\ea
That is, we include the identity as well as an operator of dimension $\hat \Delta_0$, and thus the family is labeled by this number. For generic values of this number this a bosonic type extremal solution. It includes as a special case $\hat \Delta_0=2\Df$\, where we already know that this solution is the generalized free boson correlator. We also expect that it should be somehow connected to the generalized free fermion solution as we push $\hat \Delta_0\to 1+2\Df$. Note that the OPE coefficient $a_{\Delta_0}$ is determined by the extremality conditions, and in fact, its value for the extremal solution is an upper bound valid for any CFT correlator.\footnote{This bound follows from the $\alpha_0$ sum rule of this extremal solution.\label{fn:bound}}

One interesting regime is when $\hat \Delta_0$ is close to $2\Df$. In this case the extremal solution can be constructed perturbatively and matched with the theory of a free scalar field $\Phi$ in AdS$_2$ with mass $m^2=\Df(\Df-1)$ and a $g_B\, \Phi^4$ interaction, with coupling $g_B=O(\hat \Delta_0-2\Df)$.
Instead, the strong coupling regime corresponds to $\hat \Delta_0\to 1+2\Df$. In this limit, dimensions of operators approach $1+2\Df+2n$, but this approach is non-uniform over $n$. 
We would like to understand the scale at which dimensions of operators transition from fermionic type to bosonic type, i.e. how to determine the scale $\Dg$ for this family of correlators. Let us set $g_F=1+2\Df-\hat \Delta_0$. Around strong coupling the extremal correlator is well described at low energies by a free fermion deformed by an irrelevant interaction of coupling $g_F$, schematically $(\bar \Psi \slashed{\partial} \Psi)^2$. We then have
\ba
\gamma_n^F:=\Delta_n-\Delta_n^F \underset{n\gg 1, n\ll \Dg}\sim g_F\, n^2
\ea
This effective theory breaks down at energies $\Delta\sim 1/\sqrt{|g_F|}$. Hence we can set:
\ba
\Dg:=\frac{1}{\sqrt{1+2\Df-\hat \Delta_0}} \label{eq:DgIsing}
\ea
The UV completion of this EFT description of the extremal correlator is given by the free bosonic theory. Indeed,
since the $\Phi^4$ interaction is relevant in the bulk, no matter how large we make $g_B$ we should trust perturbation theory for high energies. In particular, anomalous dimensions of operators should decay eventually, 
\ba
\gamma_n^B:=\Delta_n-\Delta_n^B\underset{n\gg \Dg}=c(g_B)\left(\frac{\Dg}{n}\right)^2
\ea
with $c(g_B)=O(1)$.

We can think of this family of extremal solutions as a simple, universal approximation to bulk RG flows involving a single scale. The flow starts off in the UV as a free boson and ends in the IR as a free fermion. The simplest example of this is a $\Phi^4$ type theory in AdS$_2$. From the bulk perspective, to define this as our theory one would need a priori to tune an infinite number of relevant couplings. This tuning is not the same as the one performed in flat space, as we are always free to add new couplings of order of the AdS scale without modifying the flat space RG flow. 
The extremal family of solutions mashes up all such couplings into just two, namely $\hat \Delta_0$ and $\Df$. This is because other couplings control how operators of the form $\phi^n$ couple to the $\phi\times \phi$ OPE, but such states cannot appear in a single correlator extremal solution -- at least not without explicitly adding them by hand as part of our fixed inputs $\tS$. The flat space limit is where $\hat \Delta_0$ is sent to $1+2\Df$. This limit should be independent of $\Df$ as long it is held fixed, or more precisely, as long as we do not scale it to infinity in taking the flat space limit. 

\begin{figure}[t]
    \centering
    \hspace{-1cm}
    \includegraphics[width=0.8\linewidth]{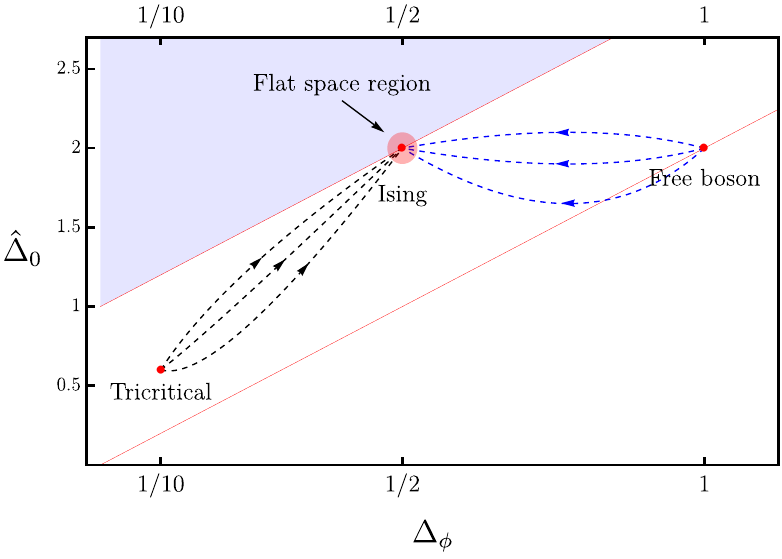}
    \caption{Possible one parameter flows. The allowed phase space is labeled by $\Df$ and $\hat \Delta_0$, where the OPE is $\phi \times \phi=\mathds 1+ \mathcal O_{\hat \Delta_0}+\ldots$.    
    The colored region is excluded by bootstrap bounds and the top/bottom red curves correspond to the generalized free fermion/boson solutions. The marked points correspond to BCFT correlators for the indicated bulk 2d CFTs and the dashed lines represent possible flows between them. For these flows the flat space limit corresponds to a region infinitesimally close to the IR fixed point.
    }
    \label{fig:flows}
\end{figure}

As an application, consider a flow from the 2d free boson CFT in the UV to the 2d Ising CFT in the IR. More precisely, we begin by considering a free massless 2d boson $\Phi$ in AdS$_2$ with Dirichlet boundary conditions. This has a boundary description equivalent to a generalized free boson of dimension $1$. We then turn on a $\Phi^4$ deformation with a coupling $g_B$ that is eventually taken to satisfy $g_B R_{\mbox{\tiny AdS}}^2\gg 1$. We can track the flow via a curve in the space $(\Df,\hat \Delta_0)$ as exemplified in figure \ref{fig:flows}.
The precise curve will depend on the choice of relevant couplings, with the mapping non-trivial. However, we do know that for each such point we can provide a unique extremal solution, which should be thought of as an approximation to the exact correlator. The flat space limit is the limit where we approach the Ising CFT in the IR with a specific choice of boundary conditions, which given our UV starting point should be the $Z_2$ preserving ones, corresponding to the so-called ordinary transition. This BCFT contains in its spectrum a primary of dimension $1/2$, whose correlator is precisely the same as a generalized free fermion -- the end point of our flow. Hence the natural conjecture is that the curve in $(\Df,\hat \D_0)$ space approaches in the flat space limit the point $(\frac 12,2)$. 

\begin{figure}
    \centering
    \includegraphics[width=1\linewidth]{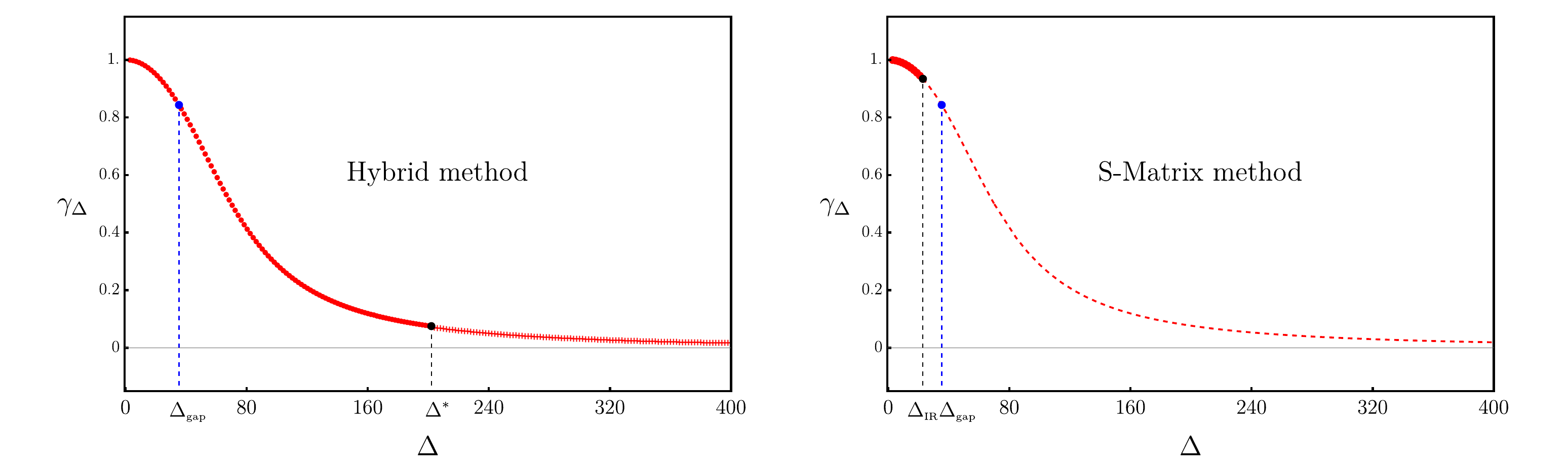}
    \caption{Comparison between the hybrid and S-matrix methods for constructing an extremal solution. Dots indicate dimensions of operators obtained numerically, while dashed lines represent analytic results for the remainder of the spectrum. The solution for $\gamma_0^F=-8\times10^{-4}$ was obtained in figure \ref{fig:nearflatgammaversus} and is reproduced on the left. On the right, we show the same solution obtained with the S-matrix method. Both curves and numerical data points match to a precision $\Dg^{-2}=O(10^{-4})$. 
       \label{fig:1dofbosoncompare}}
    \end{figure}

\begin{figure}[h]
    \centering
    \includegraphics[width=0.85\linewidth]{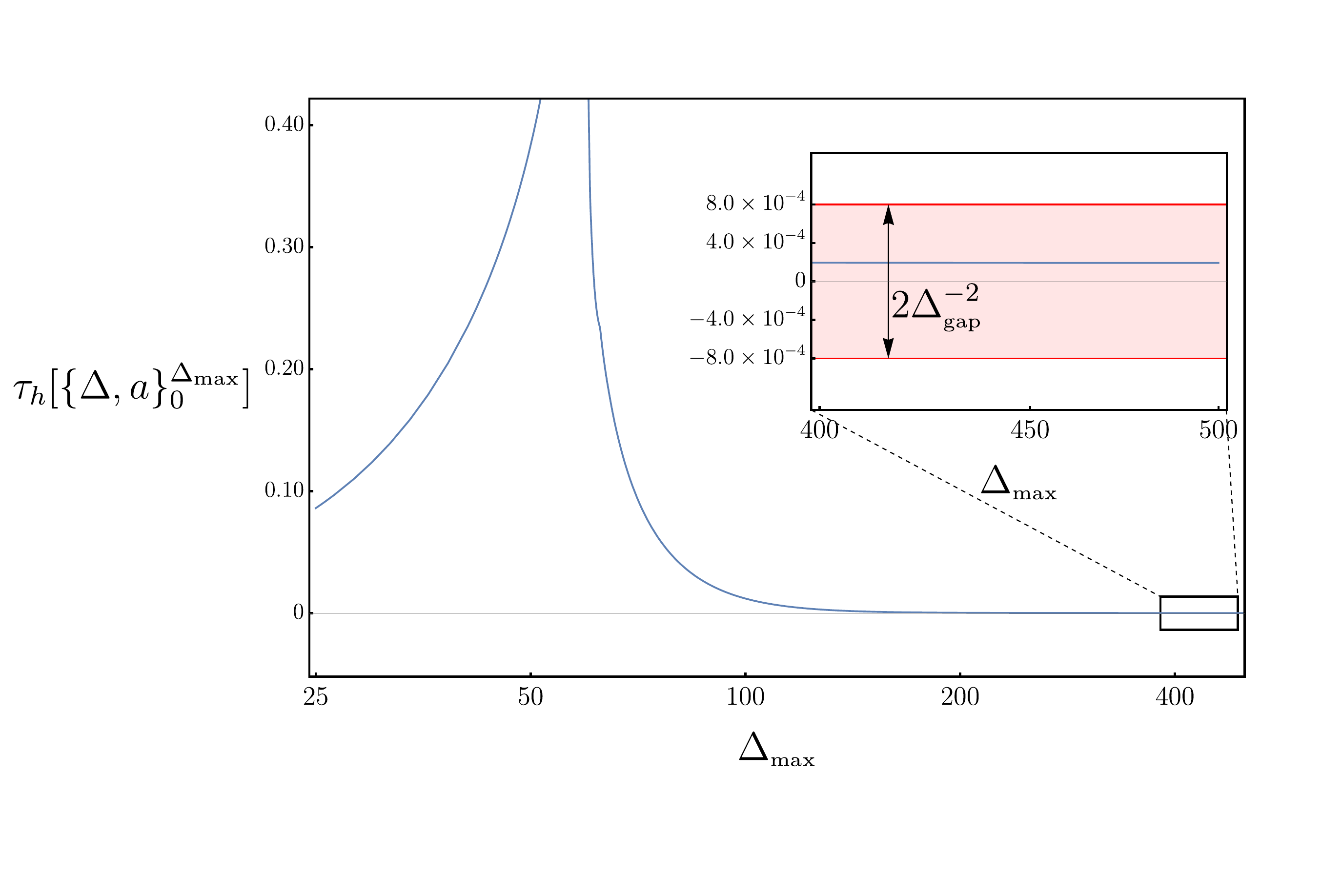}
    \caption{We act with a large$-h$ functional on a full solution: the IR CFT data is obtained numerically, while the UV CFT data comes from the flat space computations. In this case, we have 11 operators in the IR, and we act with the single zero functional labeled by $h=2\Df+n = 60$, $\Df = 1$. More details on this basis of functionals can be found in Appendix \ref{app:single}. The solution is labeled by $\gamma_0^F = -4\times 10^{-4}$. As we increase the number of operators the functional acts upon, the result initially rises as we move away from GFF. Eventually, it converges to a constant of order $\D_{\mbox{\tiny gap}}^{-2}$.}
    \label{fig:taueval}
\end{figure}

Let us finally turn to an analysis of the extremal solutions corresponding to this family of correlators in the flat space region. In section \ref{sec:appfinetune} we have already obtained these.\footnote{As discussed, we find the result is qualitatively unchanged if we shift $\Df$ by any $O(1)$ number. In the application to a flow terminating at Ising this corresponds to adding a mass term for the fermion sitting at the boundary in the BCFT.}
As a sanity check, in figure \ref{fig:1dofbosoncompare} we show a comparison of the spectrum of the extremal solution obtained using the hybrid bootstrap method versus the S-matrix method described in section \ref{sec:smatmethod}.
Both methods agree within each others accuracy, but the latter method is much more efficient, as we need only to numerically solve for a much smaller set of operators. At the end of our procedure we have produced a complete set of CFT data which we claim solves all crossing constraints. As a non-trivial test that this is indeed the case, we can take this data and plug it into a sum rule of a completely different functional basis than the one used to produce it, and check that it satisfies it. We do precisely this for a single-zero functional in figure \ref{fig:taueval}.

The spectrum of these solutions has the expected qualitative features: in the IR the spectrum approaches that of a generalized free fermion, while in the UV it transitions back to a free boson. 
Furthermore we have found that the CFT data is quantitatively described by a smooth function describing a very simple integrable S-matrix. Defining $s:=(\Delta/\Dg)^2$ with $\Dg$ given above in \reef{eq:DgIsing} this S-matrix takes the form
\ba
S(s)=\frac{s-im}{s+im} \label{eq:cddzero}
\ea
The parameter $m$ can be found in two distinct but equivalent ways by matching to the microscopic CFT data. Firstly, in the IR the fermionic EFT picture predicts a specific growth of anomalous dimensions,
\ba
\gamma_n^F\underset{n\gg 1}= \gamma_0^F c(\Df) n^2 \sim -\frac{c(\Df)}4 \, s_n\,, \qquad 
c(\Df)=\frac{4^{\Delta _{\phi }+1} \Gamma \left(\Delta _{\phi }\right) \Gamma \left(\Delta _{\phi }+1\right) \Gamma \left(\Delta _{\phi }+2\right)}{\pi  \Gamma \left(\Delta _{\phi }+\frac{3}{2}\right)
   \Gamma \left(2 \Delta _{\phi }+\frac{5}{2}\right)}\,.
\ea
Matching this to the S-matrix in the IR limit we find
\ba
\gamma_n^F=-1-\frac{1}{i \pi} \log S(s_n)\underset{s_n\ll 1}{\sim} - \frac{2}{m\, \pi}  s_n\qquad \Rightarrow\qquad m=\frac{8}{\pi c(\Df)} 
\ea
Alternatively we can use the functional sum rules to perform the matching, as explained in section~\ref{sec:smatmethod}. Acting with $\beta_0^F$ we have
\ba
\sum_{\Delta=0}^{\Dir} a_{\Delta} \beta_0^F& =-\frac{16 \Df}{\pi^2 c(\Df)\Dg^2} \int_{s_{\mbox{\tiny IR}}^2}^\infty\, \frac{\ud s}{s^2}\, \mbox{Re}[1+S(s)]\\
\Leftrightarrow 2\Df \gamma_0^F &\sim -\frac{16\Df}{c(\Df)\pi m \Dg^2}\, 
\ea
leading to the same result.

This S-matrix is then meant to describe the flat space RG flow of a free boson deformed by a $\Phi^4$ interaction, with mass term tuned so that the theory flows to the Ising CFT in the IR. It is essentially the only S-matrix we could have written down, given the sparse number of inputs in the extremal solution, the fact that it had to interpolate correctly between bosonic and fermionic free S-matrices in the UV and IR respectively, and that it had to saturate unitarity. To be clear, we do not claim that this is the exact S-matrix describing the $\Phi^4$ flow. For instance, that S-matrix is certainly not integrable. Rather, just like the extremal solution is an approximation to the exact CFT correlator by neglecting couplings to ``higher trace'' operators like $\phi^n$, the above is an S-matrix which approximates the exact one by neglecting particle production. 

There is another way to understand why we obtained the S-matrix above. In fact, it is known that there is an integrable deformation\footnote{Whether such integrable deformations can survive being placed in AdS is an open question, see \cite{Antunes:2025iaw} for a recent negative result.} of the (supersymmetric) tricritical Ising model that terminates at Ising, and in the IR the scattering of the massless goldstinos has been argued to be described precisely by the S-matrix above \cite{Zamolodchikov:1991vx}. To get an RG flow in AdS, we can start off with the tricritical CFT with the so-called (2,2)$_4$ boundary condition \cite{Antunes:2024hrt}. The 1d CFT has an operator of dimension $1/10$ which fuses into the identity as well as a primary $\psi_{3/5}$ of dimension $3/5$. Upon adding the integrable bulk deformation, plus suitable couplings of order the AdS scale, we generate a flow. It is natural to conjecture that under this flow the operator $1/10$ goes to the boundary spin operator of dimension $1/2$ in the Ising CFT, while $\psi_{3/5}$ becomes the displacement with dimension $2$, as exemplified once again in figure \ref{fig:flows}. As we approach the Ising point we must recover the exact S-matrix describing the tricritical to Ising flow, and indeed we do.\footnote{Strictly speaking the physical meaning of this S-matrix is unclear; after all the Ising CFT is not the same as a free fermion, and the goldstinos are really in the twisted sector of the theory. We thank A. Antunes for discussions on this point.}

Finally we may also determine a form factor associated to this family of solutions \cite{Levine:2024wqn}. The most interesting case corresponds to the flat space limit. Using the formula \reef{eq:formfactorflat} derived in the previous section we can determine\footnote{It is of course possible to obtain an expression valid for any complex $s$, from which we can derive the ones shown here.}
\ba
\lim_{\epsilon \to 0}\mathcal F(s\pm i \epsilon)&\propto \frac{\sqrt{s}}{\left(m^2+s^2\right)^{\frac 14}}\,\exp\left\{\frac{D(i m/s)}\pi \pm\frac 12 \log[S(s)]\right\}\,, \qquad s>0\\
\mathcal F(s)&\propto\frac{\sqrt{-s}}{\left(m^2+s^2\right)^{\frac 14}}\,\exp\left\{\frac{D(im/ s)}\pi\right\}\,, \qquad s<0
\ea
with $D(z)$ the Bloch-Wigner function\,,
\ba
D(z)=\mbox{Im}\,[ \mbox{Li}_2(z)]+\mbox{arg}(1-z)\log |z|\,.
\ea
Note we have
\ba
\mathcal F(s)&\sim \sqrt{s}&\,, \qquad s&\ll 1\\
\mathcal F(s)&=O(1)&\,, \qquad s&\gg 1
\ea
which is consistent with the expected behaviour of the form factors of operators $\bar \Psi \Psi$ and $\Phi^2$ in the IR fermionic/UV bosonic theory, see appendix \ref{app:fermion}. In figure 
\ref{fig:formfactor} we show that this predicted analytic form of the form factor perfectly matches numerical results for the BOE coefficients obtained by direct application of formulae \reef{eq:solbs} and \reef{eq:localfs}. 

\begin{figure}
    \centering
    \hspace{-2cm}
    \includegraphics[width=0.6\linewidth]{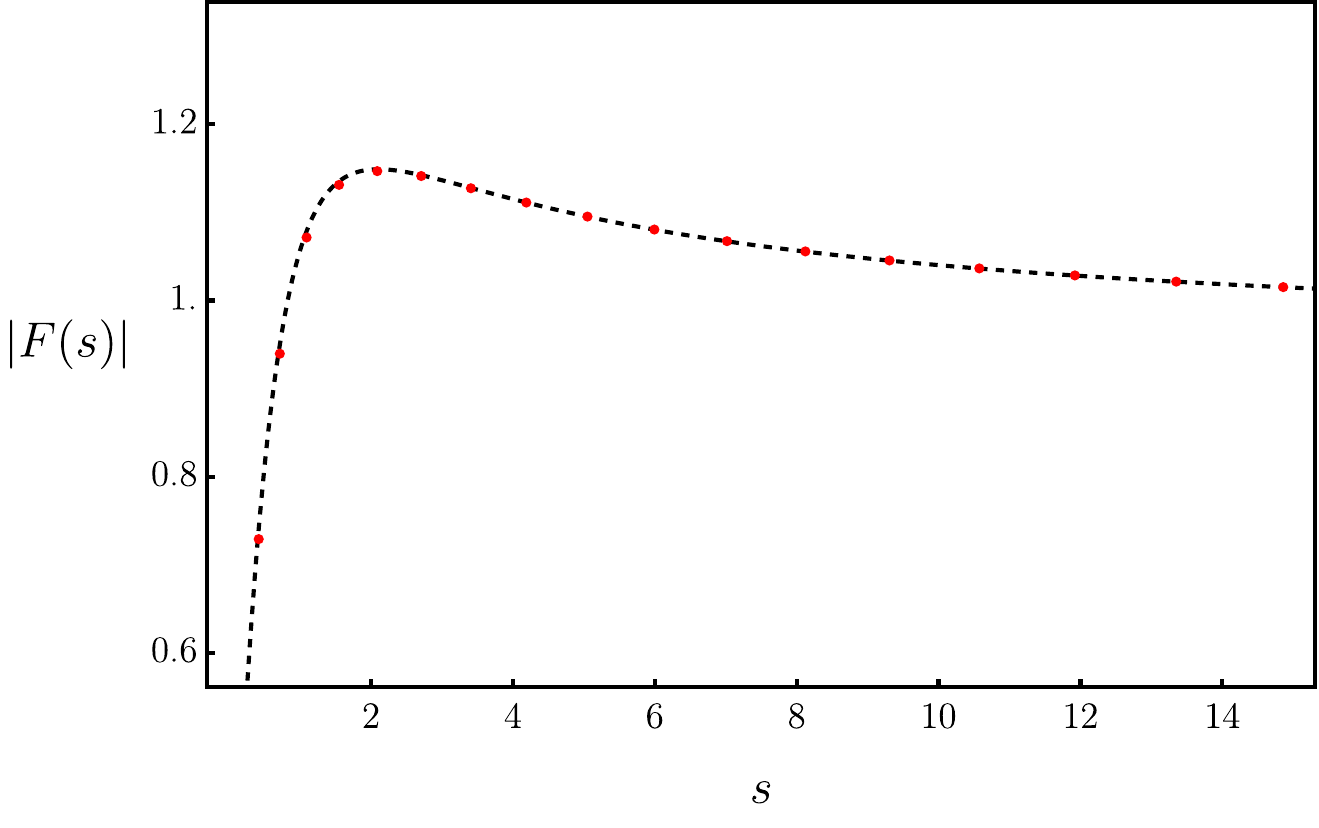}
    \caption{Form factor corresponding to the single parameter flow. Analytic prediction is the dashed curve versus direct numerical computation shown as black dots at values $s_n=\Delta_n^2/\Dg^2$ with $\Dg=100$.}
    \label{fig:formfactor}
\end{figure}

\subsection{Fermion coupled to an extra state}
\label{sec:fermextra}
We will now examine the next to simplest family of extremal solutions, where the set of inputs consists of a new operator. We take:
\ba
S=\{\mathds 1,(\hat \Delta,\hat a)\}
\ea
In this case extremal solutions are of fermionic type, and so for generic values of $(\hat \Delta,\hat a)$ the extremal spectrum rapidly converges to that of a generalized free fermion as we approach the UV. For instance, for small $\hat a$ the extremal solution can obtained by starting with a free massive Majorana fermion in AdS$_2$ and weakly coupling it to a pseudoscalar field dual to a primary of dimension $\hat \Delta$. Denoting this coupling $g\propto \hat a$ the extremal correlator is then simply
\ba
\mathcal G(z)=\mathcal G^{\tt gff}(z)+g\, \mathcal P^{F}_{\hat \Delta}(z)
\ea
Here $\mathcal P_\Delta^F$ is the Polyakov block, a crossing symmetric sum of Witten exchange diagrams in AdS$_2$ \cite{Mazac:2018ycv}. It is the same as the crossing block $\mathcal Q_\Delta^{\bD}$ introduced in \reef{eq:polyblock} specialized to the free fermionic basis. Examining the OPE of this correlator we find that the extremal spectrum contains operators with dimension 
\ba
\Delta_n=2\Delta_\phi+2n+1+\gamma_n\,, \quad \gamma_n=-\frac{\hat a}{a_n^F}\,\beta^F_{n}(\hat \Delta)\,\,,
\ea
and $\gamma_n$ decays for large $n$, e.g. $\gamma_n=O(n^{-2})$ if $\hat \Delta>1$.

\begin{figure}[t]
    \centering
    \includegraphics[width=1\linewidth]{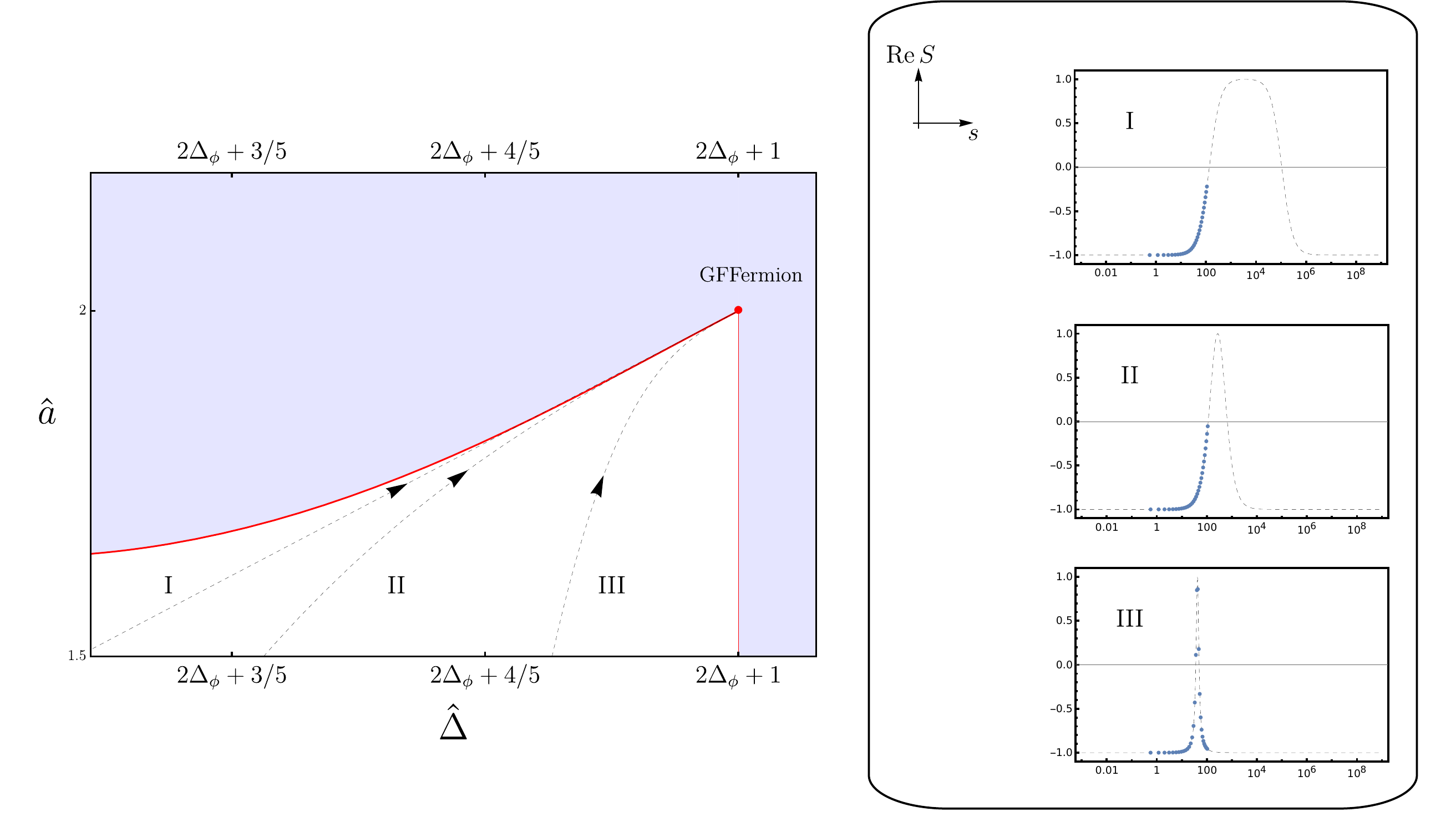}
    \caption{Phase space of extremal solutions with one input state, $\tS=\{\mathds 1,(\hat \Delta,\hat a)\}$. On the left the allowed region by unitarity. As we approach the free fermion point along different curves the spectrum of the extremal solution has different possible behaviours. These are shown on the right in terms of the S-matrix $S(s)=\exp[-i \pi \gamma(s)]$ with $s\sim n^2$. Data are shown for $\Df=1$.}
    \label{fig:phasespace}
\end{figure}

To approach the flat space limit we must perform a tuning. The first such tuning we will consider is to crank up $\hat a$ close to its maximal value. If we would set it literally to its maximal value then we would go back to the family of solutions studied in the previous section (cf. footnote \reef{fn:bound}). But those solutions always behaved as a boson in the UV. Thus by taking $\hat a$ close to but not exactly its maximal point we expect to get an extremal solution describing a flow from a fermion in the deep UV to a boson in the IR. To be more precise this is true only for generic values of $\hat \Delta$. Indeed as long as we stay away from $\hat \Delta=1+2\Df$ then the particular choice of $\hat \Delta$ shows up only as a modification of the IR boson spectrum which is invisible in the flat space limit. In contrast, tuning both $\hat a$ and $\hat \Delta$ towards their values in the generalized free fermion solution leads to interesting flows that begin and end with a free fermion, while passing through a bosonic intermediate phase. We will look into this in the next subsection. Figure \ref{fig:phasespace} helps to visualize this phase space of extremal solutions.

\subsubsection{One fine tuning}
For simplicity, let us set $\hat \Delta=2\Df$ and $\hat a$ close, but not equal to, its free value $a_0^{B}=2$ (which is also the maximum allowed value):
\ba
\hat a=2-\delta a 
\ea
For very small $\delta a$ the extremal spectrum in the IR is well described by the free bosonic one. However, we know that at some point the fermionic nature of the extremal solution will have to appear. Indeed, we can understand the deformation of the free boson \label{eq:defgff}  as being described by an irrelevant contact interaction in AdS$_2$ of the schematic form $g \partial^4 \Phi^4$. More precisely it is possible to choose a basis of contact interactions such that the corresponding deformation of the free bosonic correlator is
\ba
\delta \mathcal G^B(z)=-C^{(4)}(z)\,, \qquad C^{(4)}(z)=\sum_{n=0}^\infty \left[\delta a_n G_{2\Df+2n}(z)+a_n^{\tt gff} \gamma_n^{(4)} \partial G_{2\Df+2n}(z)\right]
\ea
where $\gamma_0^{(4)}=0$ and $\delta a_0:=\delta a$. The $\gamma_n^{(4)}$ are known in closed form \cite{Knop:2022viy}. They satisfy
\ba
\gamma_n^{(4)}\underset{n\to \infty}=\delta a\,c^B(\Df) n^2\,, \qquad c^B(\Df)=\frac{2 \Gamma \left(\Delta _{\phi }+\frac{3}{2}\right)^2 \Gamma \left(2 \Delta _{\phi }+1\right)}{\sqrt{\pi } \Delta _{\phi } \Gamma \left(\Delta _{\phi }+1\right)^2 \Gamma \left(2 \Delta _{\phi
   }+\frac{5}{2}\right)}\,.
\ea
This implies a breakdown of effective theory at a scale which can be set to
\ba
\Dg:=\frac{1}{\sqrt{2-\hat a}}\quad \left(=\frac{1}{\sqrt{\hat a_{\mbox{\tiny max}}-\hat a}}\right)
\ea
where in parenthesis we presented the result expected now for any choice of $\hat \Delta$ (not parametrically close to $1+2\Df$). At and above this scale anomalous dimensions become order one and we expect that the extremal solution goes back to a free fermion.

\begin{figure}
    \centering
    \includegraphics[width=0.7\linewidth]{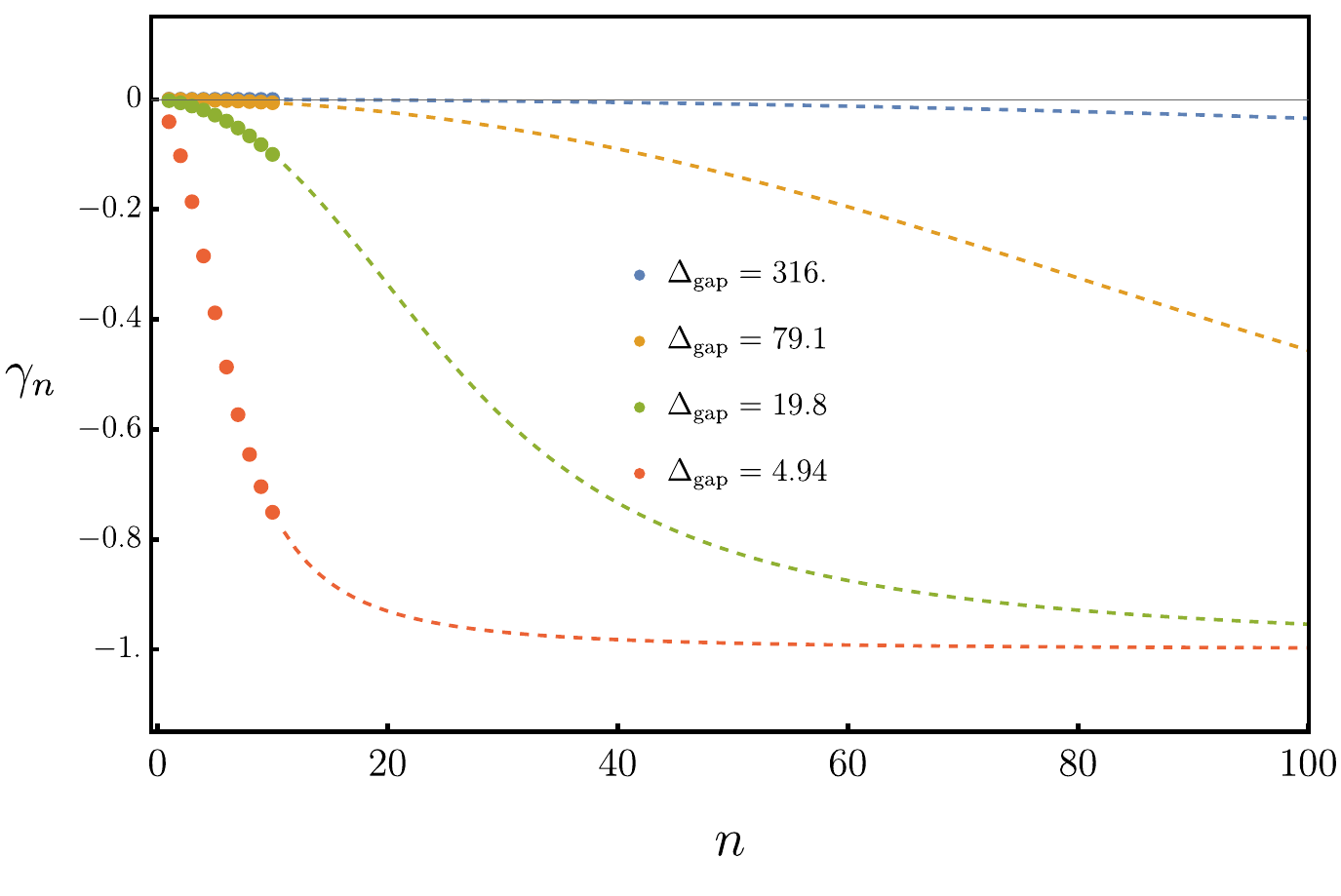}
    \caption{Anomalous dimensions $\gamma_n:=\Delta_n-\Delta_n^B$ for the single parameter family labeled by $\hat a$. Dots correspond to the IR data solved for numerically, while the dashed lines are the analytic result from (\ref{eq:cddzerominus}). $\Df$ is set to 1. In the UV the solution becomes a free fermion.}
    \label{fig:1doffermiongamma}
\end{figure}

Given that this is a simple flow between a boson and a fermion we expect that in the flat space limit the extremal spectrum is well described by the S-matrix
\ba
S(s)=- \frac{s-i m}{s+im}\,,
\label{eq:cddzerominus}
\ea
that is, the same as the one in the previous section but with an overall minus sign. Accordingly the anomalous dimensions (defined with respect to the bosonic free values) are the same as before but with a $-1$ shift. The parameter $m$ can once again be determined by matching with the microscopic CFT data:
\ba
\gamma_n^{(4)}=-\frac{\log(S(s_n))}{i \pi}\underset{s\ll 1}= -\frac {2}{\pi m}  \,s_n\,, \qquad \Rightarrow \qquad m =\frac{8}{\pi\,  c^B(\Df)}\,.
\ea
In figure \ref{fig:1doffermiongamma} we show the extremal solution spectrum for various values of $\Dg$ bootstrapped using the method described in section \ref{sec:smatmethod}. The results show perfect agreement with our expectations.

Finally, we can once again construct a form factor associated to this solution. Since the S-matrix is of the same form as before, so the canonic form factor is almost the same: it merely comes multiplied by an extra factor of $\sqrt{s}$. Note we have
\ba
\mathcal F(s)&\sim s&\,, \qquad s&\ll 1\\
\mathcal F(s)&\sim \sqrt{s}&\,, \qquad s&\gg 1
\ea
which are now the expected behaviour of form factors of $\Phi \nabla^2 \Phi$ and $\bar \Psi \Psi$ in the IR bosonic/UV fermionic theory. 

\subsubsection{Two fine tunings}

Let us now discuss the case where we finetune both the dimension $\hat \Delta$ and the OPE coefficient $\hat a$ of the operator in $\tS$ to approach those of the first operator in the generalized free fermion solution:
\ba
\hat \Delta \to 2\Df+1\,, \qquad \hat a\to a_0^F=2\Df
\ea
In this case we expect that the low energy spectrum is well described by a free fermion in AdS$_2$ with two irrelevant deformations. The two leading such deformations have two and six derivatives and lead to changes in the CFT data which we write as
\ba
\begin{array}{lcc}
g^{(2)}\partial^2 \Psi^4:& \delta a_n^{(2)}\,, & \gamma_n^{(2)} \\
g^{(6)}\partial^6 \Psi^4:& \delta a_n^{(6)}\,, &\gamma_n^{(6)}
\end{array}
\qquad \mbox{with}\quad \gamma_n^{(k)}\underset{n\to \infty}=O(n^k)
\ea
Under these deformations the CFT data changes {\em at leading order} as
\ba
\delta a_n&:=a_n-a_n^{\tt gff}=g^{(2)} \delta a_n^{(2)}+g^{(6)} \delta a_n^{(6)}\\
\gamma_n&:=\Delta_n-\Delta_n^F=g^{(2)} \gamma_n^{(2)}+g^{(6)} \gamma_n^{(6)}
\ea
In particular it is possible to choose a basis of contact terms  such that $\gamma_0^{(6)}=0$ \cite{Knop:2022viy}. The reason we emphasize this is a leading order result is that we have not discussed the relative sizes of the couplings $g^{(2)}$ and $g^{(6)}$. Thus higher order corrections in $g^{(2)}$ can a priori appear before $g^{(6)}$. To see that this is the case, let us proceed naively. We can choose our basis of contact terms such that%
\ba
g^{(2)}&=\gamma_0\\
g^{(6)}&=\delta a_0- \gamma_0 c
\ea
with $c:=\delta a_0^{(2)}$ a known constant.\footnote{Explicitly, \
\ba
     \delta a_0^{(2)} =2\Df\left[\frac{1}{2 \left(\Delta _{\phi }+1\right)}-\frac{2}{2 \Delta _{\phi }-1}-\frac{3}{2 \Delta _{\phi }+1}-H_{\Delta
   _{\phi }-\frac{3}{2}}+H_{\Delta _{\phi }}-H_{2 \Delta _{\phi }}+\log (4)\right]\,.
\ea
}
Since these are irrelevant perturbations, the EFT description breaks down for sufficiently large scaling dimension. There are a priori two possible scales, which can be written in terms of physical quantities as
\ba
\Dg^{(2)}=\frac{1}{\sqrt{-\gamma_0}}\,, \qquad \Dg^{(6)}=\frac{1}{(\hat a_{\mbox{\tiny max}}-\hat a)^{1/6}}\,, \qquad \left\{\begin{array}{l}
\gamma_0=\hat \Delta-1-2\Df <0\vspace{0.3cm}\\
\hat a_{\mbox{\tiny max}}:=2\Df+\gamma_0 c + O(\gamma_0^2)
\end{array}
\right.
\ea
We now see that in order for these scales to be of the same order we must actually have $g^{(6)}=O[ (g^{(2)})^3]$. 

As we will see shortly, the primary consequence of this fact is that the effective cutoff scale is modified whenever we try to bring $\Dg^{(6)}$ below $\Dg^{(2)}$. Ignoring this case then, there are several regimes of interest. Firstly, 
by tuning $\hat a$ very close to $\hat a_{\mbox{\tiny max}}$ and $\gamma_0\ll 1$ we can have $\Dg^{(6)}\gg \Dg^{(2)}\gg 1$. This describes a two step RG flow:
\ba
\mbox{UV} \qquad \mbox{Fermion} \quad \underset{\Dg^{(6)}}\longrightarrow \quad \mbox{Boson} \quad \underset{\Dg^{(2)}}\longrightarrow \quad \mbox{Fermion} \qquad \mbox{IR}
\ea
In the $\Dg^{(6)}\to \infty$ limit with fixed $\Delta \sim \Dg^{(2)}$ we recover the family of extremal solutions already discussed in section \ref{sec:single}, interpolating between a free fermion in the IR and a boson in the UV, while if we zoom in on the region $\Delta \sim \Dg^{(6)}$ in the same limit (or set $\Dg^{(2)}\to 0$) we recover the family of the previous subsection. Secondly, we can set both scales to be the same order. In this case we obtain a flow:
\ba
\mbox{UV} \qquad \mbox{Fermion} \quad \underset{\Dg^{(6)}\sim \Dg^{(2)}}\longrightarrow \quad \mbox{Fermion} \qquad \mbox{IR}
\ea

We can write down what we expect to get for the spectrum of the extremal solution in terms of an S-matrix. Given that we have two parameters, the natural guess is then
\ba
S(s)=-\left(\frac{s-\mu_1}{s-\mu_1^*}\right)\left(\frac{s-\mu_2}{s-\mu_2^*}\right)\,, \qquad s:=\left(\frac{\Delta}{\Dg^{(2)}}\right)^2
\label{eq:twoCDD}
\ea
This S-matrix saturates unitarity, as appropriate for describing an extremal solution, but it must be further constrained to satisfy crossing symmetry, i.e.:
\ba
\mathcal S(-s)=\mathcal S(s) \quad \Rightarrow \quad S^*(-s)=S(s)
\ea
In particular we must set
\ba
\mu_1\propto 2i \frac{1-\sqrt{1-\alpha}}{\alpha}\,, \qquad \mu_2 \propto 2i \frac{1+\sqrt{1-\alpha}}{\alpha}\,, \qquad \alpha>0\,.\label{mueqs}
\ea

Let us now go back and try to understand the fermion EFT a bit more carefully. We can think of this EFT as having a cutoff scale $\Dir$. The EFT is weakly coupled at the cutoff since $\Dir/\Dg\ll 1$. We first switch on the $\partial^2 \Psi^4$ type interaction with coefficient $g^{(2)}=\gamma_0=\Dg^{-2}$. This leads to anomalous dimensions $\gamma_n^F\sim \gamma_0 n^2$ at leading order. At loop level we necessarily need to add new irrelevant terms to the EFT. These terms must be fine-tuned as they naturally come suppressed only by powers of the cutoff $\Dir$ whereas the true cutoff of the theory is at the scale $\Dg$. For instance, at one-loop we generate a behaviour $\gamma_n \sim \gamma_0^2 n^6 \sim \Dg^2 s^3$ in contradiction with the low energy expansion of the $S$-matrix. 

In principle all such terms are generated by analysing the matching equations described in section \ref{sec:smatmethod}. If we solve them numerically then there is no difficulty, but aiming for an analytic result as we do here seems more difficult. Luckily there is a shortcut: we can use the fact that we already succeeded in performing the matching for an S-matrix described by a single CDD zero. The difference between that S-matrix (with a suitable choice of parameter $m$) and the present one first appears at $O(s^3)$, and can be accounted for by an extra $g^{(6)} \partial^6 \Psi^4$ term in the EFT with coefficient
\ba
g^{(6)}:= \hat a_{\mbox{\tiny max}}-\hat a \quad (=O[(g^{(2)})^3] )
\ea
where $\hat a_{\mbox{\tiny max}}$ is the OPE coefficient we would have obtained at two loop level had we not switched on this interaction. This new term in the EFT leads to an additional growth in the anomalous dimensions of the form
\ba
\gamma_n\bigg|_{g^{(6)}} \underset{n\gg 1}=- (\hat a_{\mbox{\tiny max}}-\hat a)\, d(\Df) n^6\,, \qquad d(\Df)=\frac{4^{3-\Delta _{\phi }} \left(\Delta _{\phi }+2\right) \Gamma \left(2 \Delta _{\phi
   }\right)^2 \Gamma \left(\Delta _{\phi }+3\right)}{\Gamma \left(\Delta _{\phi
   }+\frac{3}{2}\right)^3 \Gamma \left(2 \Delta _{\phi }+\frac{11}{2}\right)}.
\ea
This can be matched to the $O(s^3)$ terms in the S-matrix. Concretely:
\ba
-\frac{\log[S_{\mu_1,\mu_2}(s)-S_{m}(s)]}{i\pi}\bigg|_{m= \frac{\mu_1 \mu_2}{i(\mu_1+\mu_2)}}\underset{s\ll 1}=\frac{2 i}{\pi}\left(\frac{\mu_1+\mu_2}{\mu_1^2 \mu_2^2}\right)\, s^3
\ea
Thus we have two equations
\ba
\hat a_{\mbox{\tiny max}}-\hat a& =-i \frac{\gamma_0^3\,\pi}{2^7 d(\Df)}\, \left(\frac{\mu_1+\mu_2}{\mu_1^2\mu_2^2}\right)\,,\qquad 
\frac{\mu_1 \mu_2}{i(\mu_1+\mu_2)}&=\frac{8}{\pi \, c(\Df)}
\ea
which have solutions precisely of the form in \reef{mueqs}:
\ba
\mu_{1,2}=\frac{16i}{\pi c(\Df)}\,\left(\frac{1\mp \sqrt{1-\alpha}}{\alpha}\right)\,,
\ea
but where $\alpha$ now is determined by the CFT data:
\ba
\alpha=\frac{16}{\pi^2}\frac{d(\Df)}{c(\Df)^3}\, \left(\frac{\hat a_{\mbox{\tiny max}}-\hat a}{-\gamma_0^3}\right)=\frac{16\,d(\Df)}{\pi^2 c(\Df)^3}\, \left(\frac{\Dg^{(2)}}{\Dg^{(6)}}\right)^6
\,.\label{mueqs2}
\ea
Note that the prefactor in $\alpha$ is $O(1)$ for generic $\Df$.

This result for the S-matrix has several interesting regimes which line up with our previous discussion:

\begin{itemize}
\item The first regime corresponds to having $|\mu_2|\propto \Dg^{(6)}/\Dg^{(2)}\gg 1$ implying $\alpha\ll 1$. In this case the S-matrix reduces to a single CDD zero and we recover the previous families of extremal solutions that we studied. Concretely, holding $s$ fixed and sending $\mu_2\to \infty$ we recover the UV boson to IR fermion flow ; while holding $s/\mu_2$ fixed in the same limit we recover the UV fermion to IR boson flow.

\item The second regime corresponds to $0<\alpha<1$. Here we have $\Dg^{(6)}\gtrsim \Dg^{(2)}$ and $\mu_{i}=i |\mu_i|$ and the flow is from a fermion in the UV to a fermion in the IR.
\item Finally we can ask what happens after lowering $\Dg^{(6)}$ enough so that $\alpha\to 1$. Beyond this we have $|\mu_1|=|\mu_2|$ and both move towards the real axis while shrinking in absolute value. In this regime the S-matrix has a single feature located at a scale $s\sim 1/\alpha$ which thus moves towards $s\to 0$ as we continue to lower $\Dg^{(6)}$. From the EFT perspective the effective cutoff is thus no longer $\Dg^{(2)}$ but rather $\Dg^{(2)}/\sqrt{\alpha}$, so that in particular the EFT tends towards strong coupling at the scale $\Dir$ as we increase $\alpha$.
\end{itemize}

Note that there is a simple understanding of why something had to break down eventually as $\Delta^{(6)}$ and $\Delta^{(2)}$ collide. In particular consider the extreme case where $\gamma_0$ is set to zero. In this case, for any choice of $\hat a$, there is only one extremal solution, namely a free fermion at all scales. Concretely, in that case we must necessarily have $\Delta_0=\hat \Delta$ and $a_0=a^{\tt gff}-\hat a$. But then the same must also be true for any given fixed $\hat a$ as we send $\gamma_0\to 0$. This signals that in this regime it is not correct to think of the EFT as describing a single free fermion, but instead as a free fermion coupled to an extra state.

\begin{figure}
    \centering
    \includegraphics[width=0.7\linewidth]{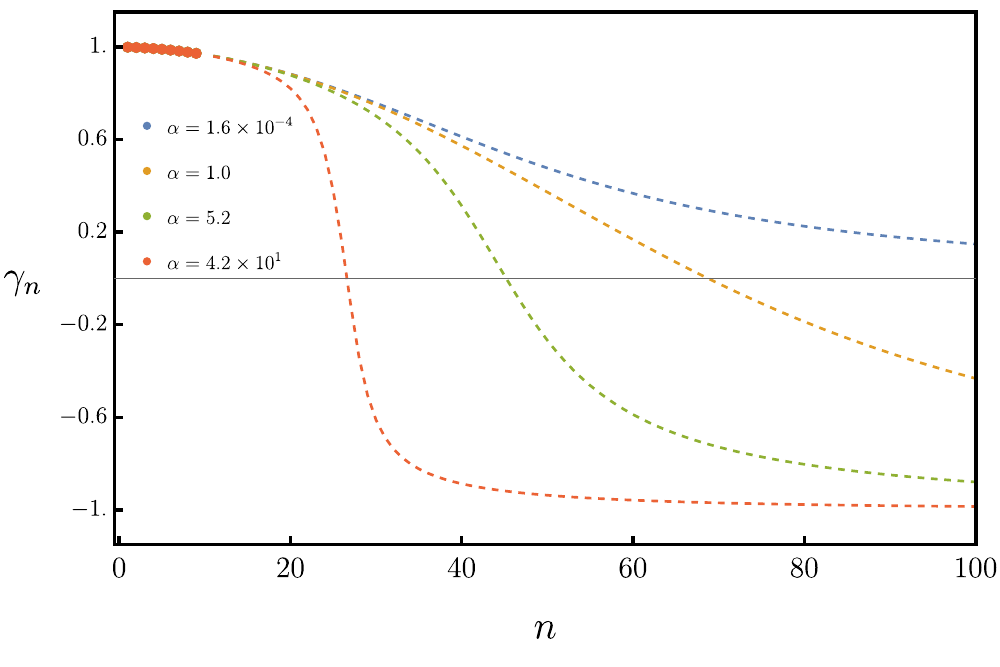}
    \caption{Anomalous dimensions for the two parameter family labeled by $(\hat \Delta,\hat a)$. The parameter $\alpha$ is related to these by equation \reef{mueqs2}. Dots correspond to the IR data solved for numerically, while the dashed lines are the analytic result from (\ref{eq:twoCDD}) and (\ref{mueqs2}). $\Df$ is set to 1 and $\gamma_0^F = -4\times 10^{-4}$. When $\alpha<1$, the data fits the bosonic solution ($a_0 =a_0^{\mbox{\tiny max}}$) for a long time, eventually transitioning to a fermion. When $\alpha > 1$, the transition is more sudden, fewer states lie near $\gamma_n = 0$. As $\alpha$ gets larger, the dip becomes sharper and moves towards the IR.}
    \label{fig:2dofgamma}
\end{figure}

\begin{figure}
    \centering
    \includegraphics[width=1.05\linewidth]{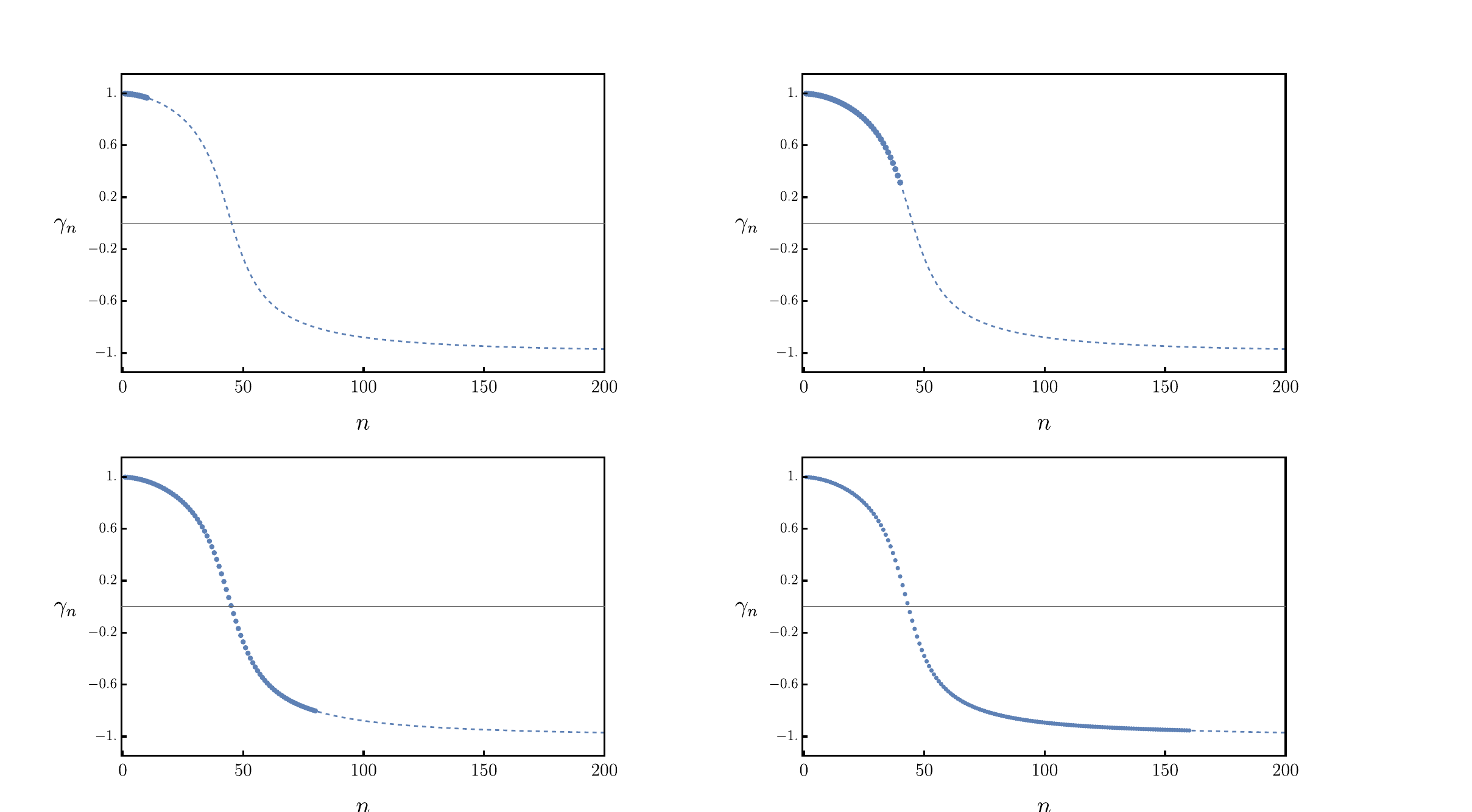}
    \caption{An extremal solution is bootstrapped with increasing number $N_{\mbox{\tiny max}}$ operators solved for numerically. $\gamma_0^F=-4\times 10^{-4}$ and $\alpha=5.24288$. From top to bottom and left to right we have $N_{\mbox{\tiny max}}=10,40,80$ and $160$.}
    \label{fig:2dofcutoffincrease}
\end{figure}

Our numerical checks of these expectations are shown in figures \ref{fig:phasespace}, \ref{fig:2dofgamma} and \ref{fig:2dofcutoffincrease}. In the first we have chosen to represent the extremal spectrum in terms of the S-matrix. We can clearly see that there are three distinct regimes. The second figure is now in terms of anomalous dimensions but represents the same physics. All results were obtained using the S-matrix method of section \ref{sec:smatmethod}. Finally in the last figure we check that our results are not modified as $\Dir$ is increased.

\section{Mixed locality/crossing bounds}
\label{sec6}
\subsection{Setup}
In this section, we explore the constraints arising from combining those from crossing with those of locality. We have seen that they may be written in the form
\ba
\sum_{\Delta} \lambda_{\Delta}^2 \alpha_n(\Delta)=0\,, \qquad \sum_{\Delta} \lambda_{\Delta}^2 \beta_n(\Delta)=0\,, \qquad \sum_{\Delta} \lambda_{\Delta} \mu^{\Psi}_{\Delta}\, \theta_n^{(\alpha)}(\Delta)=0
\ea
In order to obtain bounds on the combined OPE/BOE data we will need some kind of constraint which is quadratic in the BOE coefficients. Consider the two point function of the bulk field:
\begin{equation} \label{2ptbulk}
    \langle \Psi(y_1,x_1)\Psi(y_2,x_2)\rangle \sim \sum_{\Delta} (\mu^{\Psi}_{\Delta})^2 f_{\Delta}(\chi),
\end{equation}
where the precise form of the bulk block $f_{\Delta}(\chi)$ and the cross-ratio $\chi$ won't be necessary. For a generic bulk operator there is unfortunately no a priori constraint on this two point function. One possibility is to choose some set of linear functionals $\nu_i$ and demand
\ba
\sum_{\Delta} (\mu^{\Psi}_{\Delta})^2 \nu_i(\Delta)=C_i
\ea
Physically we have the freedom to rescale the bulk operator by any factor so  that one of these constraints merely sets the overall scale of the BOE data, but any additional ones are physically meaningful. To proceed one could for instance choose functionals which are sensitive to the short distance limit and demand an appropriate power law behaviour. It would be interesting to explore this approach, but we will not attempt to do so here. 

Instead, we will choose to consider a very special bulk operator, namely the trace of the stress tensor of our QFT in AdS$_2$ denoted $\Theta$. In this case it turns out that there are additional sum rules which can be included~\cite{Meineri:2023mps}. These take the form:
\ba \label{eqn:kappa}
\sum_{\Delta} \lambda_{\Delta} \mu^\Theta_{\Delta} \kappa^{(\tilde \alpha)}(\Delta)&=\Delta_{\phi}\,,\\
\sum_{\Delta} (\mu^{\Theta}_{\Delta})^2 \nu(\Delta)&=C_T\,.
\ea
where $\tilde \alpha$ is a free parameter and $C_T$ is the central charge.
The important point is that morally the first sum rule sets the overall normalisation of the operator, while the second now provides a non-trivial constraint on the BOE in the terms of the UV central charge of the bulk 2d QFT. The precise form of the functionals $\nu, \kappa$ read as:

\ba
&\nu(\D)=\frac{24\sqrt{\pi}\G\left(\frac{\D}{2}\right)^2}{(\D-2)^2(\D+1)^2\G(\D)}\\
&\k^{(\tilde{\a})}(\D)=\frac{4}{\sqrt{\pi}(\D+1)(2-\D)}+\frac{1}{2\sqrt{\pi}}\frac{\G(\tilde{\a})\G\left(\tilde{\a}-\frac{1}{2}\right)}{\G\left(\tilde{\a}+\frac{\D}{2}+\frac{1}{2}\right)\G\left(\tilde{\a}-\frac{\D}{2}+1\right)}\\
&\times\Bigg[\, _3F_2\left(\begin{aligned}&\qquad\quad 1,\,\tilde{\a}-\frac{1}{2},\,\tilde{\a}\\
&\tilde{\a}-\frac{\D}{2}+1,\,\tilde{\a}+\frac{\D}{2}+\frac{1}{2}\end{aligned};1\right)+\frac{\tilde{\a}}{\tilde{\a}-1}\, _4F_3\left(\begin{aligned}&\quad1,\,\tilde{\a}-1,\,\tilde{\a}+1,\,\tilde{\a}-\frac{1}{2}\\
&\tilde{\a},\,\tilde{\a}-\frac{\D}{2}+1,\,\tilde{\a}+\frac{\D}{2}+\frac{1}{2}\end{aligned};1\right)\Bigg]
\ea
Our expressions differ from~\cite{Meineri:2023mps} by an overall factor of $\G(\D/2)^2/\G(\D+1/2)$ owing to our normalisation conventions, cf. equations \reef{eq:H} and \reef{eq:N}\\

With the addition of these constraints, the most general sum rule on the combined CFT/BOE data is written as
\ba
\sum_{\Delta} (\lambda_{\Delta},\mu_{\Delta})\cdot \Omega(\Delta)
 \cdot (\lambda_{\Delta},\mu_{\Delta}) =f \Df+g C_T \label{eq:sumruleOm}
\ea
with
\ba
\Omega(\Delta):=\begin{pmatrix}
\sum_n \big(a_n\alpha_n(\Delta)+b_n \beta_n(\Delta)\big) & \frac12 \sum_n d_n \theta_n^{\mathcal A}(\Delta)+\frac{1}{2}f \kappa^{(\tilde \alpha)}(\Delta)  \\
\frac12 \sum_n d_n\theta_n^{\mathcal A}(\Delta)+\frac{1}{2}f \kappa^{(\tilde \alpha)}(\Delta)& g\, \nu(\Delta) 
\end{pmatrix} 
\ea
The strategy for finding bounds now is to choose $\Omega$ such that it is a positive semi-definite matrix for some range of $\Delta$, together with additional constraints depending on the specific bound of interest.
It is important to note that this might not be possible for asymptotically large $\Delta$. In this regime the functionals scale as follows, 
\ba
   & \omega_n(\Delta)\sim\frac{1}{ a^{\tt gff}_\Delta} \frac{1}{\Delta^3} \\
    & \nu(\Delta)\sim \frac{a^{\tt gff}_\Delta}{ (b^{\tt gff, B}_\Delta)^2}\frac{1}{\Delta^5} \sim \frac{a^{\tt gff}_\Delta}{ (b^{\tt gff, F}_\Delta)^2}\frac{1}{\Delta^3} \\
    & \theta_n^{(\alpha)}(\Delta)\sim \frac{1}{b^{\tt gff, B}_\Delta } 
     \frac{1}{\Delta^{-2\Delta_{\phi}+2\alpha+3}}\sim \frac{1}{b^{\tt gff, F}_\Delta } 
     \frac{1}{\Delta^{-2\Delta_{\phi}+2\alpha+2}}\\
    & \kappa^{(\tilde \alpha)}(\Delta)\sim \frac{1}{b^{\tt gff, B}_\Delta } \frac{1}{\Delta^{-2\Delta_{\phi}+2\tilde \alpha+1}}\sim \frac{1}{b^{\tt gff, F}_\Delta } \frac{1}{\Delta^{-2\Delta_{\phi}+2\tilde \alpha}}
\ea
where $b^{\tt gff, F}_\D = b^{\tt gff}_\D\big|_{\a=\Df+1/2}$ and $b^{\tt gff, B}_\D = b^{\tt gff}_\D\big|_{\a=\Df}$ are the $b_\D$ for the generalized free fermion and boson respectively.
Thus to ensure asymptotic positive semidefineteness we choose $\alpha$ and $\tilde \alpha$:
\ba
\alpha\geq \Delta_{\phi}+\frac 12\,,\qquad \tilde \alpha \geq \Delta_{\phi}+\frac 32\,.
\ea
More precisely, note that bases with higher values of $\alpha$ (shifted by one) can be obtained from those with lower values by taking linear combinations. Thus, while we could have chosen lower values for these parameters, in practice positivity would have landed us on these choices anyway. Let us make the comment here that had we chosen to work in the usual derivative basis for the crossing equation, then asymptotic positivity would be impossible, as such fuctionals are parametrically softer for large $\Delta$ then the $\alpha_n,\beta_n$.

\subsection{Bounds on the central charge}
\subsubsection{An exact bound $C_T\geq \frac 12$}
As a first application, we will construct an exact functional establishing a lower bound on the central charge:
\ba
C_T\geq \frac 12
\ea
under mild assumptions on the CFT spectrum. This is of course a well known result from 2d CFT, but here we do not rely on Virasoro symmetry.\footnote{A {\em numerical} lower bound on $C_T\gtrsim 1/2$ was obtained a long time ago from the 2d correlator bootstrap~\cite{ElShowk:2014}.}
The idea for constructing the functional is simple: as the lower bound is saturated by a free fermion in AdS$_2$, the functional should have double zeros on the generalized free fermion spectrum. More precisely, since it is a matrix, the functional has to have a zero eigenvector for $\Delta=\Delta_n^F$. The direction of this eigenvector is fixed by the known free fermion solution. Let us then set:
\ba
\hat \nu(\Delta):=\Omega(\Delta)\big|_{g=1}\,, \qquad \hat \nu(\Delta,r):=(r,1)\cdot \hat \nu(\Delta)\cdot (r,1)
\ea
We require
\ba
\hat \nu(\Delta_n^F,r_n)=\partial_\Delta\hat \nu(\Delta_n^F,r_n)=\partial_r\hat \nu(\Delta_n^F,r_n)=0
\ea
where 
\ba
    r_n:=\frac{\lambda_n^{F}}{\mu_n^{F}}=    
    \frac{(-1)^{1-n} 2^{1-2 n} \G(n+\Df+\frac{1}{2})\Gamma (n+\Delta_\phi +1) \Gamma (n+2 \Delta_\phi )}{(2 \Delta_\phi -1) \Gamma (2 \Delta_\phi ) \Gamma \left(n+\frac{3}{2}\right) \Gamma \left(2n+2\Delta_\phi +\frac{3}{2}\right)}
\ea
is the ratio of OPE to BOE data in the free fermion solution.
To solve these conditions we work with the fermionic basis of functionals and set $\alpha=\Delta_{\phi}+\frac 12$ and $\tilde \alpha=\Df+\frac{3}{2}$. Then we get simply: 
\begin{equation}
\begin{split}
&  a_n=\frac{\nu(\Delta_n^F)}{r_n^2},\qquad d_n=-2\frac{\nu(\Delta_n^F)}{r_n} \qquad f=-\frac{1}{\Delta_{\phi}},\\
& b_n=-\frac{1}{r_n^2}\bigg(\nu'(\Delta_n^F)+r_n\sum_m d_m \theta'_m(\Delta_n^F)+f r_n \kappa'(\Delta_n^F)\bigg),
\end{split}
\end{equation}
An easy computation now determines the action of the functional on the identity to be:
\begin{equation}
    \nu[\mathds 1]:=(1,a_T)\cdot \hat{\nu}(0)\cdot (1,a_T)=-C_T^F=-\frac{1}{2}.
\end{equation}
where $a_T:=\mu_0^T$ is the vev of $\Theta$ (which actually does not contribute since all functionals apart from $\alpha_n$ vanish at $\Delta=0$).
Overall the sum rule \reef{eq:sumruleOm} becomes 
\ba
\sum_{\Delta>0} (\lambda_{\Delta}, \mu_{\Delta})\cdot \hat \nu(\Delta)\cdot(\lambda_{\Delta}, \mu_{\Delta})=C_T-\frac 12
\ea
This equation gives us a bound as long as $\hat \nu$ has appropriate positivity properties. We find experimentally:
\ba
\hat \nu(\Delta)\succeq 0\qquad \mbox{for}\quad \Delta>\Delta_+(\Df)\,, \qquad \Df\geq1/2\,.
\ea
thus establishing $C_T\geq \frac 12$ as long as the CFT contains an OPE channel $\phi\times \phi=1+\cO$ with $\Delta_\cO>\Delta_+(\Df)$. 
One has approximately $\Delta_+(\Df)\sim 2\Df-2/5$. Below $\Df=\frac 12$ the functional we constructed develops a negative region and the bound is lost. While we were unable to find a deep reason for this statement, we show explicitly in appendix \ref{app:ce} that fermionic type solutions coupled to an extra state with dimension arbitrarily close to $1+2\Df$ support the reconstruction of a bulk stress tensor with $C_T\leq 1/2$. Since this is a valid solution of all sum rules it cannot be ruled out by the $\hat \nu$ functional, which accordingly develops a negative region. This result does not mean that there is no lower bound below $\Df=\frac 12$, but it does mean it will necessarily be lower than $C_T=1/2$ and sensitive to our choice of gap.


\begin{figure}[t]
        \centering
        \includegraphics[width=0.7\textwidth]{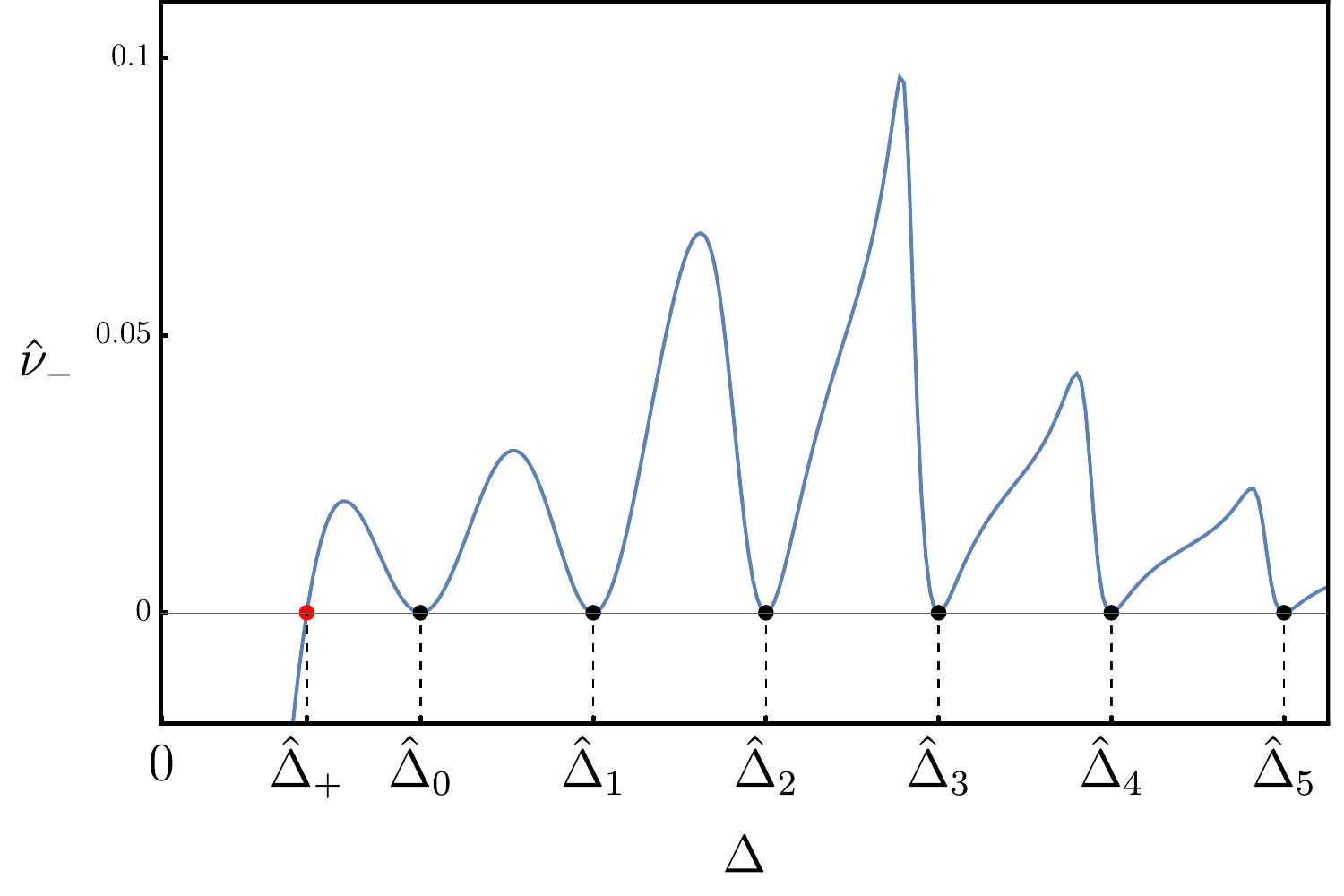}
        \caption{Positivity of the dressed functional $\hat \nu$. The lowest eigenvalue $\hat \nu_-$ of the functional is positive above some critical value $\Delta_+$.}
    \hfill
    \label{fig:CTbound}
\end{figure}

\subsubsection{Bounds assuming displacement operator}
Under further assumptions it is possible to get a tighter bound on the central charge. Let us assume that an OPE of the form 
\ba
\phi \times \phi=1+D+\cO_0\ldots\,, \qquad \Delta_D=2
\ea
That is, we assume the first operator in the OPE is the displacement, which appears if the bulk QFT is actually a CFT, followed by an operator with dimension $\Delta_0$. In this case strictly speaking the stress-tensor trace vanishes, and we must bootstrap instead the stress-tensor. As explained in~\cite{Meineri:2023mps} this can be achieved in practice by simply setting $\mu_{\Delta}^\Theta=\mu_{\Delta}^{T} \frac{(2-\Delta)}{\Delta}$. It now follows that for $\Delta_+(\Df)<\Delta_D$ and $\Delta_0>1+2\Df$ the bound on the central charge must necessarily improve, as these assumptions rule out the free fermion solution.

\begin{figure}[t]
    \centering
    \includegraphics[width=0.8\linewidth]{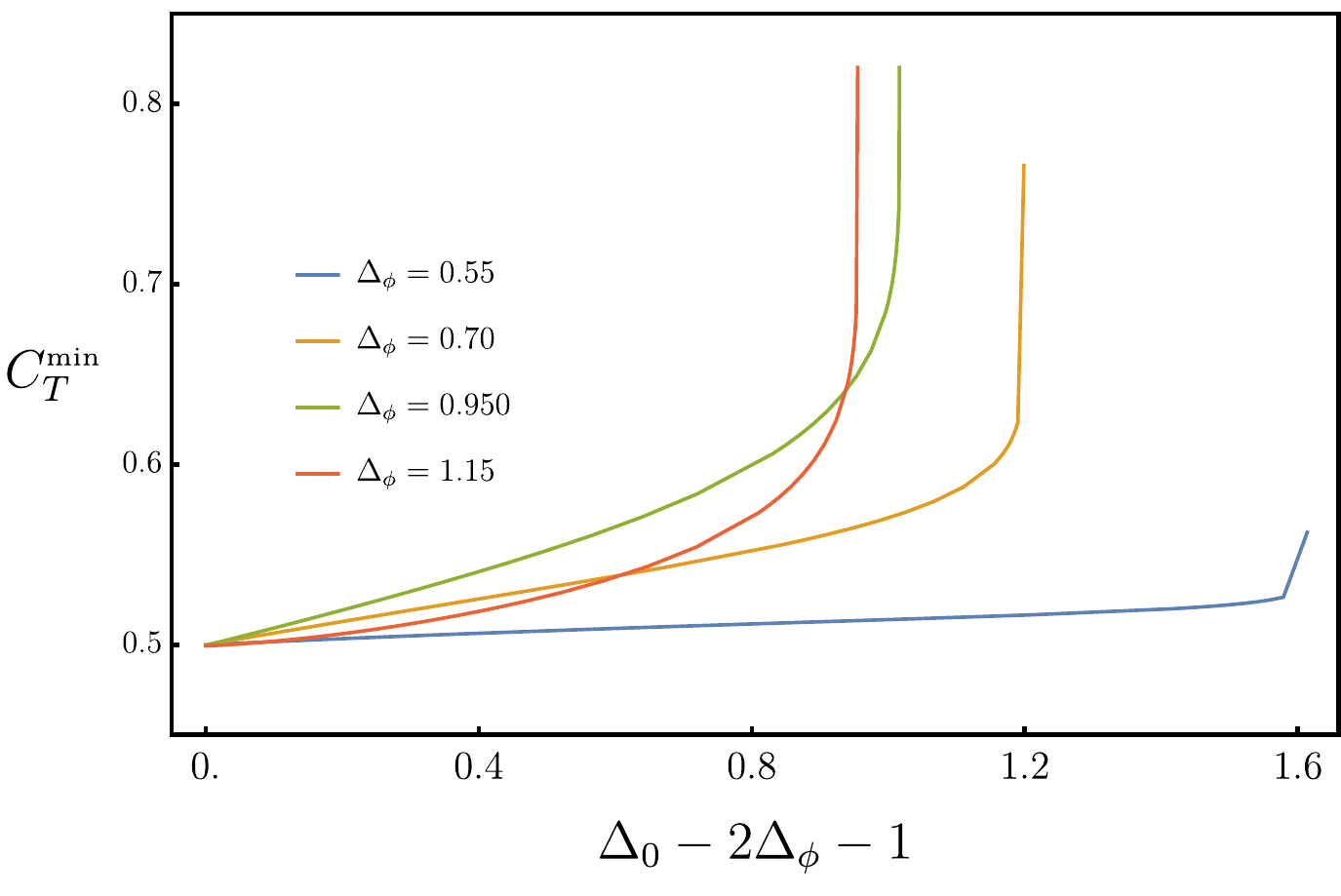}
    \caption{Central charge bounds for CFTs containing an operator $D$ and a gap to the next operator of dimension $\Delta_0$. The curves represent lower bounds on the central charge and terminate at the maximal allowed gap. The bounds are saturated by a fermionic type extremal solution coupled to the state $D$. As the gap increases towards its maximum so does the OPE coefficient of $D$ and the solution approaches the bosonic OPE maximization solution. Only the finite range  $1/2\leq\Df\leq 3/2$ is allowed. The curves were obtained with a cutoff on all functionals equal to $n_{\mbox{\tiny max}} = 20$ and have converged. }
    \label{fig:clowbound}
\end{figure}

To find this improved bound we may use a positive semidefinite solver such as SDPB~\cite{Simmons-Duffin:2015qma,Landry:2019} to construct an appropriate $\Omega(\Delta)$ satisfying
\ba
\begin{array}{ll}
\Omega(\Delta_D)\succeq 0\vspace{0.3cm}\\
\Omega(\Delta)\succeq 0 \quad \mbox{for} \quad \Delta>\Delta_0\,,
\end{array}\qquad \Rightarrow\quad  C_T\geq \frac{\Omega[\mathds 1]-f \Df}g
\ea
However, there is a simpler approach that does not require solving an optimization problem. Given our assumptions on the OPE, we expect that any such bound will be saturated by an extremal solution with %
\ba
\tS=\{\mathds 1, (\Delta_D,a_D)\}
\ea
where $a_D$ can be thought of as controlling (or being determined by) our choice of $\Delta_0$. This solution is of fermionic type and can be bootstrapped. The locality functionals $\theta,\kappa$ uniquely determine the entirety of the BOE data with the exception of a single BOE coefficient, say $\mu_D$. The $\nu$ sum rule is quadratic in $\mu_D$, from which we can extract a value of the central charge which we identify as the lower bound saturated by this solution. To construct the functional that determines this lower bound we use the fact that it must annihilate this extremal spectrum and can be thus obtained in a similar fashion to the one found for $C_T=\frac 12$. We then merely check that it indeed satisfies the right positivity conditions. 
We have found that this simple procedure leads to functionals in perfect agreement with those produced by semidefinite optimization methods using SDPB. The results for the lower bound are shown in figure \ref{fig:clowbound}.

\subsection{Bounds on OPE/BOE}
\label{sec:boundopeboe}
%

As our final application we will derive bounds on the combined OPE/BOE data associated to a specific operator assuming the existence of a local QFT bulk dual. Concretely, we assume an OPE of the form
\begin{equation}
\phi \times \phi\sim 1+\mathcal{O}_{\hat \Delta}+..
\end{equation}
and demand the existence of the (trace of) the stress tensor in the bulk.
We write the CFT data associated with this operator as 
\ba
(\lambda^{\phi \phi}_{\mathcal O_{\hat \Delta}}, \, \mu^{\Theta}_{\hat \Delta}):=(\hat \lambda,\hat \mu):= \hat \lambda \left(1,1/\hat r\right)
\ea
A generic constraint on a solution to crossing and locality is written
\ba
\Omega[\mathds 1]+\hat \lambda^2\, \Omega(\hat \Delta,\hat r)+\sum_{\Delta>\Delta_0} (\lambda_{\Delta},\mu_{\Delta})\cdot \Omega(\Delta)
 \cdot (\lambda_{\Delta},\mu_{\Delta}) =f \Df+g C_T \label{eq:sumruleOm2}
\ea
Now for a given $\hat r$ we demand
\ba
\Omega(\hat \Delta,\hat r)=1\,,\qquad \Omega(\Delta)\succeq 0 \quad \mbox{for}\quad \Delta\geq \hat \Delta \label{eq:demand}
\ea
which gives an upper bound on $\hat \lambda^2$:
\ba
\hat \lambda^2\leq f\, \Df+g C_T-\Omega[\mathds 1]
\ea
We can now search over all $\Omega$ satisfying the constraints to find the best possible bound.

Let us try to understand what result we should expect. Recall that we have already considered the problem of OPE maximization and bulk reconstruction: it corresponds to the family of extremal solutions considered in section \ref{sec:single} (more precisely, the $\alpha_0$ functional for this solution implies an upper bound on $\lambda_0^2$ valid for any CFT,  which it saturates). We also solved the constraints of locality assuming a BOE matching the extremal spectrum. In the present context there are two key differences. Firstly, we are now aiming to reconstruct a specific bulk operator, the stress-tensor, which presents us with additional constraints. Secondly we have relaxed the locality constraints: in that section we used $\theta$ functionals with $\alpha=\Df$ whereas here we require $\alpha\geq \Df+1/2$. In practice here we choose the smallest such number corresponding to a solution with bosonic/fermionic asymptotic spectrum, so either $\alpha=\Df+1/2$ or $\alpha=\Df+1$  (both these bases introduce equivalent constraints).

Let us discuss the effect of these changes and whether they can rule out the solution we previously constructed, in which case the bound must improve. Firstly, in that solution we had succeeded in determining all but one piece of BOE data, corresponding to the overall normalisation of the bulk operator. Here we have also relaxed the locality constraints, leaving us thus with not one with two undetermined BOE coefficients. However, these are respectively fixed by specifying $\hat r$ and by the $\kappa^{(\tilde \alpha)}$ sum rule. At this point we have fixed all the OPE and BOE data in our extremal solution. We can now plug the BOE data into the lefthand side of the $\nu$ sum rule in \reef{eqn:kappa} and read off the result, which we can call the ``effective'' central charge $C_T^{\mbox{\tiny eff}}$. There are two possibilities. If the $C_T^{\mbox{\tiny eff}}>C_T$ then the solution is ruled out and therefore the bound on the OPE coefficient must improve relative to the one found by using crossing alone. If instead $C_T^{\mbox{\tiny eff}}\leq C_T$  then the bound will not improve. This is because we can extend our extremal solution by an arbitrary set of operators with non-zero $\mu_{\cO}^{\Theta}$ and zero OPE $\lambda_{\phi \phi \cO}$. This leaves all sum rules unchanged except for the $\nu$ sum rule, to which such operators can be chosen to  contribute by a positive amount, thus soaking up the difference between the effective and real central charges.


Next, let us discuss what happens if the bound does indeed improve. In this case we expect that the nature of the associated extremal solution will have to change. Since before we had a bosonic type solution, it is natural to expect we will now have a fermionic one, with 
\ba
\tS=\{\mathds 1,(\hat \Delta,\hat \lambda_{\tE}^2)\}
\ea
where $|\hat \lambda_{\tE}|$ should be taken below the OPE maximum of the bosonic solution but otherwise free.\footnote{If we take it above we can still construct an extremal solution but it will not lead to a bound.} It may seem confusing that we are choosing $\hat \lambda$ here, but the point is that in this way of approaching the problem it is $\hat r$ which will end up being fixed, leading us to a curve on the $(\hat \lambda,\hat \mu)$ plane traced out by the extremal solution. Let us go through the logic. First, we can construct the fermionic solution associated to the set $\tS$, namely the OPE coefficients and dimensions for all fermionic double trace operators, as was done in section \ref{sec:fermextra}. Secondly, we use the locality functionals with $\alpha=\Df+1/2$ to fix the BOE data for all but one of those operators. This leaves us with $\hat \mu$ and $\hat \mu_0$ unfixed. These are now in turn fixed by the $\kappa$ and $\nu$ sum rules. 

At this point we have fixed all the OPE and BOE data of the extremal solution which we denote:
\ba
\{(\hat \Delta,\hat \lambda_{\tE},\hat \mu_{\tE}\} \cup \{(\Delta_n, \lambda_n, \mu_n)\}_{n=0}^\infty
\ea
with $\Delta_n$ asymptoting to the free fermion dimension $1+2\Df+2n$.
We can associate to this solution a functional $\Omega_{\tE}$:
\ba
\Omega_{\tE}(\hat \Delta,\hat r_{\tE})=1\,, \qquad \Omega_{\tE}(\Delta_n,r_n)=\partial_{\Delta}\Omega_{\tE}(\Delta_n,r_n)=\partial_{r}\Omega_{\tE}(\Delta_n,r_n)=0
\ea
If this functional satisfies the positive semidefiniteness conditions spelled out in \reef{eq:demand} then it gives a rigorous bound saturated by this extremal solution for the choice $\hat r=\hat r_{\tE}$:
\ba
\hat \lambda^2\leq f \Df+g C_T-\Omega_{\tE}[\mathds 1]=\hat \lambda^2_{\tE}
\ea
   
   

\begin{figure}[t]
        \centering
        \includegraphics[width=0.8\textwidth]{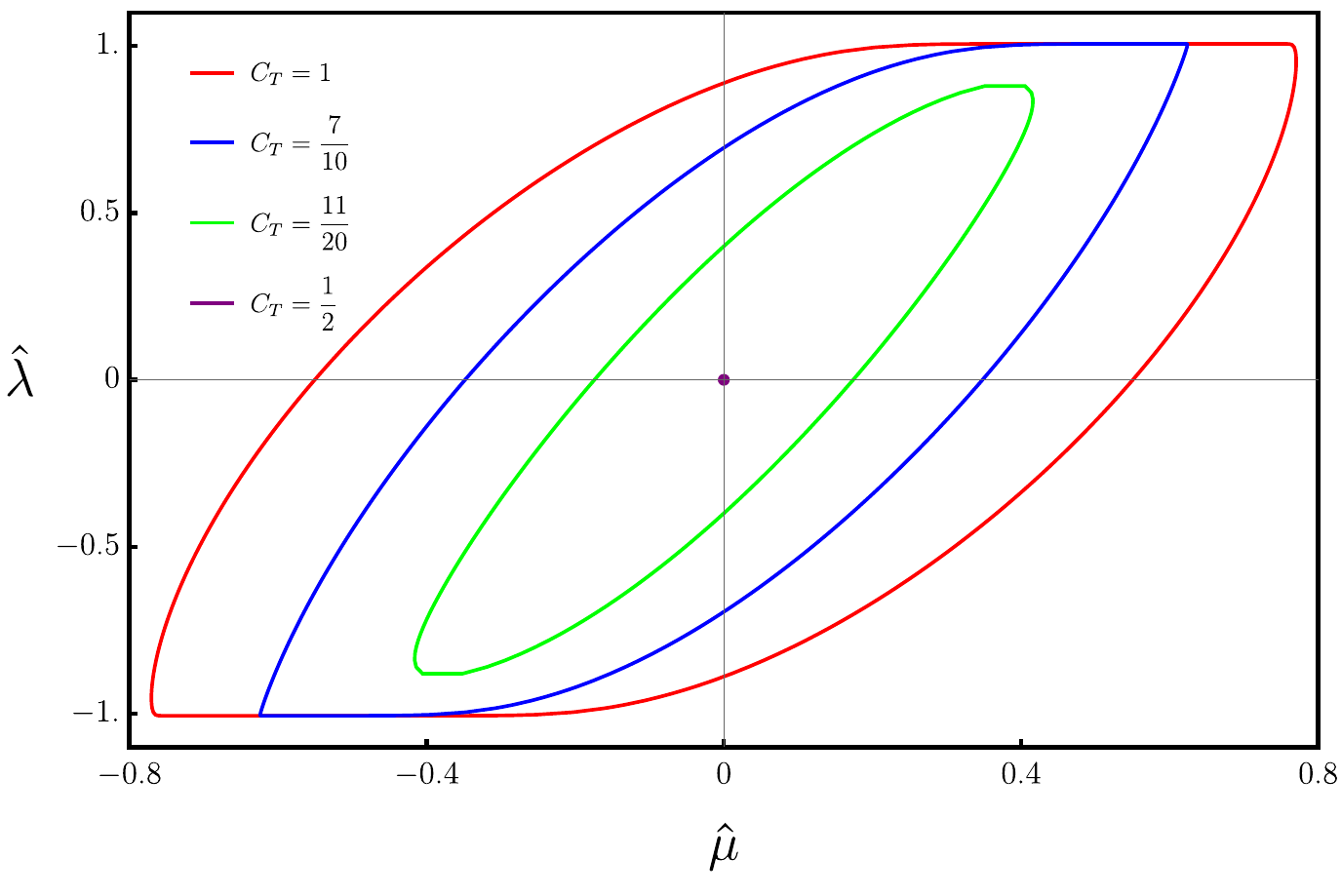} 
        \caption{Allowed regions in the $(\hat \lambda,\hat \mu)$ plane of an operator of dimension $\hat \Delta=2\Df+\frac 12$, for different values of the central charge and $\Df=\frac 12$. The allowed regions lie to the interior of the colored curves. Similar plots can be done for other values $\Df>1/2$ and $\hat \Delta$, with results qualitatively unchanged. When $C_T=\frac 12$ the allowed region collapses to the origin.       \label{fig:opeboe}
   }
\end{figure}
We show the result of this construction for $\Df=1/2$ and various values of $C_T$ in Figure (\ref{fig:opeboe}). Concretely we have constructed an extremal solution for each point on the plot and verified that we have derived a proper bound by checking positivity of $\Omega_{\tE}$. As a cross-check we also independently computed the same bound using SDPB. In general we expect a flat region in the plot corresponding to $\lambda_0=\pm\lambda_{\mbox{\tiny max}}$. In this region the solution is bosonic, and the value of the central charge depends on $\hat \mu$. However below a certain value ($C_T \sim 0.669$), any such solution has a greater central charge and this flat regions disappears. As we lower $C_T$ the allowed region progressively shrinks until for $C_T=1/2$ only the origin remains, consistently with our exact lower bound (as only the generalized free fermion solution is allowed). Note that this implies that for fixed $\hat \lambda, \hat \mu$ it is possible to derive an improved lower bound on $C_T$.

\section{Discussion}

In this paper we examined the consequences of crossing symmetry for 1d CFTs. We have argued that crossing is not all-powerful: it can at most fix the CFT data of states in an extremal solution. Furthermore we have given a characterization of such solutions: they have simple spectra consisting of a single tower of operators with spacing $\Delta_{n+1}-\Delta_n\sim 2$ \cite{Mazac:2018ycv} and eventually asymptoting to a generalized free field. An immediate consequence is that even when given the exact (exponentially dense) spectrum of a correlator, crossing cannot fix the OPE data of this spectrum completely. This is consistent with what has been found on works in 1D CFTs where the spectrum is known from integrability~\cite{Cavaglia:2024dkk,Cavaglia:2023mmu,Cavaglia:2022yvv,Cavaglia:2022qpg,Cavaglia:2021bnz}.

To improve on this state of affairs we must necessarily include extra four-point functions. A natural guess based on existing constructions of functional bases \cite{Ghosh:2023wjn} is that extremal {\em sets} of CFT correlators will have OPEs consisting of multiple `double trace' operator towers, built by fusing all possible pairs of external operators. Thus by adding more and more correlators we can obtain denser and denser OPEs, and a better and better match to the actual CFT OPE proceeding from low to high energies.

It is also natural to expect that some version of our results holds in higher dimensions. The key point is that functional bases have to be, well, bases. In particular they have to provide a complete set of constraints, and completeness poses global constraints on the allowed set of scaling dimensions, most notably on their asymptotics, as was made precise in recent work analysing the bootstrap problem of bulk locality \cite{Levine:2024wqn}. This reasoning leads us to expect that solutions which saturate bootstrap bounds in higher dimensions will have relatively simple, constant density spectra. A noteworthy point is that this picture seems to be in tension with arguments coming from the lightcone bootstrap\cite{Simmonsduffin:2017}  which would be important to clarify. In any case if it is correct than including more four-point functions is again expected to improve solutions, but not fundamentally so. This might seem puzzling as the bootstrap has been so successful in determining properties of actual CFTs, see~\cite{Chang:2024whx} for a recent dramatic example. An answer to that is at luckily CFT data at low energies is highly insensitive to the fine details of what happens in the UV, probably thanks to the exponentially fast convergence of the OPE \cite{Pappadopulo:2012jk}. Extremal solutions then merely replace a complicated chaotic UV in the exact CFT correlator with a particularly simple set of GFF like states. This picture predicts on the one hand that for any fixed low energy observable we should expect to see exponential improvements in its determination as we include more four-point functions; but on the other that no matter how many correlators we bootstrap, their high energy OPE content will always be incorrect. Thus, arbitrarily small islands with respect to low energy CFT data can hide large variations in the possible behaviour of the high energy data.

The emergent picture seems to be that the bootstrap is a method for building a series of exact solutions to crossing which approximate actual CFT correlators. These exact solutions approximate the actual solution by successively incorporating more and more ``particle production'', i.e. the presence of more and more complicated operators in the OPE associated to many particle states. This seems to be literally the same mechanism at work in the S-matrix bootstrap, which is not surprising since in principle it follows from the CFT one. At some point including more four-point correlators presents diminishing returns as we hit the barrier of an exponentially large number of states in the OPE. It is thus important to find ways to characterize and approximate the structure of these states.

An important outcome of this paper is that CFT$_1$ correlators seem to be especially well adapted to describe local QFT physics in AdS. Technically this follows from a matching of the constraints of crossing on one hand and locality on the other \cite{Levine:2024wqn}. The tight control over extremal CFT data presents us with an opportunity to extract flat space physics of QFTs of interest from a CFT analysis. In this work we have focused on single correlators which allowed us to extract flat space S-matrices with no particle production. In the future it would be important and interesting to extend our analysis to mixed correlator systems. This would allow us to go one step further and bootstrap S-matrices that do contain particle production, without including it by hand, as in e.g. \cite{Antunes:2023irg,Tourkine:2023xtu}. More importantly they may illuminate how to directly tackle this problem in flat space.

\subsection*{Acknowledgements}
It is a pleasure to acknowledge discussions with Ant\'onio Antunes, Costas Bachas, Nat Levine, Marco Meineri, Ritam Sinha, Ning Su, Balt van Rees and Philine van Vliet. We are especially grateful to Elliott Morgensztern for developing the webapp demonstrating the construction of extremal solutions (\url{https://cft.starfree.app/}) and Bastien Girault for collaboration in the initial states of this project. 
This work was co-funded by the European Union (ERC, FUNBOOTS, project number 101043588). Views and opinions expressed are however those of the author(s) only and do not necessarily reflect those of the European Union or the European Research Council. Neither the European Union nor the granting authority can be held responsible for them. KG is supported by the Royal Society under the grant RF\textbackslash ERE\textbackslash231142.

\newpage 
\appendix

\section{Asymptotics of functional actions}
\label{app:asymp}
In this appendix we describe asymptotic formulae for functionals, building on previous work \cite{Mazac:2018ycv}. As in the main text, it is convenient to trade the '$n$' label on functionals by a new parameter $h$. This parameter is defined slightly differently for bosonic and fermionic bases:
\ba
B: \quad h=2\Df+2n\,, \qquad F:\quad h=1+2\Df+2n
\ea
With this notation we can describe the asymptotic of bosonic and fermionic basis simultaneously. 
\subsection{Large $\Delta,h$}
First, let us consider the limit of large $\Delta$ and $h$ with $\Delta/h$ fixed. We set
\ba
\omega_h(\Delta)=\frac{4 \sin^2\left[\frac \pi 2(\Delta-h)\right]}{\pi^2}\, \left(\frac{a_{h}^{\tt GFF}}{a_{\Delta}^{\tt GFF}}\right)\, R_{\omega}(\Delta,h)
\ea
with $\omega$ a functional. This functional can be either $\hat \beta$ or $\check \alpha$. The latter is defined as
\ba
\check \alpha_h= \hat \alpha_h-r_h \hat \beta_h
\ea
with $r_h$ chosen so that the large $\Delta$ behaviour of $\check \alpha_h$ is improved. In particular it can be determined from knowledge of the CFT data of two and four derivative contact terms, which may be bootstrapped by such functionals. Such contact terms were constructed in reference \cite{Knop:2022viy}. One simply finds
\ba
r_h^F&=\partial_h\left[\log(a_h^{\tt GFF} \gamma_h^{F,(2)})\right]\\
r_h^B&=\partial_h\left[\log(a_h^{\tt GFF} \gamma_h^{B,(4)})\right]
\ea
The full expressions are too unwieldy to give here, but we have
\ba
r_h=-\log(4)+\frac{1+8\Df}{2h}+\frac{3+16 \Df}{8h^2}+\ldots
\ea
These coefficients differ between bosonic and fermionic bases only at $O(h^{-5})$.
We then have:
\ba
R_{\hat\beta}(\Delta,h) &\underset{\Delta,h\to \infty}{\sim} \frac{4 h^2 \Delta}{(\Delta^4-h^4)}\,,&
\qquad R_{\check \alpha}(\Delta,h) &\underset{\Delta,h\to \infty}{\sim} \frac{16 h^{5}\Delta}{(\Delta^4-h^4)^2}
\ea
\subsection{Large $\Delta$, fixed $h$}
To determine the large $\Delta$ behaviour we can use the fact that subtracting a fixed functional with a correctly chosen coefficient we can improve this behaviour to get new functionals which bootstrap higher derivative contact terms. For instance, consider
\ba
\hat \beta_h- m_h \hat \beta_{h'}
\ea
for some fixed $h'$. If we choose $m_h$ appropriately this functional will have softer behaviour at large $\Delta$. Acting on a two derivative contact term we find
\ba
m_h \propto \ahf \gamma_h^{(2)}
\ea
Hence the dependence on $h$ of the large $\Delta$ behaviour of $\beta_h$ is given by $m_h$. We can fix the overall proportionality constant by matching with the large $h,\Delta$ result from before. A similar logic applies to the $\check \alpha$ functionals but now using six derivative contact terms. We get
\ba
R_{\hat{\beta}^F}(\Delta,h)&\underset{\Delta \to \infty}{=} \frac{B_h^F}{\Delta^3}:= \frac{M_{\beta^F}}{\Delta^3}\left(\frac{\gamma_h^{F,(2)}}{\gamma_{h_0}^{F,(2)}}\right)\,\\
R_{\hat{\beta}^B}(\Delta,h)&\underset{\Delta \to \infty}{=}\frac{B_h^B}{\Delta^3}:=\frac{M_{\beta^B}}{\Delta^3} \left(\frac{\gamma_h^{B,(4)}}{\gamma_{h_1}^{B,(4)}}\right)\, \\
 R_{\check \alpha^F}(\Delta,h)&\underset{\Delta \to \infty}{=}\frac{A_h^F}{\Delta^3}:=  \frac{M_{\check \alpha^F}}{\Delta^7}
\left( \frac{\partial_h\left(\ahf \gamma_h^{F,(6)}\right)-r_h^F\, \ahf \gamma_h^{F,(6)}}{\ahf \gamma_{h_1}^{F,(6)}}\right)\\
 R_{\check \alpha^B}(\Delta,h)&\underset{\Delta \to \infty}{=}\frac{A_h^B}{\Delta^3}:= \frac{M_{\check \alpha^B}}{\Delta^7}
\left( \frac{\partial_h\left(\ahf \gamma_h^{B,(8)}\right)-r_h^B\, \ahf \gamma_h^{B,(8)}}{\ahf \gamma_{h_2}^{B,(8)}}\right)
\ea
with
\ba
M_{\hat \beta^F}&= \frac{\pi  4^{1-\Delta _{\phi }} \Gamma \left(\Delta _{\phi
   }+\frac{3}{2}\right) \Gamma \left(2 \Delta _{\phi
   }+\frac{5}{2}\right)}{\Gamma \left(\Delta _{\phi }\right)
   \Gamma \left(\Delta _{\phi }+1\right) \Gamma \left(\Delta
   _{\phi }+2\right)}  \\
M_{\hat \beta^B}&= \frac{\pi  4^{-\Delta _{\phi }} \left(4 \Delta _{\phi }+1\right) \Gamma \left(\Delta _{\phi }+3\right) \Gamma \left(2 \Delta _{\phi
   }+\frac{5}{2}\right)}{\Gamma \left(\Delta _{\phi }+\frac{3}{2}\right){}^2 \Gamma \left(\Delta _{\phi }+\frac{5}{2}\right)} \\
M_{\check \alpha^F}&=\frac{6 \sqrt{\pi } \left(4 \Delta _{\phi }+3\right) \Gamma \left(\Delta _{\phi }+\frac{3}{2}\right) \Gamma \left(\Delta _{\phi
   }+\frac{9}{2}\right) \Gamma \left(2 \Delta _{\phi }+\frac{11}{2}\right)}{\Gamma \left(\Delta _{\phi }+3\right) \Gamma \left(\Delta _{\phi
   }+4\right) \Gamma \left(2 \Delta _{\phi }+1\right)}  \\
M_{\check \alpha^B}&=  \frac{3 \pi  2^{-2 \Delta _{\phi }-1} \left(\Delta _{\phi }+1\right) \left(4 \Delta _{\phi }+3\right) \left(4 \Delta _{\phi }+5\right) \Gamma
   \left(\Delta _{\phi }+6\right) \Gamma \left(2 \Delta _{\phi }+\frac{11}{2}\right)}{\Gamma \left(\Delta _{\phi }+\frac{5}{2}\right) \Gamma
   \left(\Delta _{\phi }+\frac{7}{2}\right) \Gamma \left(\Delta _{\phi }+\frac{9}{2}\right)}
\ea

\subsection{Large $h$, fixed $\Delta$}
 In this case we get:
\ba
\frac{\hat \beta_h(\Delta)}{a_h^{\tt GFF}}&=-c_0 \frac{\xi(\Delta)}{h^2}+
\left(\frac{\Gamma(2\Df)}{\Gamma(2\Df-\Delta)}\right)^2\, \frac{1}{h^{2\Delta}}\,\frac{\tan\left[\frac \pi 2\Delta\right]}\pi+\ldots\\
\frac{\check \alpha_h(\Delta)}{a_h^{\tt GFF}}&=4 c_0\frac{\xi(\Delta)}{h^3}-
\left(\frac{\Gamma(2\Df)}{\Gamma(2\Df-\Delta)}\right)^2\, \frac{1}{h^{2\Delta}}+\ldots\,
\ea
These formulae are only the leading terms in the large $h$ expansion. There are further corrections of the form $h^{-k}$ as well as $h^{-2\Delta-k}$. The coefficient $c_0$ is given by
\ba
B:\qquad c_0&=\frac{4^{4 \Delta _{\phi }-2} \Gamma \left(\Delta _{\phi }-\frac{1}{2}\right) \Gamma \left(\Delta _{\phi
   }+\frac{1}{2}\right){}^3}{2\pi ^2\Gamma \left(4 \Delta _{\phi }-2\right)}\\
F:\qquad c_0&=\frac{4^{\Delta _{\phi }-1} \left(1-2 \Delta _{\phi }\right){}^2 \Gamma \left(\Delta _{\phi }-1\right) \Gamma \left(\Delta _{\phi
   }\right){}^2}{\pi  \Gamma \left(\Delta _{\phi }-\frac{3}{2}\right) \Gamma \left(2 \Delta _{\phi }-\frac{1}{2}\right)}
\ea
Furthermore, $\xi^{B,F}(\Delta)$ can itself be identified with a functional action. More precisely, it is a kind of prefunctional, as it arises by acting with a functional kernel with insufficiently soft asymptotics, i.e., a `non-swappable' functional. We have
\ba
B:\quad \xi(\Delta)&:=\tilde \beta^B_0(\Delta)\\
F:\quad \xi(\Delta)&:=\tilde \alpha^F_{-1}(\Delta)
\ea
where these functionals satisfy:
\ba
\tilde \beta_0^B(\Delta_m^B)&=0\,,& \qquad \partial_\Delta \tilde \beta_0^B(\Delta_m^B)&=\delta_{0,m}\\
\tilde \alpha_{-1}^F(\Delta_m^F)&=\partial_{\Delta} \tilde \alpha_{-1}^F(\Delta_m^F)=0\,,& \qquad \tilde \alpha_{-1}^{F}(\Delta_{-1}^F)&=1
\ea
Explicit expressions for these functionals are given in a {\tt Mathematica} notebook attached to this paper's arXiv submission. They can be derived by exploiting the relation between the kernels of these functionals to the kernels of known ordinary bosonic and fermionic functionals. We have
\ba
\tilde \beta_0^B(\Delta|\Df)=\sum_{p=0}^2 c_p(\Delta) \hat \beta_0^F(\Delta-2+2p|\Df-\tfrac 32)
\ea
with
\ba
c_0(\Delta)&=1\,,\\
c_1(\Delta)&=-\frac{\Delta^2-\Delta-1}{2(2\Delta-3)(1+2\Delta)}\,,\\
c_2(\Delta)&=\frac{\Delta^4+2\Delta^3+\Delta^2}{16(2\Delta-1)(1+2\Delta)^2(3+2\Delta)}\,.
\ea
Similarly,
\ba
\tilde \alpha_{-1}^F(\Delta|\Df)=N(\Df)\left[-\hat \beta_0^F(\Delta|\Df)+\sum_{p=0}^2 d_p(\Delta) \hat \beta_0^F(\Delta-2+2p|\Df-1)\right]
\ea
with
\ba
N(\Df)&=\frac{16\,\Df(\Df-1)}{3-4 \left(\Delta _{\phi }-1\right) \Delta _{\phi }
   \left(4 \Delta _{\phi }-5\right)}\,,\\
d_0(\Delta)&=1\,,\\
d_1(\Delta)&=\frac{7\Delta^2-7\Delta-5}{2(2\Delta-3)(2\Delta+1)}\,\\
d_2(\Delta)&=\frac{\Delta^2(1+\Delta)^2}{16(2\Delta-1)(2\Delta+1)^2(2\Delta+3)}\,.
\ea
\subsection{Analytic results for $\gamma_n$}
\label{app:analg}
In this section we describe how to obtain analytic results for the anomalous dimensions in the UV. The hybrid method equations to second order are given by:
\ba
a_h\gamma^{(1)}_h&=-\hat \beta_h[\tL]\\
a_h\gamma^{(2)}_h&=-\sum_{\Delta>\Delta^*}^\infty a_{\Delta} \left[\gamma_\Delta^{(1)}\right]^2 \partial_\Delta^2 \beta_h(\Delta)\\
\ea
Using the asymptotic expressions for the functional actions we get
\ba
\gamma_h^{(1)}=-\frac{\hat \beta_h[\tL]}{a_h^{\tt gff}}&\underset{h\to \infty}\sim \frac{c_0}{h^2}\,\sum_{\Delta\geq 0}^{\Delta^*} a_{\Delta} \xi(\Delta)+\Sigma_{\Delta\leq 1}
\ea
with
\ba
\Sigma_{\Delta\leq 1}:= \sum_{\Delta\leq 1}\, \frac{1}{h^{2\Delta}}\,\left(\frac{\Gamma(2\Df)}{\Gamma(2\Df-\Delta)}\right)^2\,\frac{\tan\left[\frac \pi 2\Delta\right]}\pi
\ea
Thus $\gamma_h^{(1)}$ decays as $h^{2}$ generically, or as $h^{-2\Delta_1}$ with $\Delta_1$ the lowest dimension in the spectrum if it is below unity.
To second order we get:
\ba
\gamma_h^{(2)}&=-\frac 12 \dashint_{\Delta^*}^\infty \ud \Delta \left[\gamma_\Delta^{(1)}\right ]^2 \frac{4\Delta h^2}{\Delta^4-h^4}
\ea
As long as $\Delta_1$ is above $1/2$ we can take the large $h$ limit inside the integral, leading to a term $O(h^{-2})$ which amounts to $\sum_{\Delta\geq \Delta^*} a_\Delta  \xi(\Delta)$. The leftover piece is then $O(h^{-4\Delta_1})$, or $O(h^{-6})$ if $\Delta_1>1$. Thus we find
\ba
\gamma_h^{(1)}+\gamma_h^{(2)}=\frac{c_0}{h^2}\xi[\tL+\tH^{(1)}]+\Sigma_{\Delta\leq 1}
\ea
is correct up to $O(h^{-2})$.

\section{Central charge near free fermion}
\label{app:ce}

In this appendix we show  that for $\Df<1/2$ crossing and locality are consistent with extremal solutions supporting a bulk stress tensor with $C_T<1/2$ for $\hat \D$ sufficiently close to $2\Df+1$. To see this we consider a fermionic type solution with
\ba
\tS=\left\{\mathds 1,(\hat\D=\D_0^F-\epsilon,\hat a)\right\}
\ea
i.e. a generalized free fermion coupled to an extra state. To leading order in $\epsilon$, crossing is solved by setting 
\ba
\g_0 = \frac{\hat a}{a_0}\epsilon\,,\quad \delta a_0 =-\hat a
\ea
with all remaining CFT data unchanged. Here $a_n\equiv a_n^F$ and $\gamma_n\equiv \gamma_n^F\equiv \Delta_n-\Delta_n^F$. One can think of this solution as taking the generalized free fermion and splitting the first operator into two, one slightly below $2\Df+1$, and one slightly above, with their OPE coefficients adding up to the original GFF value $a_0 = 2\Df$.

We now use locality equations to solve for all $b_n$ except $\hat b$, using $\a=\Df+1/2$ and $\tilde{\a} = \Df+3/2$ (we will keep these implicit for readibility). The result is:
\ba
&\delta b_0=-\hat b + \frac{\k^{'}_0}{\k_0}\,x\, \hat a \epsilon\\
&\delta b_n= -\frac{\k^{'}_0\theta_{n,0}-\k_0\theta^{'}_{n,0}}{\k_0}\,x\, \hat a\epsilon
\ea

Where we introduced the variable $x:=\frac{\hat b}{\hat a}-\frac{b_0^{\tt gff}}{a^{\tt gff}_0}$. The central charge is a second degree polynomial in $x$:
\ba
\delta C_T=\hat a g_0x^2+\hat a \epsilon (g_1x^2+g_2x+g_3) 
\ea
To leading order in $a_0$ and $\epsilon$, the $g_i$ read as:
\ba
&g_0 = 1\\
&g_1 = -\nu^{'}_0\\
&g_2 =\frac{2b_0^{\tt gff}}{a^{\tt gff}_0}\frac{\nu_0\k^{'}_0-\nu^{'}_0\k_0}{\k_0} -\frac{2}{\k_0}\sum_{n=1}^{\infty}\frac{b^{\tt gff}_n}{a^{\tt gff}_n}\nu(\hat{\D}_n)(\k^{'}_0\theta_{n,0}-\k_0\theta^{'}_{n,0})\\
&g_3 = -\left(\frac{b^{\tt gff}_0}{a^{\tt gff}_0}\right)^2\frac{\nu_0\k^{'}_0}{\k_0}\\
\ea
We can minimize the central charge varying $x$, to find
\ba
C_T^{\mbox{\tiny min}}=\frac{1}{2}+\hat a (\epsilon g_3- \frac{\epsilon^2g_2^2}{4(g_0+\epsilon g_1)})=\frac{1}{2}+\hat a \epsilon g_3
\ea
$g_3$ is positive for $\Df>1/2$ and negative for $\Df<1/2$. Consequently, in the latter case, it is possible to obtain a central charge smaller than $1/2$ by coupling the fermion to an extra state, sufficiently close to $2\Df+1$.

\section{Majorana fermions in $AdS_2$}
\label{app:fermion}
In this section, we summarize the form taken by free correlators for massive majorana fermions in $AdS_2$. These are not new results and have already been derived in other works \cite{Giombi_2022}. We use the Poincaré patch coordinates $(u,x)$, $u\geq 0$, with the metric given by
\ba
\ud s^2=\frac{\ud u^2+\ud x^2}{u^2}
\ea
and define:
\ba
\xi(X_1,X_2) \equiv (X_1-X_2)^2 = \frac{(x_1-x_2)^2+(u_1-u_2)^2}{u_1u_2}
\ea
We introduce the lightcone coordinates $z,\bar{z}=u\pm ix$, and the corresponding derivatives $\partial \equiv \partial_x-i\partial_u$, $\bar{\partial} \equiv \partial_x+i\partial_u$. The action for a pair of massive Majorana fermions in $AdS_2$ reads as:

\ba
S=-\frac{1}{2\pi}\int_{AdS_2} \frac{\ud u\ud x}{u^2}\left(u\bar{\psi}\partial \bar{\psi}+u\psi\bar{\partial}\psi+2m\bar{\psi}\psi\right)\,,\qquad \psi^* = \bar{\psi}
\ea
Boundary condition have to be imposed at $u=0$. There are two independent choices, corresponding to
\ba
\psi(0,x) = \pm i\bar{\psi}(0,x)
\ea
The basis of gamma matrices we take is $
\gamma_0 = \sigma_2\,, \gamma_1=\sigma_1$. 
We introduce the two point function $G(X_1,X_2)$:
\ba
G(X_1,X_2) = \begin{pmatrix}
    \langle \psi(X_1)\bar{\psi}(X_2)\rangle && \langle \psi(X_2)\psi(X_1)\rangle \\
    \langle \bar{\psi}(X_1)\bar{\psi}(X_2)\rangle&& \langle\psi(X_2) \bar{\psi}(X_1)\rangle
\end{pmatrix}
\ea
It verifies the equation of motion:
\ba
(\gamma \cdot \nabla+m)G(X_1,X_2) =(u\gamma_0\partial_u+u\gamma_1\partial_x-\frac{1}{2}\gamma_0 +m)G(X_1,X_2)= -\delta(X_1,X_2)
\ea

The solution can be expressed in terms of two hypergeometric functions:
\ba
G(X_1,X_2)=-N_{\D} \begin{pmatrix}
            \frac{-u_1-u_2-i(x_1-x_2)}{\sqrt{u_1u_2}}\alpha(\xi)&& \frac{x_1-x_2-i(u_1-u_2)}{\sqrt{u_1u_2}}\beta(\xi)\\
            \frac{x_1-x_2+i(u_1-u_2)}{\sqrt{u_1u_2}}\beta(\xi)&&\frac{-u_1-u_2+i(x_1-x_2)}{\sqrt{u_1u_2}}\alpha(\xi)
        \end{pmatrix}
\ea
where
\ba
 \alpha(\xi) &= -\text{sign}(m)\frac{1}{\xi^{\D-1/2}(\xi+4)}~ _2F_1\left(\D-\frac{1}{2},\D-\frac{1}{2},2\D,-\frac{4}{\xi}\right)\\
        \beta(\xi) &= \frac{1}{\xi^{\D+1/2}}~ _2F_1\left(\D-\frac{1}{2},\D+\frac{1}{2},2\D,-\frac{4}{\xi}\right)
\ea
and we set
\ba
\D=|m|+\frac{1}{2}\,, \qquad N_{\D} = \frac{4 ^{-\D}\Gamma(2\D)}{\Gamma(\D)^2 }.
\ea\newpage
The trace of the stress tensor is $\Theta=\frac{2m}{\pi}:\bar{\psi}\psi:$. Its two-point function can be obtained from $G$:
\ba
&\langle\Theta(X_1)\Theta(X_2)\rangle = -\frac{4m^2}{\pi^2}\text{det}(G(X_1,X_2)) \\
&= \frac{4m^2}{\pi^2}N_{\D}^2\left(\frac{1}{\xi^{2\D}}~_2F_1\left(\D-\frac{1}{2},\D+\frac{1}{2},2\D,-\frac{4}{\xi}\right)^2-\frac{1}{\xi^{2\D-1}(\xi+4)}~ _2F_1\left(\D-\frac{1}{2},\D-\frac{1}{2},2\D,-\frac{4}{\xi}\right)^2\right)
\ea

Finally, we compute the form factor between a bulk field $\Theta(X_1)$ and two boundary fermions $\hat{\psi}_{2,3}(x_{2,3})$, defined as
\ba
\hat{\psi}_{\pm}(x) = \lim_{u\to 0}u^{-\D}\frac{\psi(u,x)\pm i\bar{\psi}(u,x)}{2N_{\D}^{1/2}}
\ea
Their two point function read as: 
\ba
\langle \hat{\psi}_{\pm}(x_1)\hat{\psi}_{\pm}(x_2)\rangle =\frac{x_1-x_2}{(x_{12}^2)^{\D+1/2}}\frac{1\pm \text{sign}(m)}{2}
\ea
Then the form factor is
\ba
\langle\Theta(X_1)\hat{\psi}_{\pm}(x_2)\hat{\psi}_{\pm}(x_3)\rangle = \frac{2|m|}{\pi}N_{\D}(x_3-x_2)\left(\frac{w}{x_{23}^2}\right)^{\D+1/2}\frac{1\pm \text{sign}(m)}{2}
\ea

\section{Single zeros functionals}
\label{app:single}

We summarize briefly the properties of the single zero functionals, denoted as $\ta_n$. They are constructed through the functional kernel $f_n(z) = z^{-n-1}$, with $n\geq 1$ integer. Conveniently they are known in closed form for any $\Df$ \cite{Paulos:2019gtx} :
\ba
\ta_n(\D)=&\frac{\sin\left[\pi(\D-h_n)\right]}{\pi}\Bigg[\frac{\G(2\D)\G(h_n)^2\G(\D-h_n)}{\G(\D)^2\G(\D+h_n)}\\
&+(-1)^{n+1}\frac{\G(h_n)\G(\D-2\Df+1)}{\G(\D+n+1)}  \, _3F_2\left(\begin{aligned}\D,\D&,\D-2\Df+1\\
2\D&,\D+n+1\end{aligned};1\right)\Bigg]
\ea
where $h_n = 2\Df+n$. They are dual to both generalized free fermion and boson in the following sense:
\ba
\ta_n(\hat\D_m)=\delta_{m,n}\quad m\geq n
\ea
A orthonormal basis verifying $\tilde{\tau}_n(\hat\D_m)=\delta_{m,n}$ for all $m\geq 1$ can be constructed with the usual Gram-Schmidt algorithm. Although no closed form is known for generic $\Df$, it is possible to find one for any given $2\Df$ integer. Following the conventions of Appendix \ref{app:asymp}, the $\tilde{\ta}_n$ functional have the following large $\D$ and $h$ behavior:
\ba
&\tilde{\ta}_{n}(\D) = \frac{\sin\left[\pi(\D-h_n)\right]}{\pi}\left(\frac{a_{h_n}^{\tt GFF}}{a_{\Delta}^{\tt GFF}}\right)\, R_{\ta}(\Delta,h_n)\\
& R_{\tau}(\Delta,h) \underset{\Delta,h\to \infty}{\sim} \frac{4 h^4}{\D(\Delta^4-h^4)}
\ea
Similarly to section \ref{sec:smatext}, one can write down crossing as integral equations. The relation involving the difference of two consecutive $\tilde{\ta}_n$ yields $\rho(s) = 1$. Then the analog of (\ref{eq:functionaldisp}) is:
\ba
\sum_{\Delta} \frac{a_{\Delta}}{a_{h}^{\tt GFF}} \tilde{\tau}_h(\Delta)=0 \Leftrightarrow \frac{1}{\pi}\,\dashint_{0}^\infty \ud s \frac{2 s_h^2}{s(s^2-s_h^2)} \mbox{Im}\big(S(s)\big)=\mbox{Re}\big(S(s_h)-S(0)\big)
\ea
which is equivalent to (\ref{eq:dispS}).

\newpage

\bibliography{bib}
\bibliographystyle{JHEP}

	\end{document}